\documentclass[aps,pre,twocolumn,groupedaddress,showpacs,showkeys]{revtex4}

\pdfoutput=1

\usepackage[latin1]{inputenc}

\usepackage{graphicx}
\begin{document}

\title{Credit risk --- a structural model with jumps and correlations}


\author{Rudi Schäfer}
\email[]{Rudi.Schaefer@matfys.lth.se}
\author{Markus Sjölin}
\author{Andreas Sundin}
\author{Michal Wolanski}
\author{Thomas Guhr}
\email[]{Thomas.Guhr@matfys.lth.se}

%\thanks{}

\affiliation{Department of Mathematical Physics, LTH, Lund University, Sweden}

\date{\today}

\begin{abstract}
We set up a structural model to study credit risk for a portfolio
containing several or many credit contracts. The model is based on a
jump--diffusion process for the risk factors, i.e.~for the company
assets. We also include correlations between the companies. We discuss
that models of this type have much in common with other problems in
statistical physics and in the theory of complex systems.  We study a
simplified version of our model analytically. Furthermore, we perform
extensive numerical simulations for the full model. The observables
are the loss distribution of the credit portfolio, its moments and
other quantities derived thereof.  We compile detailed information
about the parameter dependence of these observables. In the course of
setting up and analyzing our model, we also give a review of credit
risk modeling for a physics audience.
\end{abstract}



\maketitle

\section{Introduction}

Economics attracts the interest of a quickly growing community in
physics. A large part of the research addresses the financial
markets. Attempts are being made to better understand various
phenomena such as the fat tails of the stock price distributions by
relating them to physics systems, see
Refs.~\cite{mant00,bouch00,voit01} and references therein. Physicists
have also joined the activities of economists and computer scientists
in agent--based models~\cite{axel01,laff01}, and applied their
long--standing experience in complex systems, see Ref.~\cite{schw02}.

As far as the field of finance is concerned, the vast majority of
studies put forward by physicists has been devoted to market risk.
The market risk is due to the unknown time evolution of the asset
prices. In general, one is faced with a large spectrum of different
risk types. One also distinguishes, for example, operational risk (due
to failure of internal systems), political risk (due to political
decisions that affect the capital markets) and legal risk (due to
fraud and discontinued contracts). In this contribution, we address
credit risk. It is due to the failure of a counterpart to make a
promised payment. At present, risk managers and researchers are more
familiar with market risk than with any of the other risk types and
the corresponding mathematical description is highly developed. It is
of considerable practical interest to improve the knowledge about and
the modeling of the other risk types.  In the case of credit risk, the
probability that a promised payment is not made is usually small and
difficult to estimate. Nevertheless the amount of money involved and
thus the associated loss can be enormous and even jeopardize the
existence of the financial institution which issued the credit.

Only recently, physicists started applying their specific tools to
credit risk~\cite{ros04,mol05,kitsukawa06,kuehn03,neu04,hatchett06}.  
The interesting point for
practitioners and researchers, especially statistical physicists, is
the highly asymmetric form of the loss distribution and the resulting
peculiar features.  This distinguishes credit risk from market risk,
although the former clearly depends on the latter.  In investment
theory the standard deviation, referred to as volatility, of the
relative asset price change is taken as a measure of how risky a
certain investment is. If more uncertainty is incorporated in the
investment, i.e.~if the volatility is larger, then the demanded
earnings are higher. Due to this fact, investors are traditionally
risk averse. In other words, a potential loss is considered to be more
punitive than a potential gain is beneficial, even if they are equally
probable and large.  The asymmetric character of the loss
distributions makes risk measures other than the volatility also
important in credit risk management.

In this study, we set up and analyze a structural model for credit
risk, based on a jump-diffusion process for the risk factors. Our study is related to, 
but different from the work in Ref.~\cite{zhou01}.  
These models are particularly appealing to physicists, because
their starting point is, in physics terminology, microscopic and
dynamical.  This gives them a rather general character which makes
them also suited for other problems in physics and in the theory of
complex systems.  With this contribution, we pursue two goals. First,
we systematically explore the interplay between the different
parameters of our structural model, particularly the role of
leverage, jumps and correlations. 
In contrast to the existing literature, our main focus is on the full loss distribution 
of the credit portfolio, 
its moments and its tail behavior.
Second, we review credit risk
modeling and keep the whole presentation pedagogical, because we want
to make this topic more accessible to the physics audience.

The paper is organized as follows. We review the present status of
credit risk modeling in Sec.~\ref{sec:crm}. In Sec.~\ref{sec:ourmodel}
we introduce our model. We discuss a simplified version of it
analytically in Sec.~\ref{sec:analytical} and the full model
numerically in Sec.~\ref{sec:numerics}. Summary and conclusion are
given in Sec.~\ref{sec:sumcon}.

\section{Credit risk modeling}
\label{sec:crm}

After defining debt instruments in Sec.~\ref{crm1}, we discuss what
one means by default and credit ratings in Secs.~\ref{crm2}
and~\ref{crm3}, respectively. The credit risk measures are introduced
in Sec.~\ref{sec:measures}. The impact of correlations is discussed in
Sec.~\ref{crm5}. The r\^ole of the probability density function for
credit losses is explained in Sec.~\ref{crm6}. We conclude this short
review by presenting the most important credit risk models in
Sec.~\ref{crm7}. Reviews in the financial literature on credit risk
modeling can be found in Refs.~\cite{bluhm02,giesecke02}

\subsection{Debt instruments}
\label{crm1}

\begin{figure}
\resizebox{75mm}{!}{\includegraphics{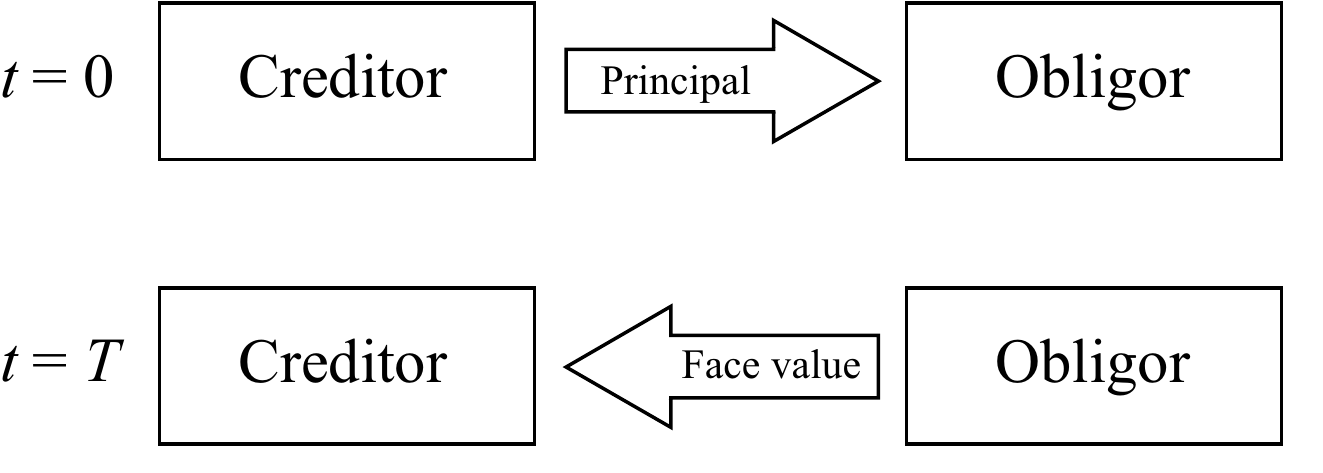}}
\caption{\label{fig1} The cash flow of a zero--coupon bond.}
\end{figure}

A debt instrument is simply a written promise to repay a debt.  There
is a wide range of such contracts. A debt instrument has two
positions: a lending side (the creditor) and a borrowing side (the
obligor). Bonds are very common. A bond is issued for a period of one
year or more with the purpose of raising capital by borrowing. The
government, states, cities, corporations, and many other types of
institutions can sell bonds. Generally, a bond is a promise to repay
the principal along with interest (coupons) on a specified date
(maturity). The principal is the amount borrowed or the part of the
amount borrowed which remains unpaid (excluding interest). Some bonds
do not pay interest.  In this study we focus on the {\it zero--coupon
bond}. The cash flow of the zero--coupon bond is limited to two dates:
the date of issue $t = 0$ and maturity $t = T$. At the issue date the
creditor lends a specified amount of money to the obligor. At
maturity, the obligor has to repay the face value of the bond. The
face value is the amount borrowed plus interest and additional yield
compensating for the risk. Fig.~\ref{fig1} shows the cash flow of the
zero--coupon bond.

\subsection{Default event}
\label{crm2}

The key issue which separates credit risk from, for example, market
risk, is the concept of default. Its definition is not homogeneous
throughout the industry of credit risk management. Usually default
means that there has been a missed or delayed payment of interest
and/or principal within a grace period, or that an obligor files for
bankruptcy~\cite{moodys02}.  Although default is a truly rare event,
creditors can lose large amounts of money.  A good example is the
bankruptcy of {\it Enron Corporation} in Dec.\ 2, 2001.  At an asset
value of US \$49.53 billion, this was the largest bankruptcy filing in
US history to that date~\cite{allen02}. The actual loss for the
creditors was US \$9.9 billion~\cite{moodys02}.

\subsection{Credit ratings}
\label{crm3}

A way to quantify credit risk is to determine the credit worthiness of
a potential customer from the historical performance of the obligors.
There is a wide range of rating systems for credit worthiness, systems
used internally by the credit institute as well as external ratings
available for public. The credit rating of a company is often directly
linked to the probability that the company defaults within a fixed
time horizon, usually one year. Two frequently used rating systems are
Standard \& Poor's and Moody's.  The probability that a company
changes its credit rating is expressed in terms of a rating transition
matrix which contains the probabilities that a company with a certain
rating migrates to another category, usually within a year.  The
credit rating transition matrix is based on the historical migration
frequencies of corporate bonds. It is observed that the most probable
future event is that the company remains in the same rating
category. This is valid for all rating categories. Moreover, the
probability of a downgrade is generally higher than the probability of
an upgrade \cite{jpmorgan97}.

\subsection{Credit risk observables}
\label{sec:measures}

There are several ways of quantifying credit risk. We distinguish between standalone risk and portfolio risk~\cite{kmv01}. 
The most frequently used standalone observables are the 
default probability (DP), the loss given default
(LGD) and the migration risk. The conventional portfolio risk observables
are default correlations and exposure, i.e.~the size, or proportion,
of a credit portfolio exposed to default risk.

\subsection{Correlations}
\label{crm5}

The credit worthiness of obligors often involves mutual
correlations. For example, defaults are more frequent in times of
regression in the surrounding economy. Furthermore, one can see that
companies in the same country and/or industry can affect each others
rating migrations, up as well as down~\cite{bluhm02}.  There are
different ways to incorporate correlations between default events of
obligors. For example, one uses the correlations between the equity
values of the companies, i.e.~the stocks, to describe the dependency
of credit migrations. Another way to examine this is to look at how
the obligors depend on the current state of the economy~\cite{bis99}.
Modeling correlations is often difficult because only limited data are
available for the indicators chosen. Moreover, because of the huge
number of correlations involved in a normal sized credit portfolio
(which can contain a few thousand bonds), it is necessary to make
simplified assumptions. A common procedure is to categorize the
companies into different groups or branches with specified
correlations. For example, one can assume correlations which are
country--specific, industry--specific etc. Even though these
simplifications lead to a more manageable model, it is still a
complicated task to decide the structure of the categorization and to
estimate the group--specific correlations.

\subsection{Probability density function of credit losses}
\label{crm6}

\begin{figure}
\resizebox{75mm}{!}{\includegraphics{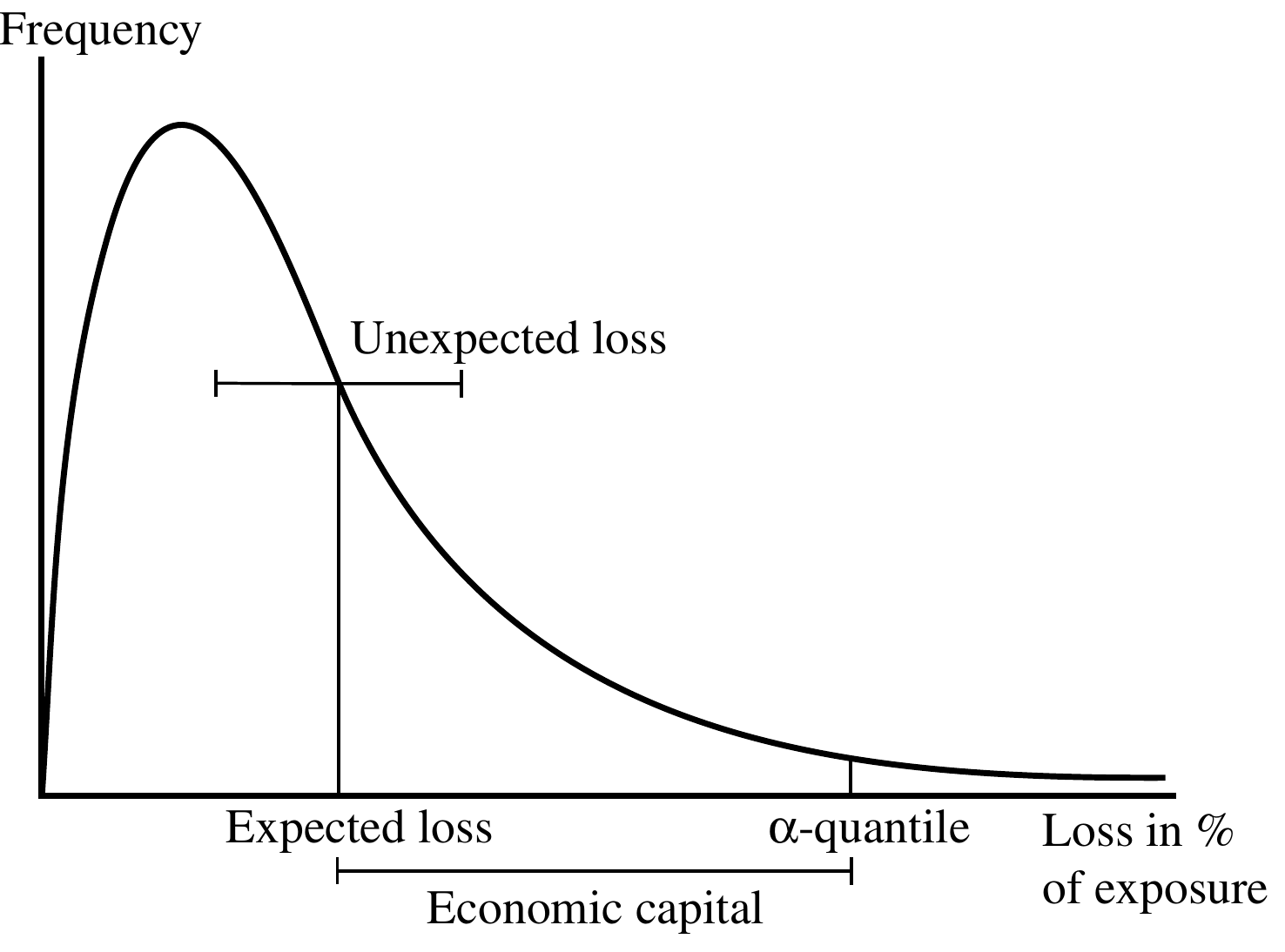}}
\caption{\label{fig2} A schematic loss distribution with the expected
loss EL, the unexpected loss UL and the economic capital EC.}
\end{figure}

The primary output of a credit risk model is the probability density
function (PDF) of credit losses for a given portfolio. Adapting to the
more common physics terminology, we refer to this function as the
loss {\it distribution}. Here we make the assumption that the loss is a 
continuous random variable, so that we can work directly with the PDF. 
From the loss distribution one determines
the expected loss EL, the unexpected loss UL, the required risk
capital and further quantities. The expected loss EL is the mean of
the loss distribution and the unexpected loss UL is its standard
deviation. It is important to notice that UL, not EL, measures risk.
However, to cover potential losses it is not enough to have a
``cushion'' within the standard deviation. The probability that losses
exceed the UL is significant and it is therefore necessary to have
another measure of risk capital. To quantify risk capital one usually
uses the economic capital EC which is also known as Capital at Risk
CaR or as the Value at Risk VaR. The economic capital EC is defined as
the difference between the expected loss EL and the $\alpha$--quantile
for a certain level of confidence.  Figure~\ref{fig2} shows a typical
loss distribution with EL, UL and EC.  The general appearance of such
a loss distribution is different from the distributions in, for
example, market risk.  While the distributions of market risk are
typically Gaussian, the distributions generated by credit risk are
{\it skewed} and {\it leptokurtic}~\cite{bluhm02}.

To measure the important tail behavior, one uses the kurtosis
excess. We recall that the kurtosis $\beta_2$ is defined as the fourth
central moment divided by the squared second central moment.  The
kurtosis excess $\gamma_2$ is the part of the kurtosis that exceeds
the kurtosis of the normal distribution (which is equal to 3),
i.e.~$\gamma_2 =\beta_2 - 3$.  A distribution is often referred to as
fat--tailed if it is leptokurtic, i.e.\ $\gamma_2 > 0$. Fat--tailed
distributions have higher quantiles than a normal distribution and
thus require more attention from a credit risk manager.

To generate a loss distribution of a credit portfolio, one can evaluate it 
for a structural model either analytically (in some simple cases) or numerically by 
Monte Carlo simulation. It is also possible to approximate an actual 
portfolio distribution by some
known analytical distribution. In the latter case, one maps the actual
portfolio with unknown distribution to an equivalent portfolio with a
known distribution. A frequently used distribution which has the right
features, i.e.~skewed and fat--tailed, is the beta
distribution~\cite{bluhm02}.
A better and more sophisticated approach than parametric approximations 
is offered by reduced form models. Some of these models lead to analytically 
tractable loss distributions, see e.g.~Ref.~\cite{errais07} and references therein.

\subsection{Credit risk models}
\label{crm7}

% Structural versus reduced form models
Current credit risk models can be divided into two main categories:
 structural models and  reduced form models. 

In the {\it structural models} one makes assumptions about the time
evolution of the risk factors, i.e.~mostly the asset or stock prices
of the companies, as well as about the liabilities. Whenever the asset
value falls below some specified threshold, like the book value of the
liabilities, the firm defaults. The structural credit risk modeling
approach has its roots in the Black and Scholes theory for option
pricing, and the Merton model, see Ref.~\cite{bluhm02}. The Black and
Scholes theory is based on the assumption of a friction--less market
where the stock or asset price is described by a geometric Brownian
motion. Merton viewed the equity value of a company as a call option
of the asset value with a strike price equal to the face value $F$ of
the debts.  It is assumed that the company has a certain amount of
zero--coupon debt due at a future time $T$. In consequence, it is
possible to apply the whole Black and Scholes machinery to the credit
risk modeling problem~\cite{bluhm02}. This is the ``microscopic''
viewpoint which makes structural models suitable for physics
approaches.

The {\it reduced form models} for credit risk are based on the
assumption of a functional relationship between the obligors' expected
default rates and different background factors. These background
factors may represent either observable or unobservable variables.
Observable variables typically depend on the general state of the
economy, and the unobservable variables often represent some random
risk factors \cite{bis99}. The event of default is often modeled by an
intensity process, e.g.~a Poisson process.  Unlike the structural
approach, the reduced form approach is not directly based on a
dynamical description of the economy. Reduced form models are often
implemented as black--box models, where the accuracy of the model
outcome is more important than an intuitive economical interpretation
of the mechanisms included in the black box.

% Default mode versus mark--to--market
One of the key issues when setting up a credit risk model is to define
an event that leads to an actual loss. Usually, credit risk modelers
use either of two definitions of credit loss:  the default mode
paradigm, or the mark--to--market paradigm. Within the {\it default mode}
(DM) paradigm, a credit loss occurs only when an obligor defaults
within the bonds maturity time. Within the {\it mark--to--market} (MTM)
paradigm a credit loss can occur without an actual default. The
creditor can lose money whenever an obligor's credit worthiness
deteriorates.  Both of these two modeling approaches are common in
current vendor credit risk frameworks. Typical arguments for using one
of them are the simplicity of the DM model and the multi--state nature
of the MTM model. The DM model suits better for creditors which only
are interested in a buy--and--hold portfolio, while the MTM model is
more adequate for pricing decisions of more liquid credits
\cite{bis99}.

% Industry models
There is a wide range of different credit risk management tools
available in the financial industry. Many of those models seek to
estimate the full distribution to be able to calculate different
statistics measures~\cite{bis99}. Examples of vendor credit risk
models are CreditMetrics (by RiskMetrics Group), PortfolioManager (by
KMV), CreditPortfolioView (by McKinsey \& Co), and CreditRisk$^+$ (by
Credit Suisse Financial Products), see Ref.~\cite{giese03} and a
review in Ref.~\cite{crou00}.

\section{A model with jumps and correlations}
\label{sec:ourmodel}

We set up our structural credit risk model in Sec.~\ref{ourmodel1}. In
Sec.~\ref{ourmodel2}, we discuss its generality by relating it to a
few other scenarios where structural models can find application.

\subsection{Setup of the model}
\label{ourmodel1}

% The jump diffusion process
We model the time evolution of the asset value of every company 
by a stochastic differential equation of the form
\begin{equation} \label{eq:jumpdiff}
\frac{dV}{V} = \mu dt + \sigma \varepsilon \sqrt{dt} + dJ \ .
\end{equation}
Apart from the jump term $dJ$, this is a geometric Brownian motion
with a deterministic term $\mu dt$ describing the exponential growth
of the asset value and a stochastic term $dW= \varepsilon \sqrt{dt}$
representing the fluctuations as a Wiener process. 
Here, $\mu$ is the drift, $\sigma$ the volatility (constant) 
and $\varepsilon$ an independently distributed
random number in each time step. We add the jump term $dJ$ which is
not contained in Merton's original model. 
%%%
In previous works, jump-diffusion models have been considered for stock returns 
(see e.g.~Ref.~\cite{hanson02b}) and also for credit risk~\cite{zhang03,zhou01,duffie01,hilberink02}. 
%%%
The economical
interpretation of the jump term is that a major setback of an asset
value is possible at any time. These setbacks are larger than the
volatility admits and may be explained by events labeled as crises,
originating from legal, operational, political or other external or
internal factors.  The probability that an economical setback takes
place during the lifetime of a bond is typically very small. Well
known examples of such economical setbacks for a large group of
companies are the great stock market crash of 1929, the oil crisis of
the mid--seventies and the ``Black Monday'' crash in 1987.  We model
the jumps by a {\it Poisson process} with intensity $\lambda$. We
recall that, in such a process, the probability function for the event
to occur $n$ times between zero and the time $t$ is given by
\begin{equation}\label{pp}
p_n^{{\rm Poisson}}(t) = 
  \frac{(\lambda t)^n}{n!}\exp\left(-\lambda t\right) \ .
\end{equation}

\begin{figure}
\resizebox{65mm}{!}{\includegraphics{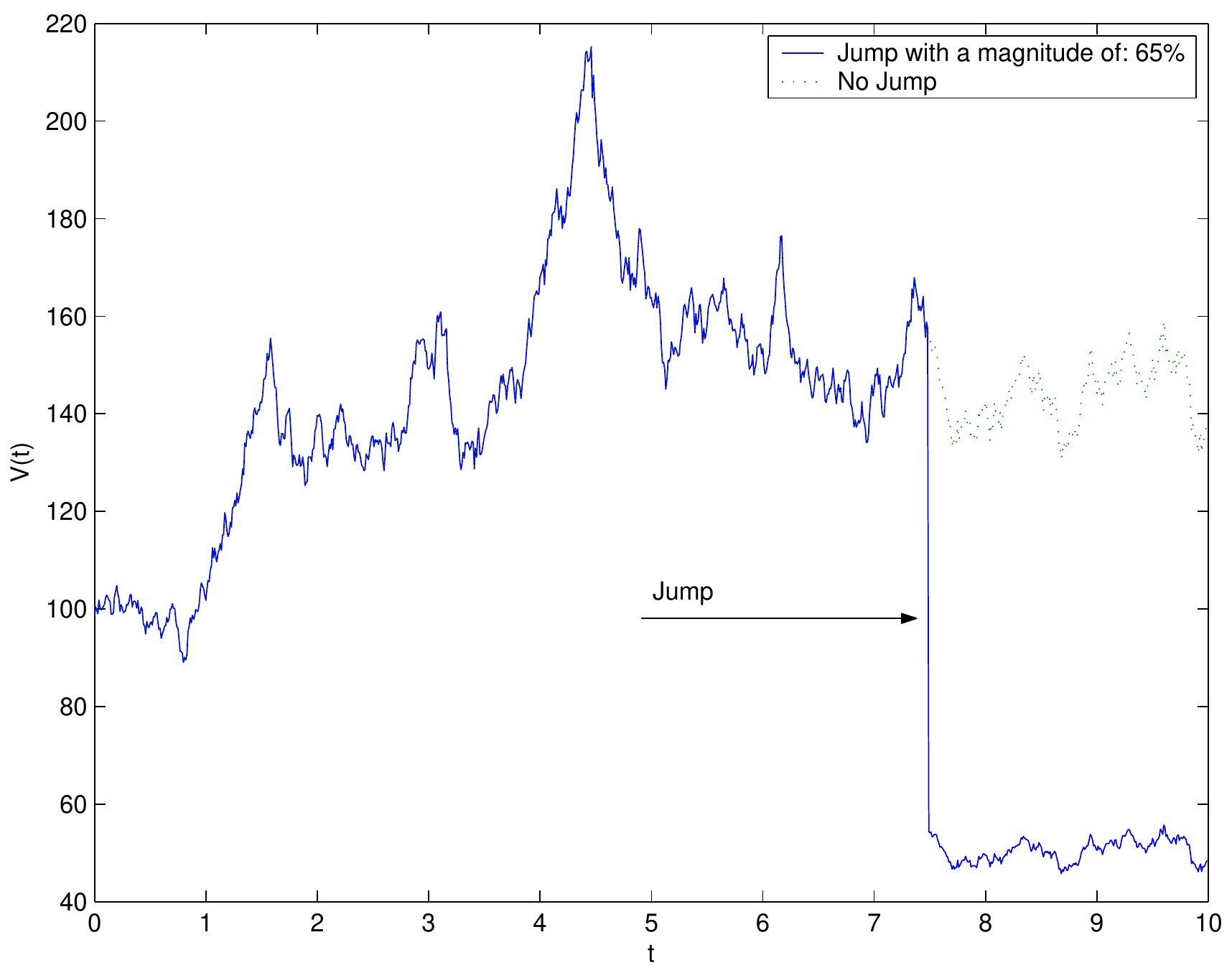}}

\caption{\label{fig3} Two time series for the asset value $V(t)$ versus
time $t$. They are equivalent except that one of them includes a large
negative jump given by the jump term $J(t)$. The lower
time series is damped because the factor $V$ in
Eq.~(\ref{eq:jumpdiff}) is much smaller after the jump.}
\end{figure}

The size $\Lambda$ of the jump, measured in units of the current asset
value $V(t)$, is a random variable with a distribution which we have
to specify.  Jumps can be positive or negative. The largest possible
negative jump is 100\% of the current asset value. Based on this
information, a possible distribution of the jump size $\Lambda$ is a
shifted lognormal distribution, $\Lambda + 1 \sim {\rm
LN}(\mu_J+1,\sigma_J)$, with mean $\mu_J$ and standard deviation
$\sigma_J$. A time series for the asset value including a negative
jump is shown in Fig.~\ref{fig3}.
Without the jump term, the distribution of the asset price $V(t)$ is
log--normal. The jumps render the tails of the asset price
distribution fatter. Fat tails are empirically
observed~\cite{mant00}. As this clearly affects the loss distribution,
we find it important to include such jumps.
% RS: added remark:
The parameters of the jump process can be adjusted in order to match the tail behavior of a given empirical time series of the asset value.

%

\begin{figure}
\resizebox{75mm}{!}{\includegraphics{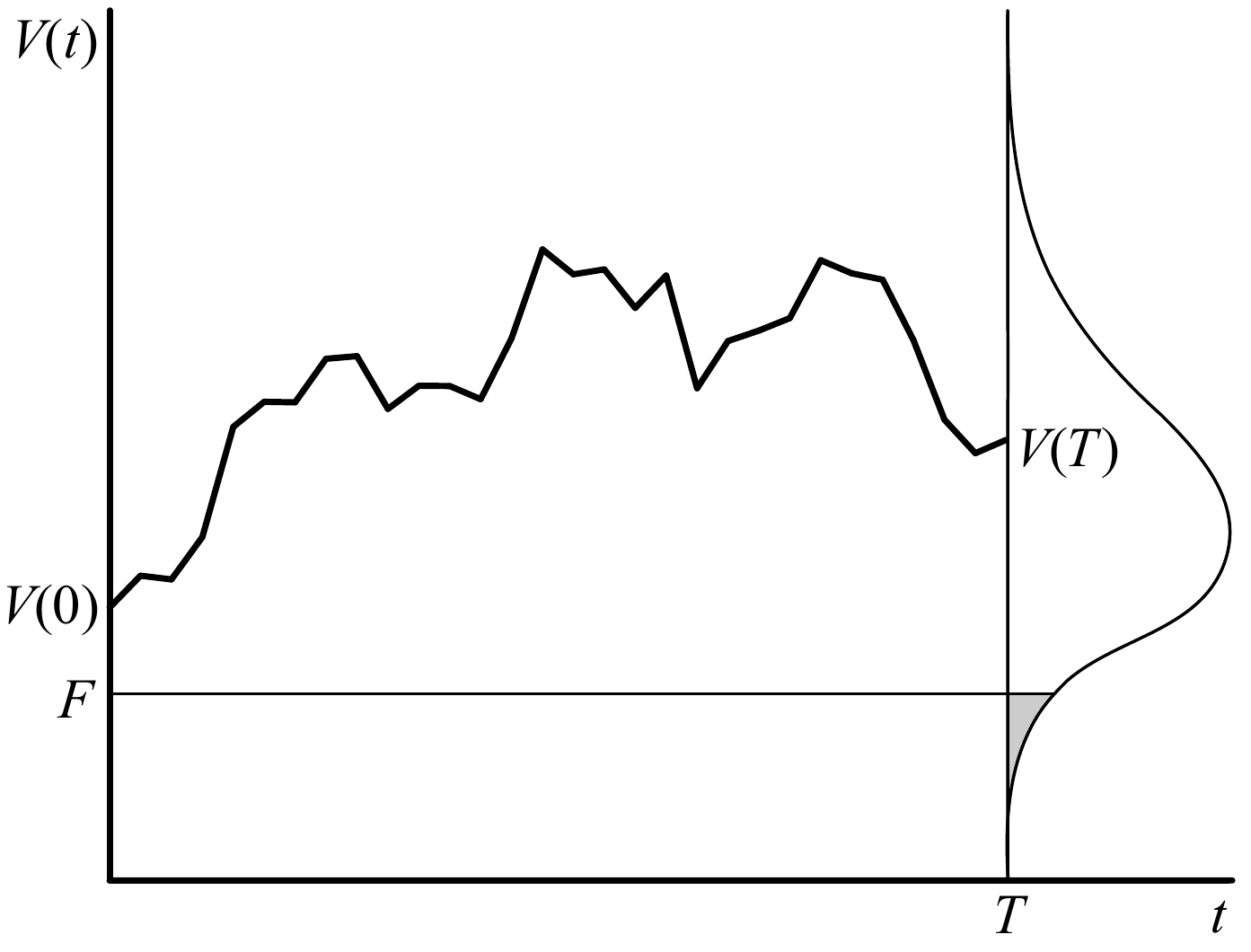}}
\caption{\label{fig4} A visualization of the underlying asset value
process, involving the time series of the asset value $V(t)$ versus
time $t$ with initial value $V(0)=V_0$, face value $F$ and maturity
$T$. The curve to the right is the distribution of the asset value at
maturity. The shaded area corresponds to the default probability.}
\end{figure}

% Default
To determine if default has occurred, we compare the asset value at
maturity $T$ to the company's financial obligations, i.e.~the face
value $F$ of the bond. 
%%%
This is in distinction to the so-called first-passage models, see e.g.~Refs.~\cite{zhang03,zhou01,schoenbucher00}, where default occurs 
as soon as the asset value $V(t)$, with $0 \le t \le T$, falls below some specified threshold function $D(t)$.
%%%
%
For simplicity, we do not choose
such a function and work with the face value $F$. In Fig.~\ref{fig4},
the asset value process with the threshold is visualized.  We employ
the default mode paradigm, which means that an actual loss only occurs
in case of a default. The size of the loss in case of default is given
by the face value of the loan minus the total asset value at
maturity. We normalize the loss to the face value $F$,
\begin{equation} \label{eq:lgd}
L = \frac{F - V(T)}{F} \, ,
\end{equation}
implying that we have $0 \le L \le 1$ for the normalized loss $L$.
%%%
Many studies assume, for the sake of simplicity, that everything is lost in case of default, i.e.~$L$ can only take the values 0 and 1, see e.g.~Refs.~\cite{lucas01,schoenbucher00}. In this simple case, many distinctive features of the loss distribution, as seen e.g. in Fig.~\ref{fig17}, cannot be reproduced, and, particularly, the tail behavior of the loss distribution is different.
Zhou~\cite{zhou01} studied a first-passage model for credit risk based on a jump-diffusion process. This leads to a recovery rate which depends solely on the jump process, whereas in our model it is $V(T)/F$, i.e.~it is determined by the whole asset process. In Ref.~\cite{zhou01} default probability and credit spread are studied in dependence of maturity $T$. The present study is different, because we we extensively examine the full loss distribution of the credit portfolio.
%%%

% Correlations
It is important to include correlations between the companies in our
model.  We mention that financial correlations presently find
considerable interest in the econophysics community. These
correlations are noise dressed if measured for finite time series.
The impact of noise dressing on portfolio optimization and methods how
to reduce the noise to find the true correlations are much discussed
in the
literature~\cite{lal99,ple99,man99,bon00,gop01,ple02,gia01,gia02,kalber03}.
The influence of noise dressing on credit risk has very recently been
studied in Ref.~\cite{ros04}.  To model the correlations, we employ
Noh's model \cite{noh00} which belongs to the class of factor and
arbitrage pricing models~\cite{ross76,schoenbucher00}. Noh's model produces
correlated normalized time series $M_k(t)$.  We use the conventions of
Ref.~\cite{kalber03}.

To achieve realistic portfolio correlations we divide the
companies into different branches $b = 1,2,\ldots ,B$. The total
number of branches represented in a portfolio is given by the
number $B$. The different companies in the portfolio are given by
the index $k = 1,2,\ldots ,K$ and the branch index $b$ is a
function of the company index, i.e.\ $b = b(k)$. The number of
companies in a specific branch $b$ is given by $\kappa_b$. For the
companies that are in no branch we have $b = 0$, and the number of
those companies is given by $\kappa$. This means that we can
write the total number of obligors $K$ in a portfolio as
\begin{equation}\label{kap}
K = \kappa + \sum_{b=1}^{B}\kappa_b.
\end{equation}
Within a branch $b$ the $\kappa_b$ companies are assumed to be
correlated with a specified correlation coefficient $C_b$. To achieve
this, the one--factor model adds a part of the branch specific time
series $\eta_{b(k)}(t)$ to the branch independent time series
$\varepsilon_k(t)$. A sum of these time series is used to construct
the asset returns for the companies in the portfolio.  The correlated
time series $M_k(t)$ has the form
\begin{equation}\label{serlin}
M_k(t) = \sqrt{\frac{p_{b(k)}}{1 + p_{b(k)}}} \eta_{b(k)}(t) +
\sqrt{\frac{1}{1 + p_{b(k)}}} \varepsilon_{k}(t).
\end{equation}
The entries in both $\eta_{b(k)}(t)$ and $\varepsilon_k(t)$ are
uncorrelated and standard normal distributed. The weights $p_{b(k)}$
measure the correlation and satisfy $p_{b(k)}\ge 0$. In particular, we
have $p_0=0$ for those companies which are in no branch.  The
normalized, correlated time series $M_k(t)$ can be arranged in a $K
\times T$ matrix $M$, where $K$ is the number of obligors and $T$ the
length of the time series.  The corresponding $K \times K$ correlation
matrix $C(T)$ is defined as
\begin{equation}\label{CT}
C(T) = \frac{1}{T}MM^\dagger = \langle M_k(t) M_l(t) \rangle_T \ ,
\end{equation}
where $M^\dagger$ is the transpose of $M$.  The index $T$ on the
brackets indicates that the average depends on the length $T$ of the
time series. If the time series are infinitely long, i.e.\ $T
\rightarrow \infty$, the correlation coefficient $C_{kl}(\infty)$ for
company $k$ and $l$,
\begin{equation}
C_{kl}(\infty) = \frac{1}{1 + p_{b(k)}} \left( p_{b(k)}
\delta_{b(k)b(l)} + \delta_{kl} \right).
\end{equation}
This value should resemble the best estimate for the correlation and
is referred to as the true correlation.  The correlation matrix
$C(\infty)$ consists of $B$ square blocks of dimension $\kappa_b
\times \kappa_b$ at the diagonal. The off--diagonal elements are $C_b
= p_b/(1+p_b)$ for branch $b$, and the no--branch companies are
represented by an identity matrix. All diagonal elements of
$C(\infty)$ are equal to one, and all entries that are not mentioned
above are zero. However, for finite length $T<\infty$, the true
correlations are noise--dressed, because every matrix element carries
a random number as offset and the block structure is obscured.

Implementing Noh's model, we are able to analyze a wide range of
different portfolio compositions. We emphasize that we automatically
include noise, because we work with time series of finite lengths.
For this investigation, we find this more realistic because it seems
to match the present usage by practitioners.
%%%
We notice that linear correlations between the asset processes are the simplest 
form to model the correlations between defaults. They do not account for 
credit risk contagion across firms and periods of default clustering, 
see e.g.~Refs.~\cite{kuehn03,neu04,giesecke04a,giesecke06,hatchett06}. 
%%%

% Model summary
Tab.~\ref{tab:modelparam} summarizes the model parameters and
Fig.~\ref{fig4} shows a visualization of the underlying asset
value process.
%RS: added
The model comprises a large number of parameters.
 However, this is indispensable if one aims at setting up a realistic 
 model. Credit risk is a problem that involves a high degree of 
 complexity. Importantly, none of our parameters is a hidden
 parameter. All of them have a direct interpretation and are measurable
 observables.  Thus it is certainly possible to calibrate our model 
 by fitting all parameters to real market data. Admittedly, this task
 might be time-consuming or even difficult, but it is definitely
 feasible. A key purpose of the present study is to investigate the tail 
 behavior of the loss distribution. Thus, even if it is difficult to determine
 some of the parameters sharply, that is, only within some uncertainties, our 
 model yields detailed information on how the tail depends on these 
 parameters.

\begin{table}
\caption{\label{tab:modelparam}Input/output of the model.}
\begin{ruledtabular}
\begin{tabular}{lcr}
\textbf{Input} & \textbf{Description} & \textbf{Unit} \\
\hline
$K$ & Number of obligors in the portfolio. & --- \\
$T$ & Timespan of the bond (maturity). & Year \\
$\mu$ & Asset drift. & [Year]$^{-1}$ \\
$\sigma$ & Asset volatility. & [Year]$^{-1/2}$ \\
$\lambda$ & Jump intensity. & [Year]$^{-1}$ \\
$\mu_J$ & Mean of the jump size. & Percent \\
$\sigma_J$ & Standard deviation of the jump size. & Percent \\
$F$ & Face value of the bond. & \$ \\
$V_0$ & Asset value at the issue date $t = 0$. & \$ \\
$C$ & Correlation matrix. & --- \\
\hline \hline
\textbf{Output} & \textbf{Description} & \textbf{Unit} \\
\hline
$P_D$ & Probability of default. & Percent \\
$p(L)$ & loss distribution $L$. & ---
\end{tabular}
\end{ruledtabular}
\end{table}

\subsection{Related scenarios}
\label{ourmodel2}

The task of finding the probability to hit a certain threshold for an
object, a particle, say, whose motion is described by a stochastic
process closely relates to the much discussed problem of the stopping
time distribution. The stopping time is the time at which the particle
hits the threshold. The corresponding distributions are non--trivial
objects, because they tend to have non--generic features~\cite{ito65}.
This explains the considerable interest which they attract in
mathematical statistics.

An application in physico--chemistry is the following: A (Brownian)
particle moves stochastically in a suspension confined by a
container. It is absorbed when it eventually hits the wall of the
container, or it reacts there, or it penetrates the wall if the latter
is semipermeable and separates two suspensions. Obviously, the
associated loss distribution is very similar to the one in credit
risk, if we consider a first passage model. The numerical simulations 
can be easily adjusted to such a scenario. 
One could even imagine a force acting between the stochastic
particles which would imply a certain correlation between them.
However, we do not elaborate on this further, because the purpose 
of the present discussion is only to establish a qualitative connection.

Another example is a farm where different types of crops are grown.
The time evolution of their quality and their future value has a
deterministic and a stochastic component. The analogue of default for
a given type of crop arises when its value falls below a certain
threshold. This could be due to bad weather, diseases, pests, fire,
etc., also motivating the inclusion of jump processes. The time series
for the crops are correlated, for example, because the diseases are
contagious. The knowledge of a distribution for the loss similar to
the one in credit risk would be most helpful for the farmer.

The last example shows that the structural model for credit risk can
find application in a large number of logistic problems where various
objects or quantities have to be available in a certain state at some
time in the future.

\section{Analytical discussion}
\label{sec:analytical}

We investigate a simplified version of our credit risk model
analytically. The purposes of this discussion are, first, to
illustrate the general mechanisms and, second, to look at the tail
behavior. In doing so we demonstrate how and with which speed the loss
distribution converges, under certain assumptions, to a universal limit as the
number $K$ of companies is made large. Thus, in this case study, we do
not include the jump term in the diffusion process. Moreover, we also
make the assumption that the obligors are uncorrelated.  We
distinguish between individual and portfolio losses in
Secs.~\ref{ana1} and~\ref{sec:portfolio}.

\subsection{Individual losses}
\label{ana1}

The asset value $V(t)$ for every company $k$ with $k=1,\ldots,K$
follows a geometric Brownian motion; its distribution is log--normal. With the
initial value $V(0)=V_{0}$ at $t=0$, the distribution of $V(T)$ at maturity
$t=T$ reads
\begin{eqnarray} \label{eq:pvt}
&& p_k(V(T)) = \frac{1}{V(T) \sqrt{2 \pi \sigma^2 T}} \nonumber\\
&& \quad  
\exp\left(- \frac{\left(\ln(V(T)/V_0) -(\mu - \sigma^2/2)T\right)^2}
                 {2 \sigma^2 T}\right) \, .
\end{eqnarray}
We use the company index only for the distribution $p_k$, and we
suppress it in the quantities of the asset value process.  Due to
$0\leq V(T)\leq F$ at maturity in the case of default, we have to
truncate Eq.~(\ref{eq:pvt}). Using Eq.~(\ref{eq:lgd}) we map the
distribution of $V(T)$ to that of the loss given default $L$ for this
company,
\begin{eqnarray}
\label{eq:analyticpdf}
&& p_k(L) = \frac{1}{P_D(1-L)\sqrt{2 \pi \sigma^2 T}}\nonumber\\
&& \quad  
\exp\left(- \frac{\left(\ln(F(1-L)/V_0) -(\mu - \sigma^2/2)T\right)^2}
                 {2 \sigma^2 T}\right) \, .
\end{eqnarray}
The factor needed to restore the normalization,
\begin{equation}\label{defprob}
P_D = \frac{1}{2} + 
      \frac{1}{2} {\rm erf\,}
            \left(\frac{\ln(F/V_0) -(\mu - \sigma^2/2)T}
                       {\sqrt{2\sigma^2 T}} \right) \, ,
\end{equation}
is the default probability for this company $k$, where we employ
the error function according to the definition
\begin{equation}\label{errfunc}
{\rm erf\,}(x) = \frac{2}{\sqrt{\pi}}\int\limits_0^x\exp(-\xi^2)d\xi \ .
\end{equation}
The number $P_D$ is the probability that the asset process $V(t)$ is
below the threshold {\it at} maturity $T$, i.e.~for having $V(T)\le
F$.  The default probability is shown in Fig.~\ref{fig4} as shaded
area. 

\begin{figure}
\resizebox{75mm}{!}{\includegraphics{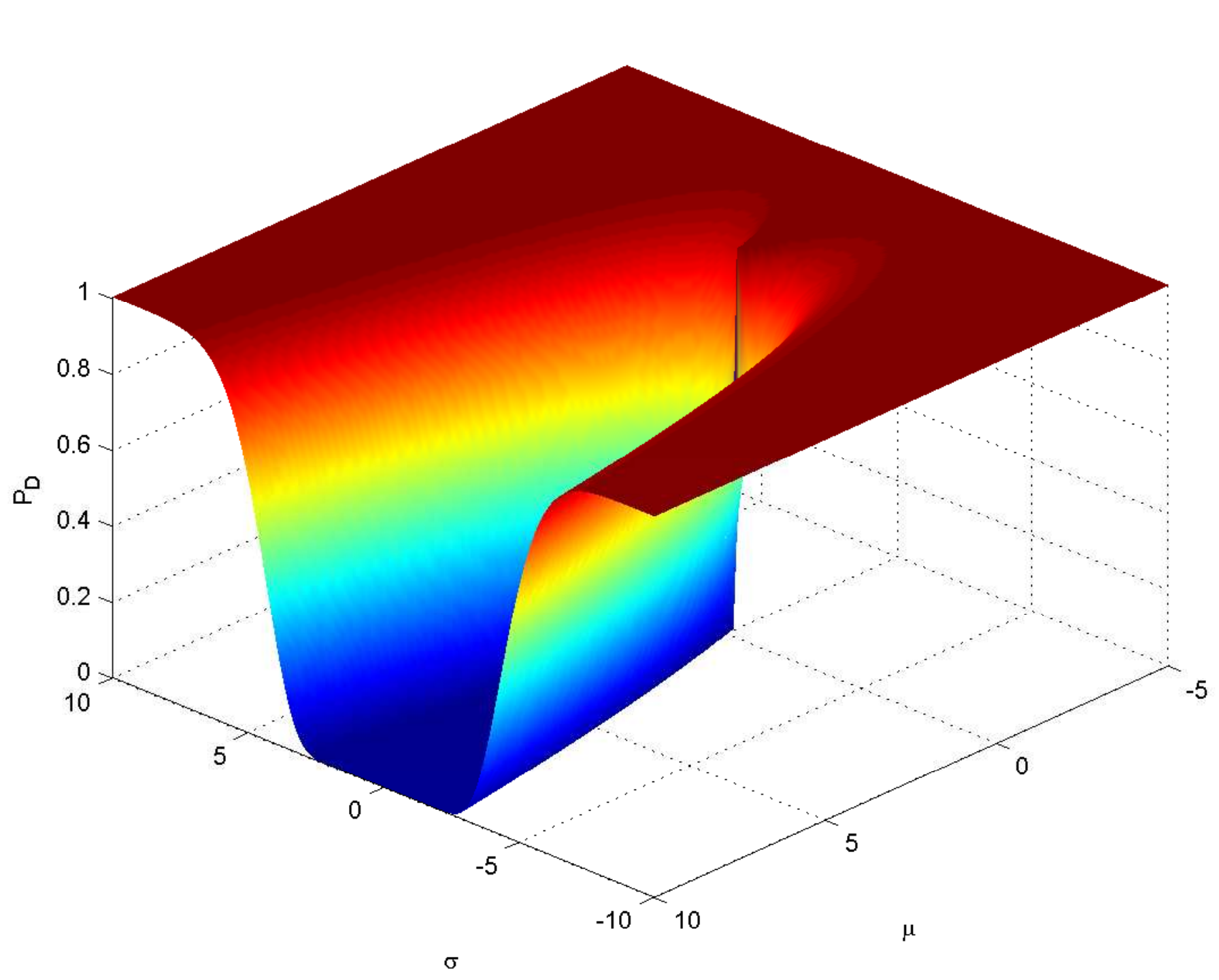}}
\caption{\label{cap:PD3D}The default probability~(\ref{defprob}) as a function of the parameters
$\mu$ and $\sigma$ for the values $V_{0}=100$, $F=75$ and $T=1$. The default probability tends to increase rapidly
when choosing parameter values that are too large. Also one can see
that it is an even function in $\sigma$, indicating that it is only
the absolute value that matters. }
\end{figure}

The impact of the drift term $\mu$ and the volatility
$\sigma$ on the default probability is shown in Fig.~(\ref{cap:PD3D}).
The range in which realistic default
probabilities are generated is vary narrow, both in the $\mu$ and
$\sigma$ direction. For our simulations we choose $\mu=0.05$ and $\sigma=0.15$, which
results in a default probability of $P_{D}\approx 0.0148$.

For later purposes, we calculate the $n$--th moment
\begin{equation}\label{momdef}
\langle L_k^n \rangle = \int\limits_0^1 L^n p_k(L) dL
\end{equation}
of the distribution~(\ref{eq:analyticpdf}). By using the index $k$ in
our notation, we want to underline that the $\langle L_k^n \rangle$
are the moments of the distribution for the {\it individual}
loss given default. We arrive at
\begin{eqnarray}\label{momres}
&& \langle L_k^n \rangle = \frac{1}{2P_D} \sum_{j=0}^n (-1)^j
\Bigl( \begin{array}{c}
                        n \\
                        j
                         \end{array} \Bigr)\nonumber\\
&& \exp\left(j(j-1)\frac{\sigma^2}{2}T+
              j\left(\mu T-\ln\frac{F}{V_0}\right)\right)\nonumber\\
&& \left(1+{\rm erf\,}\left(\frac{(1-2j)\sigma^2 T/2-\mu T+\ln(F/V_0)}
          {\sqrt{2\sigma^2 T}}\right)\right) \
\end{eqnarray}
after a straightforward calculation.

\subsection{Portfolio losses}
\label{sec:portfolio}

The loss $L$ of the total portfolio is the arithmetic mean
of the individual losses,
\begin{equation}\label{pfl}
\label{eq:lpf} L = \frac{1}{K}\sum_{k=1}^{K} L_k I_k \ ,
\end{equation}
where $L_k$ is the loss given default for the individual bond with
index $k$ and $K$ is the total number of bonds in the portfolio. 
As we are interested in the distribution of the portfolio loss, we
have to introduce the default indicator $I_k$ for company $k$ by
\begin{equation}\label{defaultIndicator}
I_k = \left\{
\begin{array}{ll}
1 & , \quad \text{if} \quad V(T) < F \quad \text{(default)} \\
0 & , \quad \text{if} \quad V(T) > F \quad \text{(no\ default)}
\end{array}
\right.
\end{equation}
As the distribution of the default indicator for company $k$,
we choose
\begin{equation}\label{eq:ptilde} 
\tilde{p}_k(I_k) = (1 - P_D) \delta(I_k) + P_D \delta(I_k-1) \ . 
\end{equation}
The default probability~(\ref{defprob}) depends on the specific
parameters which go into the asset value process for company $k$. To
obtain the distribution of the portfolio loss we have to average over
all distributions of the individual losses $p_k(L_k)$, given by
Eq.~(\ref{eq:analyticpdf}), and over the indicator
distributions~(\ref{eq:ptilde}),
\begin{eqnarray}\label{pfdis}
 p(L) & = &
\int\limits_{-\infty}^{+\infty} dI_1
\tilde{p}_1(I_1) \cdots \int\limits_{-\infty}^{+\infty} dI_K \tilde{p}_K(I_K) \nonumber \\
& & \quad \int\limits_0^1 dL_1 p_1(L_1)
\cdots \int\limits_0^1 dL_K p_K(L_K) \nonumber \\
& & \quad \delta \left(L - \frac{1}{K} \sum_{k=1}^{K}L_k
I_k\right) \ .
\end{eqnarray}
We notice the subtle difference between the loss distribution and the
distribution of the loss given default. This difference is best understood
by considering Eq.~(\ref{pfdis}) for $K=1$, which yields
\begin{equation}\label{pfdis1} 
p(L) = (1 - P_D) \delta(L) + P_D p_1(L) \ .
\end{equation}
This is the weighted sum of the distributions for the case of no default
and for the case of default.

We calculate an asymptotic approximation to $p(L)$ for a large number
$K$ of companies, i.e.~for large portfolio size. Here, we make the
further assumptions that the face values and the parameters of the
geometric Brownian motion are the same for all companies $k$. These
additional assumptions are not strictly necessary for the feasibility
of the calculation, but they make the resulting expressions very
compact. Details of the computation are given in Appendix~\ref{appendix}. We
obtain
\begin{eqnarray} \label{eq:analyticportapprox} 
p(L) & \approx &
\frac{1}{2\pi} \int\limits_{-\infty}^{+\infty}d\omega 
\exp\left(-i\omega\left(L- P_D\langle L_k \rangle\right)\right) \nonumber \\
& & \quad \exp\left(- \frac{\omega^2}{2K} \left( P_D \langle L_k^2 \rangle
             - P_D^2 \langle L_k \rangle^2 \right)\right) \nonumber \\
& & \quad \exp\biggl(-\frac{i\omega^3}{6K^2} 
              \bigl(P_D\langle L_k^3\rangle \bigr. \biggr. \nonumber \\
& & \qquad \biggl. \bigl. + 3P_D^2 \langle L_k \rangle \langle L_k^2\rangle 
                  + 2P_D^3 \langle L_k \rangle^3\bigr)\biggr) \ ,
\end{eqnarray}
as asymptotic approximation to order $1/K^2$.  If we skip the $1/K^2$
term and settle with an $1/K$ expansion, we find the shifted Gaussian
\begin{eqnarray} \label{approxgauss} 
p(L) & \approx & \sqrt{\frac{K}{2\pi(P_D\langle L_k^2\rangle -
P_D^2 \langle L_k\rangle^2)}} \nonumber \\
& & \exp\left(-\frac{K(L-P_D\langle
L_k\rangle)^2}{2(P_D\langle L_k^2\rangle - P_D^2\langle
L_k\rangle^2)}\right) \, .
\end{eqnarray}
Thus, the Central Limit Theorem applies if the number $K$ of companies
is very large.  The expected loss EL, i.e.~the peak position, is the
mean value of $L_k$, weighted with the default probability $P_D$. This
is so, because we defined the portfolio loss $L$ as the arithmetic
mean~(\ref{pfl}). The unexpected loss UL quantifying the risk is given
by the square root of $(P_D\langle L_k^2\rangle - P_D^2\langle
L_k\rangle^2)/K$ which becomes smaller with $K$. Even in the Gaussian
limit, the dependence of $p(L)$ on the parameters $F$, $V_0$, $\mu$,
$\sigma$ and $T$ is non--trivial due to the rather involved
expressions~(\ref{defprob}) for the default probability
and~(\ref{momres}) for the moments.  We will return to this point.  
Due to the specific nature of the approximation that leads to Eq.~(\ref{eq:analyticportapprox}), the first moments of the exact distribution are preserved up to the highest order of $\omega$ which is considered in the exponential. Therefore, Eq.~(\ref{eq:analyticportapprox}) contains the correct first, second and third moment, while the Gaussian limit in Eq.~(\ref{approxgauss}) still has the correct mean and variance.

Further, it is important to note that expression (\ref{eq:analyticportapprox}) holds for any structural model for uncorrelated and homogenous portfolios, i.e.~with all face values and parameters of the geometric Brownian motion being the same for all companies $k$.. It does not depend on the choice of random processes for the asset values $V(t)$, and it is sufficient to know the default probability and the first three moments of the individual loss given default distribution.
It is straightforward to generalize Eq.~(\ref{eq:analyticportapprox}) to the case of inhomogenous portfolios (see Appendix~\ref{appendix2}).

\subsection{Exact loss distribution versus approximations}

The approximation in the previous subsection has the advantage of being easily extendable to more general portfolios and exactly conserving the first moments of the distribution up to the order of approximation. However, it does not reproduce the shape and, in particular, the tail behavior very well.

A better approximation of the loss distribution is possible, if we restrict ourselves again to homogenous portfolios. Using the default indicator function defined in Eq.~(\ref{eq:ptilde}), we can then rewrite Eq.~(\ref{pfdis}) as a combinatorial sum % (see Appendix~\ref{} for details)
\begin{equation}\label{eq:comb_sum}
p(L) = \sum\limits_{j=0}^{K} {K \choose j} (1-P_D)^{K-j} \; P_D^j \; F_j(L)
\end{equation}
where we define the function $F_j(L)$ as
\begin{eqnarray}\label{eq:F_j}
F_j(L) & = & \int\limits_0^1 dL_1 p_1(L_1)
\cdots \int\limits_0^1 dL_j p_j(L_j) \\
& & \quad \delta \left(L - \frac{1}{K} \sum_{k=1}^{j}L_k \right) \nonumber \\
& \approx & \frac{1}{2 \pi} \int\limits_{-\infty}^\infty d\omega 
\exp\left( - i\omega\left(L- \frac{j}{K} \langle L_k \rangle\right)\right) \nonumber \\
& & \quad \exp\biggl( - \frac{\omega^2 j}{2K^2} \left( \langle L_k^2 \rangle
             - \langle L_k \rangle^2 \right)\biggr. \nonumber \\
& & \qquad \biggl. - \frac{i\omega^3 j}{6K^3} 
              \bigl(\langle L_k^3\rangle - 3 \langle L_k \rangle \langle L_k^2\rangle 
                  + 2 \langle L_k \rangle^3\bigr)\biggr) \, . \nonumber
\end{eqnarray}
The approximation for $F_j(L)$ follows the same line of arguing which is outlined in Appendix~\ref{appendix}.
Note that each term of the sum in Eq.~(\ref{eq:comb_sum}) corresponds to the event that exactly $j$ defaults occur. In particular, for $j=0$ the delta peak $(1-P_D)^K \, \delta(L)$ is obtained exactly in this approximation. 

\begin{figure}
\includegraphics[width=0.8\linewidth]{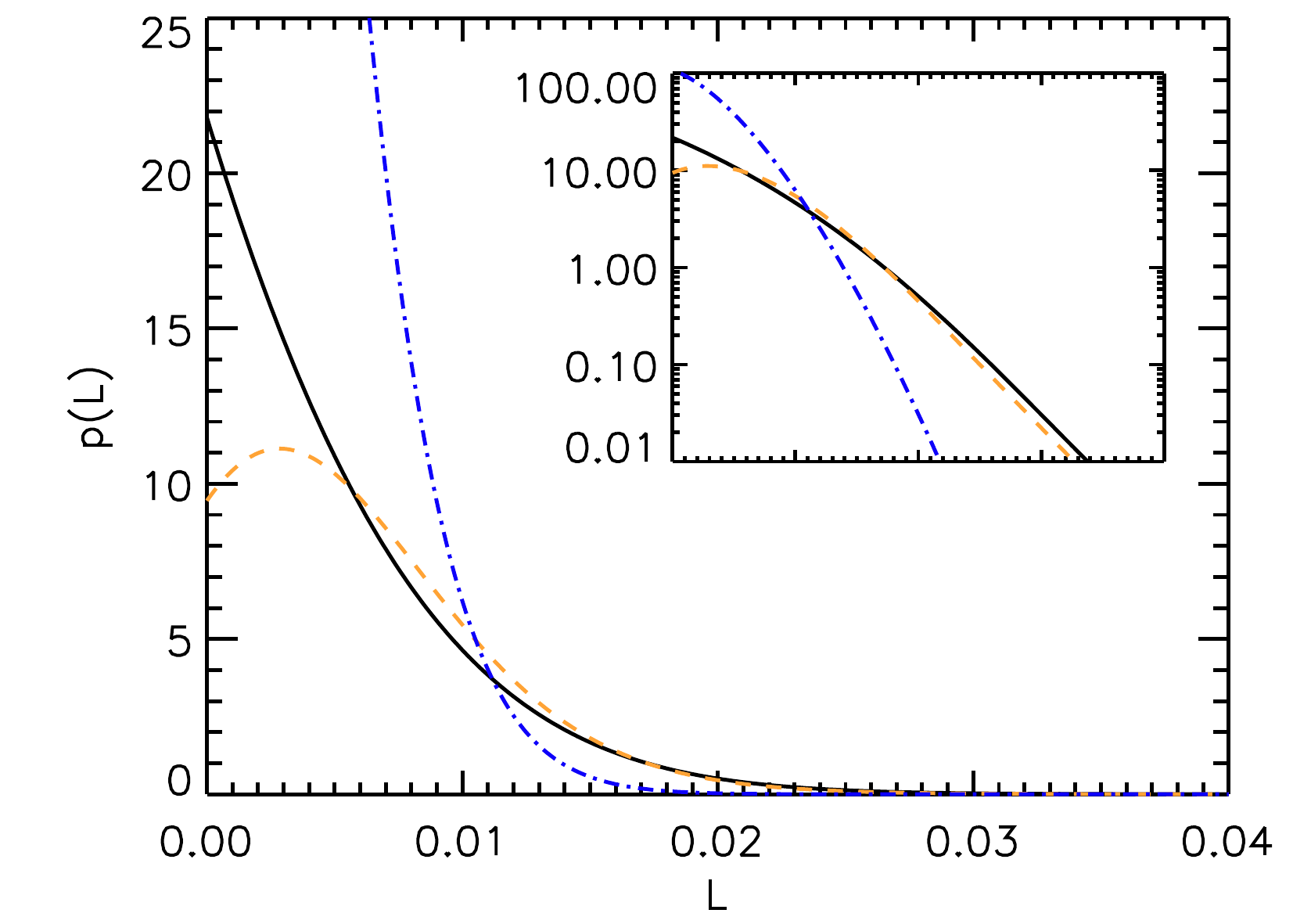}
\includegraphics[width=0.8\linewidth]{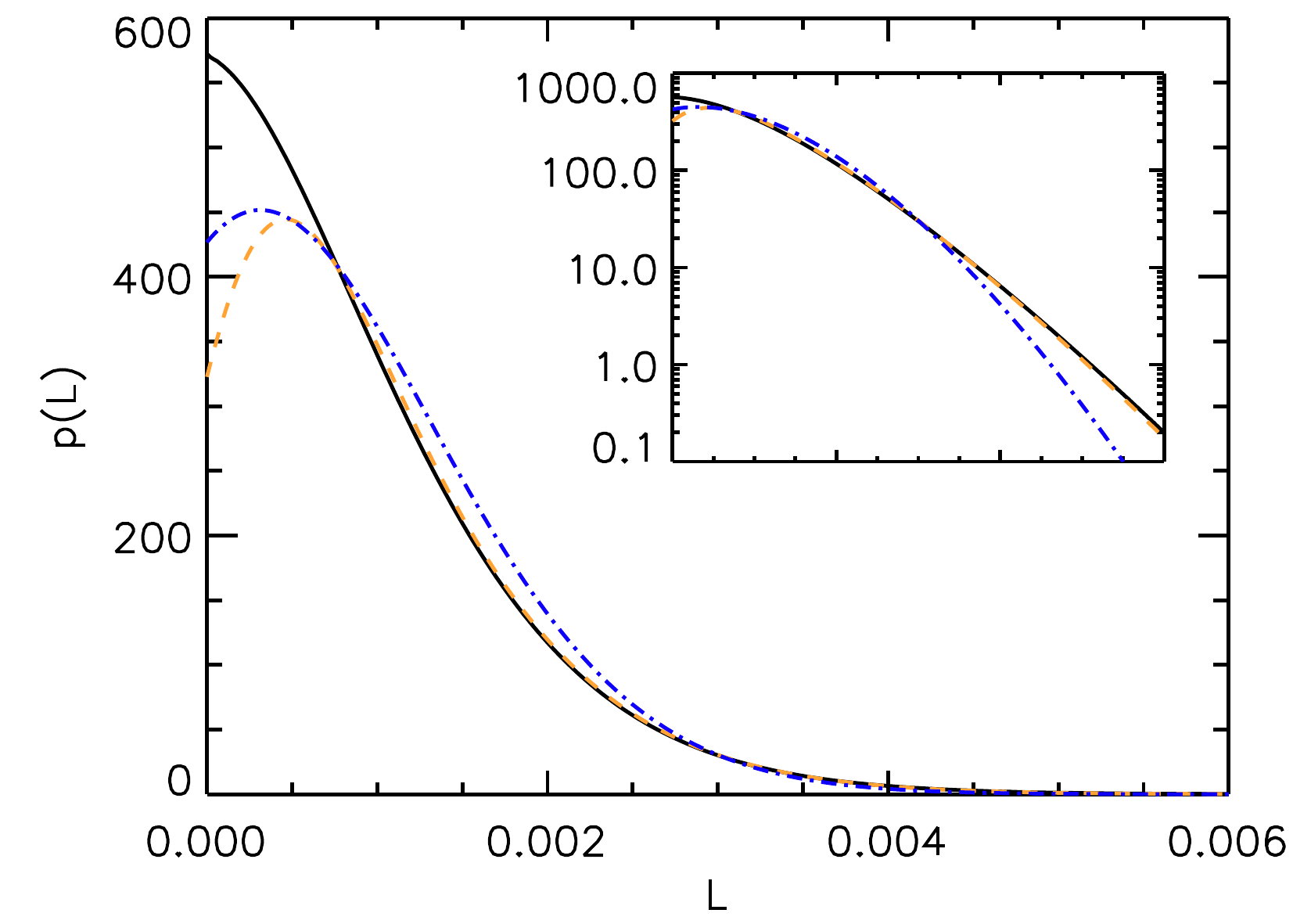}
\includegraphics[width=0.8\linewidth]{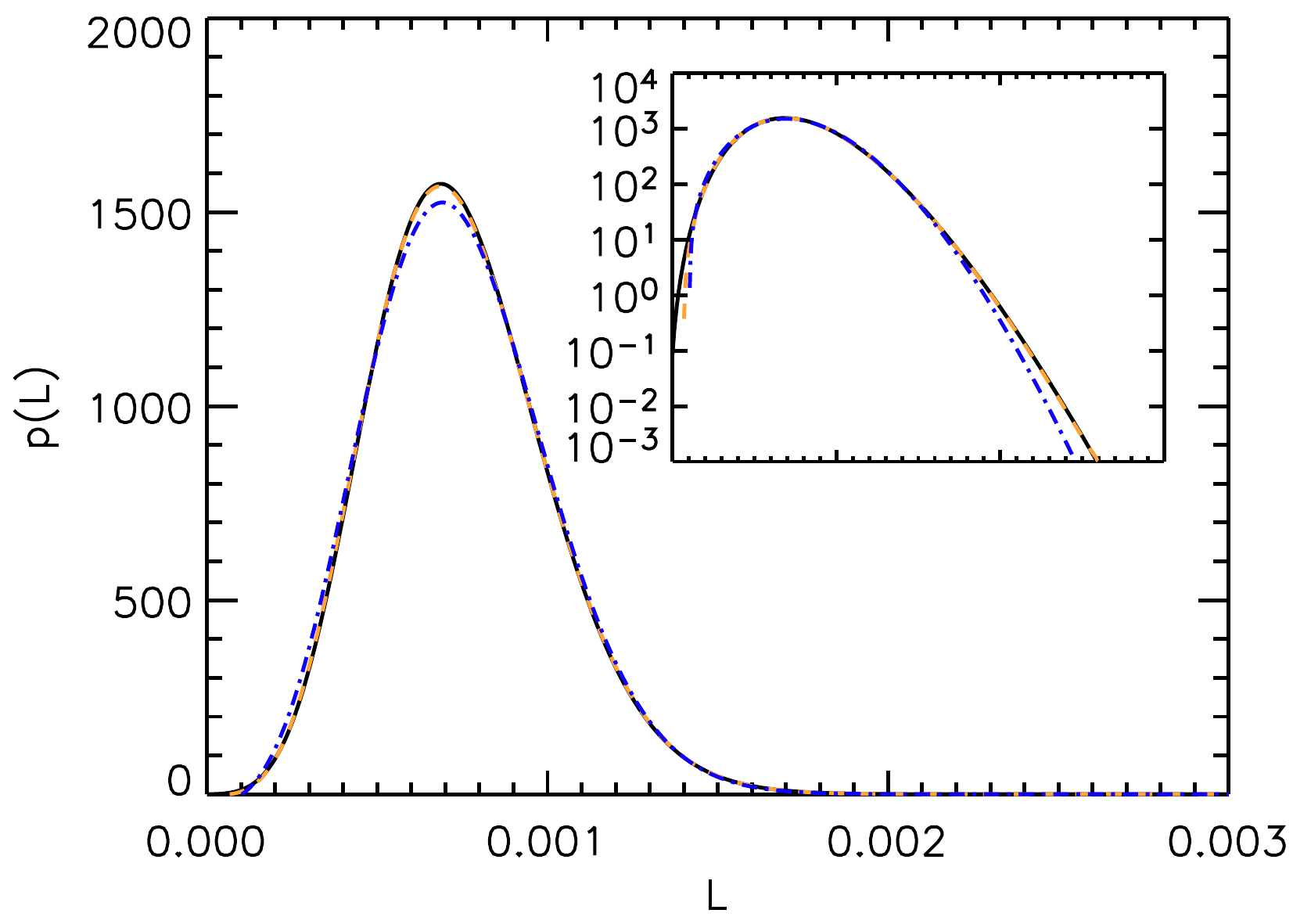}
\caption{\label{fig7a} Loss distribution for three different portfolio sizes $K=10,100,1000$,
respectively. The insets show logarithmic plots. Black solid lines show the analytical results, while the blue dashed-dotted lines correspond to Eq.~(\ref{eq:analyticportapprox}), and the orange dashed lines to the approximation of Eq.~(\ref{eq:comb_sum}).
The model parameters are $\mu = 0.05$, $\sigma = 0.15$,
$T = 1$, $V_0 = 100$ and $F = 75$.}
\end{figure}

In Figure~\ref{fig7a} we compare the exact loss distributions for $K=10$,100 and 1000 to the asymptotic approximation~(\ref{eq:analyticportapprox}) to order $1/K^2$, and to Eq.~(\ref{eq:comb_sum}) with the approximation in Eq.~(\ref{eq:F_j}). In both cases, the analytical approximations are evaluated numerically.
The exact loss distributions are obtained by calculating numerically the convolutions in Eq.~(\ref{eq:F_j}), using the characteristic function of $p_k(L)$, and inserting these $F_j(L)$ into Eq.~(\ref{eq:comb_sum}).
As numerical values for the model parameters we choose $\mu = 0.05$,
$\sigma = 0.15$, $T = 1$, $V_0 = 100$ and $F = 75$. 

For $K=10$ the delta peak  $(1-P_D)^K \, \delta(L)$ (not shown in the plots) is quite dominant and leads to a very poor result of approximation~(\ref{eq:analyticportapprox}), while the approximation of Eq.~(\ref{eq:comb_sum}) already yields a quite reasonable result.
For $K=100$, features of the Gaussian limit are already present. 
Pictorially speaking, the Gaussian moves from the left into the picture. 
The approximation of Eq.~(\ref{eq:comb_sum}) captures the tail behavior of the distribution quite nicely.

For $K=1000$, the Gaussian limit~(\ref{approxgauss}) is almost reached, but, interestingly, the
distribution is still slightly asymmetric. 
The agreement with the asymptotic approximation~(\ref{eq:analyticportapprox}) is convincing, although the tail behavior is not as well described as by the approximation of Eq.~(\ref{eq:comb_sum}).

Figure~\ref{fig7a} yields a good intuition for the tail behavior and
for the speed of convergence to the Gaussian, or any other universal
limit.  Remarkably, even for $K=1000$ slight deviations from the
Gaussian are seen. This means that, importantly, only really large
portfolio sizes imply universal shapes. 

A measure for the tail behavior of the loss distribution is the kurtosis,
which is defined as a normalized form of the forth central moment $\mu_{4}$
\begin{equation}
\beta_{2}=\frac{\mu_{4}}{\mu_{2}^{2}} \, ,\label{Kurtosis_def}\end{equation}
where  $\mu_{2}$ is the second central moment.
Since the kurtosis of the normal distribution is equal to three,
the \emph{kurtosis excess} is defined as
\begin{equation}
\gamma_{2}=\frac{\mu_{4}}{\mu_{2}^{2}}-3 \, .\label{Kurtosis_excess_def}\end{equation}
It can be shown analytically that the kurtosis excess 
for homogenous and uncorrelated portfolios is proportional to $1/K$. 
Such considerations are important in modern portfolio theory, for
example, when one tries to minimize risk by diversification. This can
require an enlargement of the portfolio.

We notice that the symmetric shape of the
Gaussian can only be reached without correlations between the asset
processes: with correlations, the probability of default is enlarged
without compensation on the positive side, because in the best case,
none of the obligors defaults. Thus, the distribution is asymmetric.

\subsection{Drill down risk}
Consider a portfolio consisting of $K$ companies. The
moments of the loss given default distribution are assumed
to be known. In which way will the properties of the portfolio change
if one decides to add or remove a company? If one had this information
an optimal strategy could be developed to drill down the parts
within the portfolio which are the most risky, or to decide
which new company should be added, so that the additional
risk is minimal.

Since we want to allow the individual face values $F_{k}$ and the initial values $V_{0k}$ to be different now, it would be unsuitable to use the dimensionless 
definition~(\ref{pfl}) of the losses. 
Instead we use a more general definition of the individual loss 
\begin{equation}
\Gamma_{k}\equiv L_{k}F_{k}=F_{k}-V_{k}(T)   \,, \label{lossDef2}
\end{equation}
which now has the dimension of dollars. The \emph{total loss} for
the entire portfolio of size $K$ is then
\begin{equation}
L^{(K)}=\frac{\sum_{k=1}^{K}\Gamma_{k}I_{k}}{\sum_{k=1}^{K}F_{k}}  \,,\label{totalLossDef}
\end{equation}
which again is a normalized quantity with $0\leq L^{(K)}\leq1$. $I_{k}$
is the indicator function for company $k$, as defined in Eq.~(\ref{defaultIndicator}).
If \emph{all} face values $F_{k}$ are the same, $F_{k}=F$, the loss
will again reduce to Eq.~(\ref{pfl}), since
\begin{equation}
L^{(K)}=\frac{\sum_{k=1}^{K}\Gamma_{k}I_{k}}{\sum_{k=1}^{K}F_{k}}=\frac{F_{k}\sum_{k=1}^{K}L_{k}I_{k}}{KF_{k}}=\frac{1}{K}\sum_{k=1}^{K}L_{k}I_{k}  \,. \label{totalLossDef2}
\end{equation}

Now, is it possible to express the moments of the loss distribution
for $K$ companies, in terms of the corresponding loss distribution
for $K-1$ companies? Indeed it is. See Appendix~\ref{drillDownRiskAppendix} for
details. The $n$:th moment can be expressed as:
\begin{eqnarray}
\left\langle \left(L^{(K)}\right)^{n}\right\rangle _{K} & = & \left(\frac{F^{(K-1)}}{F^{(K)}}\right)^{n}\nonumber \\
 & \times & \sum_{\nu=0}^{n}\left(\begin{array}{c}
n\\
\nu\end{array}\right)\left\langle
 \left(L^{(K-1)}\right)^{n-\nu}\right\rangle _{K-1}\nonumber\\
& \times & \frac{\left\langle I_{K}^{\nu}\right\rangle \left\langle \Gamma_{K}^{\nu}\right\rangle
 }{(F^{(K-1)})^{\nu}}  \,,
\end{eqnarray}
where $\nu$ is an integer and $F^{(K)}$ is the sum over all the $K$ face values:
\begin{equation}
F^{(K)}=\sum_{j=1}^{K}F_{j}  \,.  \label{F_KDef}
\end{equation}
For example, the two first moments are, for $n=1$
\begin{eqnarray}
\left\langle \left(L^{(K)}\right)\right\rangle _{K} & = &
\frac{F^{(K-1)}}{F^{(K)}}\left\langle
\left(L^{(K-1)}\right)\right\rangle_{K-1}\nonumber \\
& + & \frac{1}{F^{(K)}}\left\langle I_{K}\right\rangle \left\langle
\Gamma_{K}\right\rangle   \,,  \label{L_K-moment_n1}
\end{eqnarray}
 and for $n=2$
\begin{eqnarray}
\left\langle \left(L^{(K)}\right)^{2}\right\rangle _{K} & = &
 \left(\frac{F^{(K-1)}}{F^{(K)}}\right)^{2}\left\langle
 \left(L^{(K-1)}\right)^{2}\right\rangle _{K-1}\nonumber \\ 
& + & 2\frac{F^{(K-1)}}{\left(F^{(K)}\right)^{2}}\left\langle I_{K}\right\rangle \left\langle \Gamma_{K}\right\rangle \left\langle \left(L^{(K-1)}\right)\right\rangle _{K-1}\nonumber \\
 & + & \frac{1}{\left(F^{(K)}\right)^{2}}\left\langle
 I_{K}^{2}\right\rangle \left\langle \Gamma_{K}^{2}\right\rangle  \,.
 \label{L_K-moment_n2}
\end{eqnarray}

\section{Numerical discussion}
\label{sec:numerics}

The numerical analysis of our model is done with Monte Carlo
simulations. To achieve results that are statistically reliable, we
simulated between $10^4$ and $10^5$ asset processes.  An important
remark is in order.  Given the rich variety of credit contracts, we
decided not to calibrate our model with real market data, because this
would match only one particular scenario.  However, we have chosen the
parameter values in such a way~\cite{moodys02} that the model outcome
is economically realistic. 
%RS:  added remark:
In particular, the parameters of the jump process have been chosen in order to reproduce a realistic tail behavior.
For homogenous and uncorrelated portfolios 
the loss distributions obtained by the Monte Carlo simulations agree
very well with the analytical expectation.
In Sec.~\ref{num3} we study the impact of maturity, drift and volatility.  
Leverage, jumps and correlations are addressed in
Secs.~\ref{num5}, \ref{sec:jumps} and~\ref{num7}, respectively.  We
discuss the competition between jumps and correlations in
Sec.~\ref{num8}.

\subsection{Maturity, drift and volatility}
\label{num3}

To give the reader a better feeling for the sensitivity of the loss
distribution, we first take a look at the model in its simplest
form, i.e~without jumps and correlations, and study its dependence on
maturity $T$, drift $\mu$ and volatility $\sigma$. This is again done by calculating 
Eq.~(\ref{eq:comb_sum}) numerically, while in the remainder of the paper we 
use Monte Carlo simulations.

\begin{figure}
\includegraphics[width=0.75\linewidth]{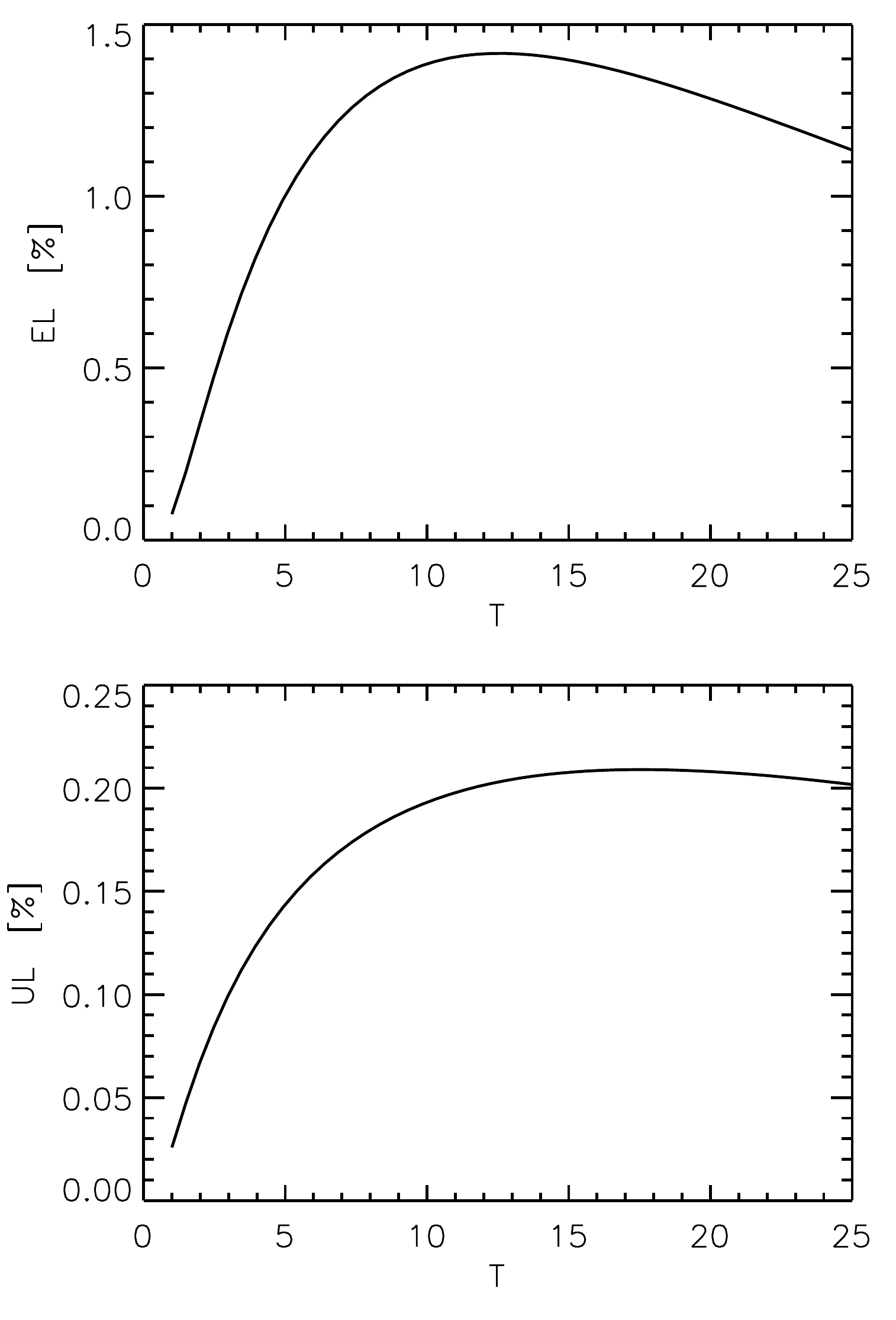}
\caption{\label{fig9}  Expected loss EL and unexpected loss UL 
as a function of the maturity $T$ for $K=1000$.}
\end{figure}

We begin with the
maturity $T$ and keep $\mu = 0.05$, $\sigma = 0.15$, $V_0 = 100$ and
$F = 75$.  It is quite common that loans span over a period longer
than $T=1$ year which has been considered so far.  In Fig.~\ref{fig9},
the expected loss EL and the unexpected loss UL are shown as a function of maturity $T$. 
The
expected loss EL is the mean value $P_D\langle L_k\rangle$ and the
unexpected loss UL squared is the variance $(P_D\langle L_k^2\rangle -
P_D^2\langle L_k\rangle^2)/K$, where the default probability $P_D$ and
the moments $\langle L_k^n\rangle$ are given by Eqs.~(\ref{defprob})
and~(\ref{momres}). 
Closer inspection shows that EL grows
monotonically in $T$ to its saturation value unity, if, for fixed
(positive) drift $\mu$ and volatility $\sigma$, the condition
$\mu<\sigma^2/2$ is met. However, if $\mu>\sigma^2/2$, EL has a
maximum for some finite value of $T$ and the saturation value of EL is
zero for $T\to\infty$. For the parameter choices $\mu = 0.05$ and
$\sigma = 0.15$, the maximum is at $T\approx 12.56$, as seen in Fig.~\ref{fig9}. 
Similar considerations apply to UL, with the maximum at $T\approx 17.55$.

In the sequel, we put $T=1$ and investigate the dependence on drift
$\mu$ and volatility $\sigma$. We choose $0.05 \le \mu \le 0.15$ and
$0.15 \le \sigma\le 0.35$ which is motivated by economical data. The
expected loss EL and the unexpected loss UL for $K=1000$ are depicted in
\begin{figure}
\includegraphics[width=0.75\linewidth]{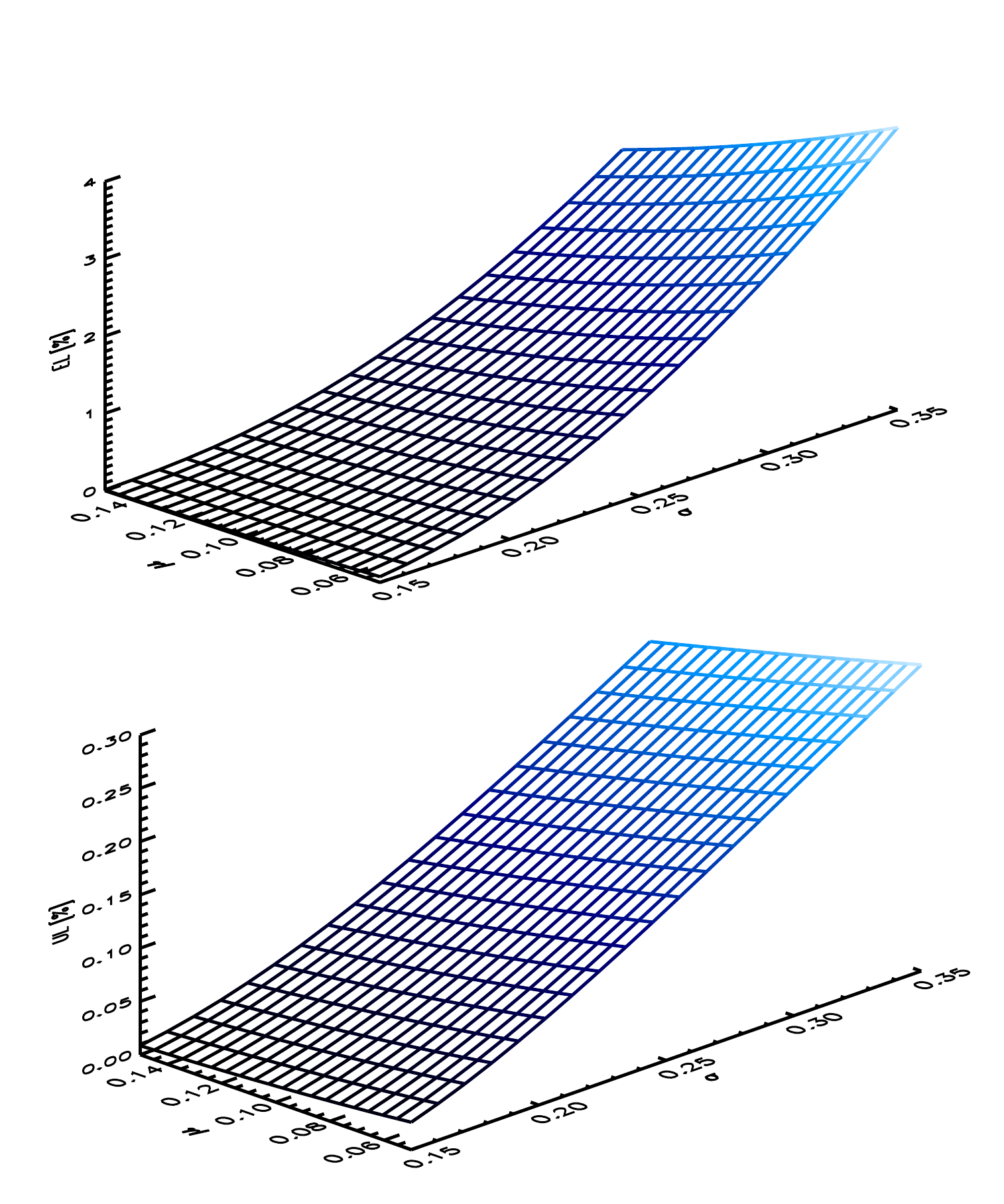}
\caption{\label{fig12} Expected loss EL and unexpected loss UL 
as a function of the drift $\mu$ and the volatility $\sigma$ for 
$K=1000$.}
\end{figure}
Fig.~\ref{fig12}. 
For $K=1$ (not shown),  both EL and UL are proportional to the volatility and 
only weakly dependent on the drift.
For portfolio sizes $K>1$ and fixed volatility $\sigma$, both EL and
UL increase with decreasing $\mu$. This reflects that the
probability to default increases with smaller $\mu$. Importantly, the
loss distribution becomes more sensitive to this effect with growing
portfolio size.  This is so, because the portfolio loss is an average
of individual losses.  The expected loss EL and the unexpected loss UL
increase with volatility. As for the drift, the sensitivity of the
loss distribution to the volatility grows with the portfolio size.  A
larger volatility implies larger asset returns which cause a greater
activity of the asset value. This, in turn, produces excessive
losses.

We conclude that the qualitative behavior of the loss distribution as
function of drift and volatility is very similar. Economically, this
means that a downwards change in the general trend of the asset prices
during a quiet, not volatile period affects the credit markets in a
similar way as an upwards change in the market activity,
i.e.~volatility, in times of a stable general trend in the markets.
According to Eq.~(\ref{defprob}), the default probability is almost
proportional to both drift and volatility in the parameter range
chosen.

\begin{figure}
\resizebox{65mm}{!}{\includegraphics{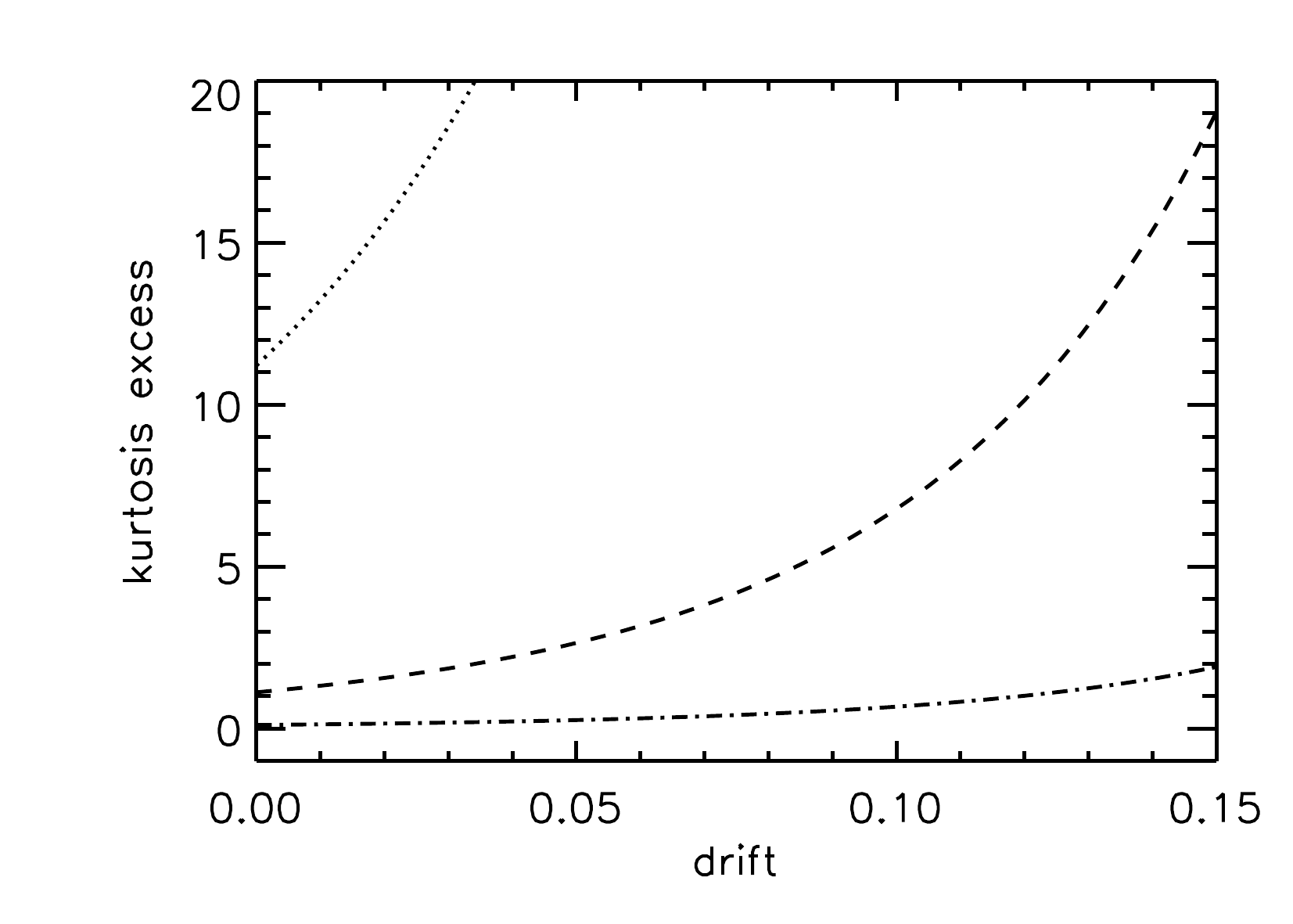}}
\caption{\label{fig14} The kurtosis excess as a function of the
asset drift $\mu$ for portfolio sizes $K=10$ (dotted), 
$K=100$ (dashed) and $K=1000$ (dashed-dotted).}
\end{figure}

\begin{figure}
\resizebox{65mm}{!}{\includegraphics{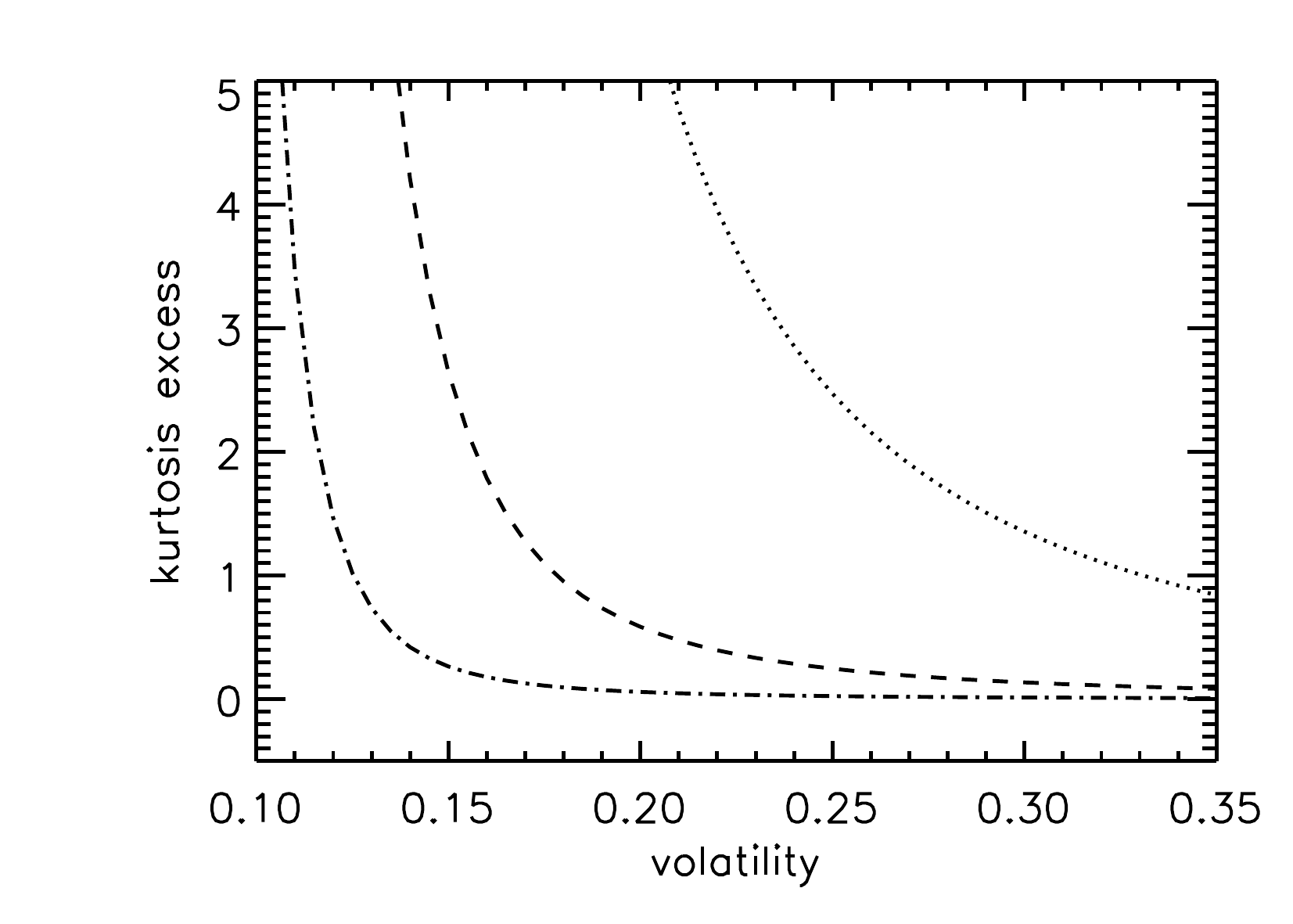}}
\caption{\label{fig15} The kurtosis excess as a function of the
asset volatility $\sigma$ for portfolio sizes $K=10$ (dotted), 
$K=100$ (dashed) and $K=1000$ (dashed-dotted).}
\end{figure}

Finally, we study how drift $\mu$ and volatility $\sigma$ affect the
tail behavior of the loss distribution. 
We recall that the universal limit is in the present case
always Gaussian for very large $K$.  
It can be shown analytically that the kurtosis excess of uncorrelated portfolios
scales as $1/K$.
Figures~\ref{fig14}
and~\ref{fig15} show the kurtosis excess as functions of drift and
volatility, respectively. In Fig.~\ref{fig14} one sees that a higher
drift gives loss distributions with fatter tails. This is more
pronounced for large portfolio sizes. Figure~\ref{fig15} shows that
the kurtosis excess approaches zero, i.e.~the distribution becomes
more Gaussian, with growing volatility. Moreover, one also observes
that the speed of convergence to a Gaussian is higher for larger
portfolio size, which reflects the Central Limit Theorem.

\subsection{Leverage}
\label{num5}

\begin{figure}[htb]
\includegraphics[width=\linewidth]{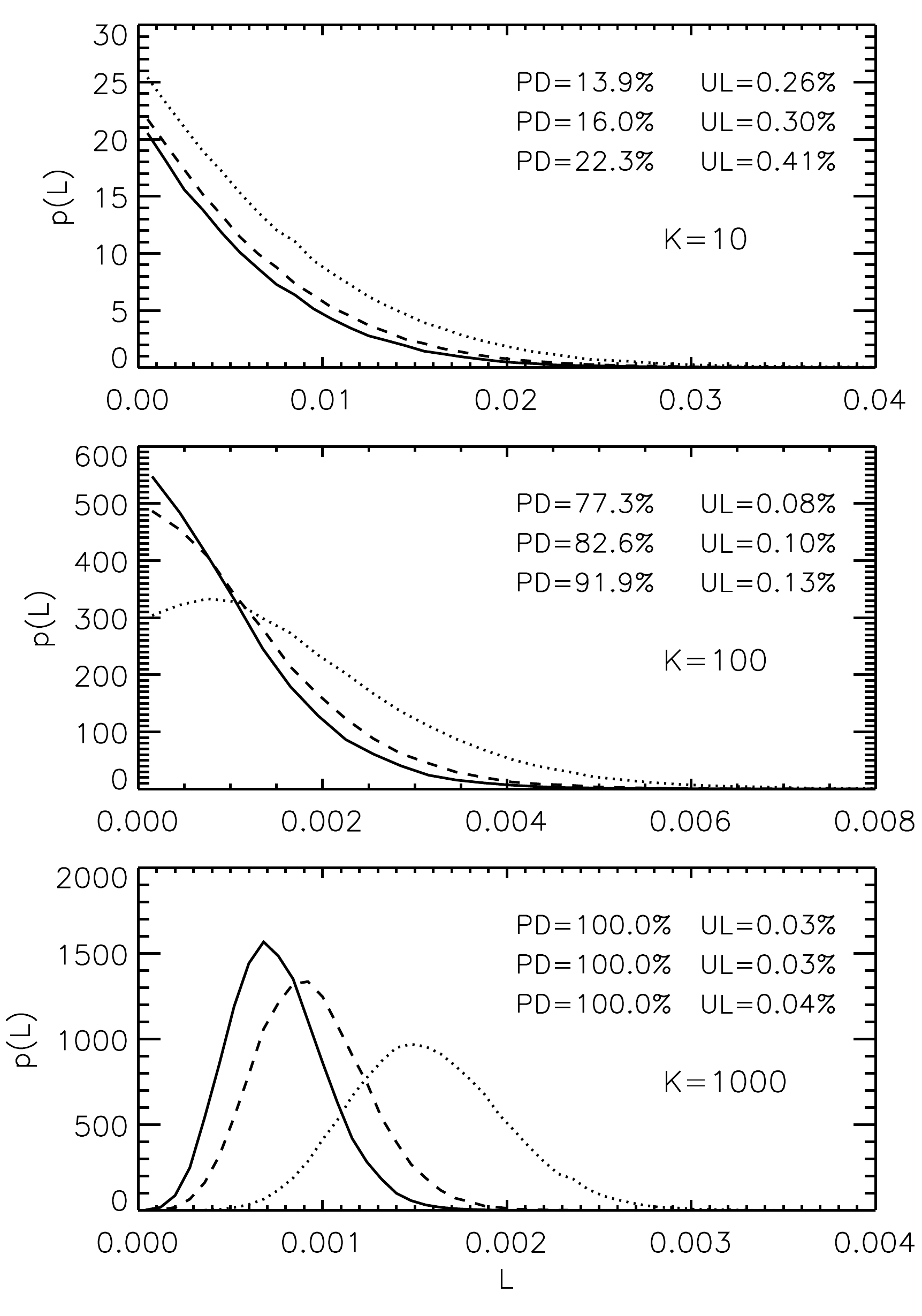}
\caption{\label{fig16a} Loss distributions for portfolios in which the
leverage is uniformly distributed. The face values are random in a
window of width $\Delta F$ around $F=75$ with $\Delta F=0$ (solid
lines), $\Delta F=10$ (dashed lines) and $\Delta F=20$ (dotted
lines), for three different portfolio sizes $K=10,100,1000$,
respectively. The expected loss EL is 0.076\%, 0.095\% and 0.157\% for
the three different leverages (independent of $K$). The insets show, for every portfolio
size, the portfolio default probability PD and the unexpected loss UL from top to
bottom for $\Delta F=0,10,20$.}
\end{figure}

The ratio $F/V_0$ between the face value of the bond $F$ and the
initial asset value $V_0$ is referred to as leverage. It is very
important for credit risk managers to know its influence on the loss
distribution.  In our model we are able to set both the face value of
the bond and the initial asset value. However, as it suffices to study
only the leverage, we keep $V_0 = 100$ fixed and only vary the face
value $F$.   
The portfolio default probability PD, expected loss EL and the
unexpected loss UL grow with the leverage. This is so, because the
higher the leverage, the more likely are early defaults. It is a
remarkable result that large portfolios are extremely sensitive to the
leverage.
As every realistic portfolio contains bonds with different leverage,
it is of considerable interest to study how the loss distribution
changes if the leverages are distributed. Here, the (relative) loss
$L$ is defined as the sum of the absolute losses $F_kL_kI_k$ divided
by the sum of the face values $F_k$ of the companies $k=1,\ldots,K$.
Here $L_k$ is the (relative) loss given default for company $k$ and
$I_k$ the corresponding default indicator.  For convenience, we choose
uniform distributions with width $\Delta F$ centered around
$F=75$. The results are shown in Fig.~\ref{fig16a}. Expected loss EL
and unexpected loss UL grow with $\Delta F$.

\subsection{Jumps}
\label{sec:jumps}

\begin{figure}
\includegraphics[width=\linewidth]{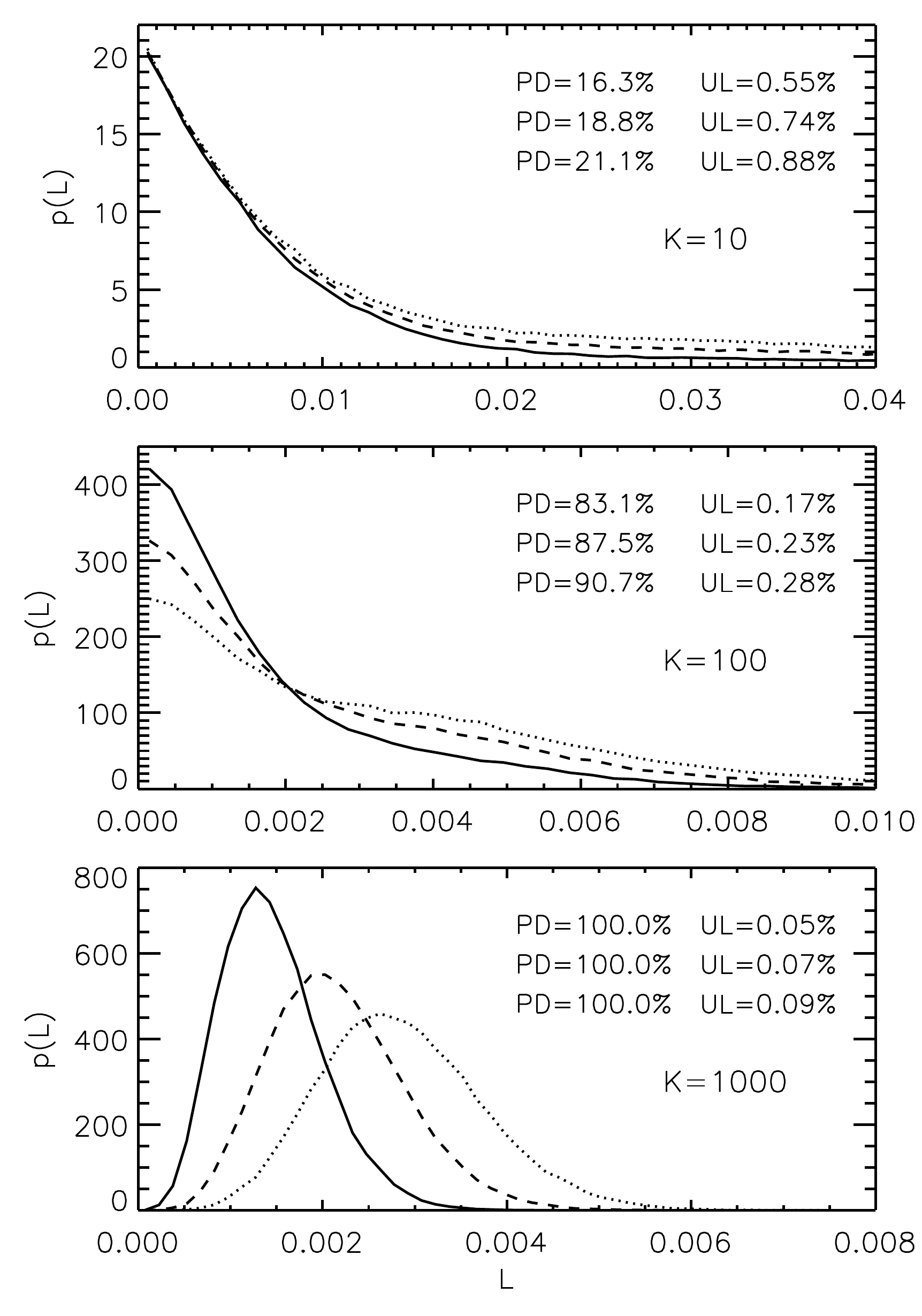}
\caption{\label{fig17} Loss distributions for portfolios with different jump
intensities $\lambda=0.005$ (solid lines), $\lambda=0.01$
(dashed lines) and $\lambda=0.015$ (dotted lines) for four
different portfolio sizes $K=10,100,1000$, respectively. 
The expected loss EL is 0.15\%, 0.22\% and 0.29\% for
the three different jump intensities. 
The insets show, for every portfolio
size, the portfolio default probability PD and the unexpected loss UL
from top to bottom for $\lambda=0.005,0.01,0.015$.}
\end{figure}

\begin{figure}
\includegraphics[width=0.85\linewidth]{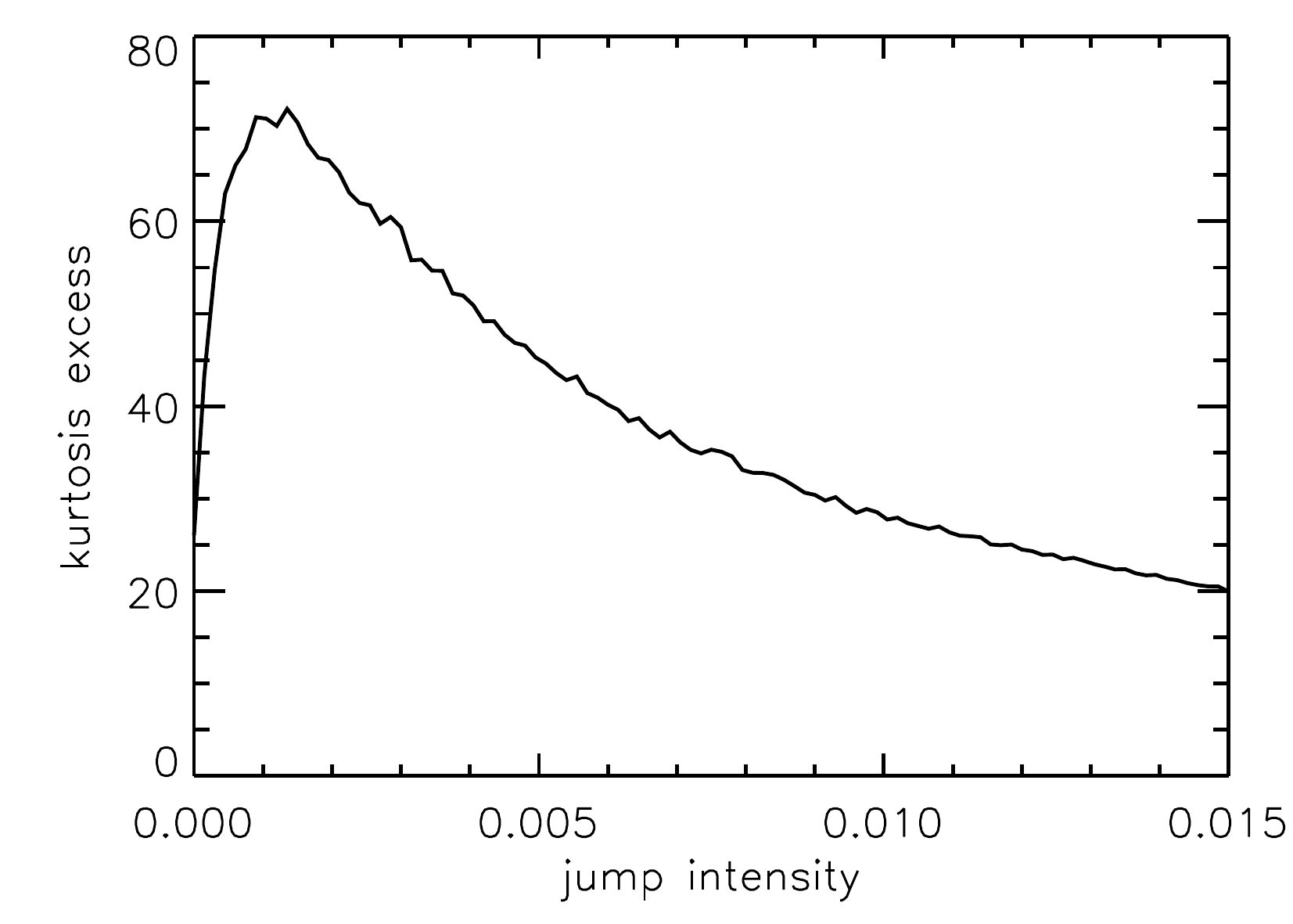}
\caption{\label{fig20} The kurtosis excess as a function of the
jump intensity $\lambda$. The result is shown for $K=10$, and it scales with $1/K$.}
\end{figure}

The need to include jumps in a realistic model has been pointed out in
Sec.~\ref{ourmodel1}. Jumps are important despite the fact that the
jump intensity is typically very small. Reasonable economical values
of the jump intensity are around $0.01$ jumps per year. If not otherwise stated, we use the following values for the jump process: jump intensity $\lambda=0.01$, mean jump size $\mu_J=-0.4$, and standard deviation $\sigma_J=0.3$. 
In Fig.~\ref{fig17} we plot the loss distributions for three different jump
intensities, $\lambda=0.005,0.01,0.015$. 
Not surprisingly, PD, EL and UL increase with the jump intensity. 
As for any uncorrelated portfolio, EL is independent of the portfolio size, 
while UL scales with $1/\sqrt{K}$. For $K=1000$, PD has already saturated at 100\%.
In Figure~\ref{fig20} the kurtosis excess is depicted
as a function of the jump intensity $\lambda$.  The kurtosis excess
grows with the jump intensity up to a certain value of the intensity,
then it decreases again and approaches zero asymptotically. For larger
values of $\lambda$, the loss distribution is more smeared out,
because the losses produced by the jumps dominate the defaults due to
the standard geometric Brownian motion. In other words, the emergence
of the maximum can be explained as follows.  The kurtosis is high for
a distribution that has a sharp peak around the mean and a fat
tail. When the intensity $\lambda$ grows starting from $\lambda=0$,
the tail becomes fatter, because the kurtosis increases.  But at some
value of $\lambda$, the tail becomes so heavy that it interferes with
and ultimately destroys the sharp peak of the distribution. Hence, the
kurtosis excess must have a maximum value for some intensity
$\lambda$. Furthermore, we notice that the kurtosis excess scales 
as $1/K$ with the portfolio size.

\begin{figure}
\includegraphics[width=\linewidth]{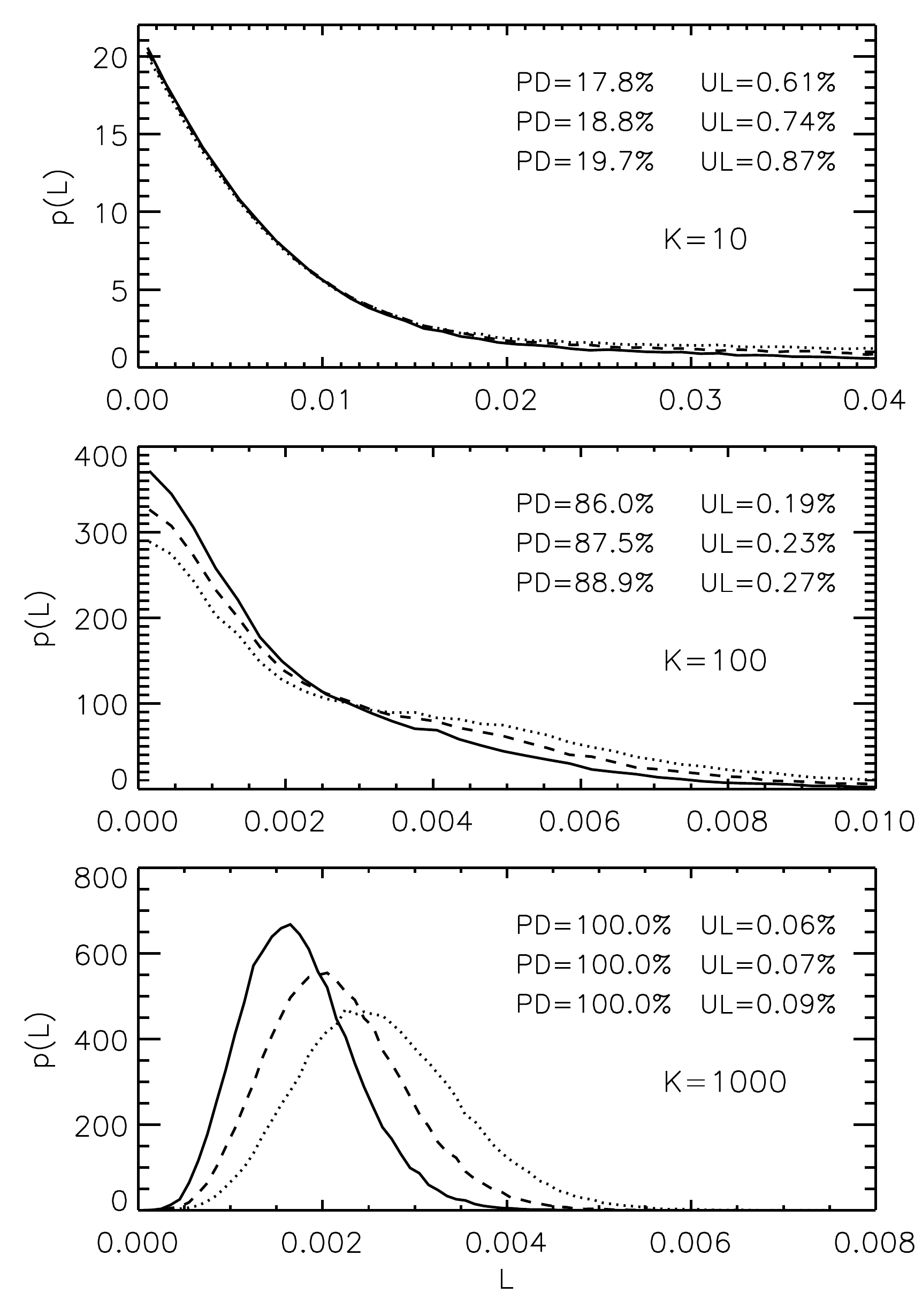}
\caption{\label{fig20e}Loss distributions for portfolios with different mean
values of the jump size, $\mu_J=-0.3$ (solid lines), $\mu_J=-0.4$
(dashed lines) and $\mu_J=-0.5$ (dotted lines) for four
different portfolio sizes $K=10,100,1000$, respectively.  The
standard deviations is always $\sigma_J=0.3$. 
The expected loss EL is 0.18\%, 0.22\% and 0.26\%, and the portion of
negative jumps are 86\%, 91\% and 94\% for the three $\mu_J$
values, respectively. The insets show, for every portfolio
size, the portfolio default probability PD and the unexpected loss UL 
from top to bottom for $\mu_J=-0.3,-0.4,-0.5$.}
\end{figure}

\begin{figure}
\includegraphics[width=\linewidth]{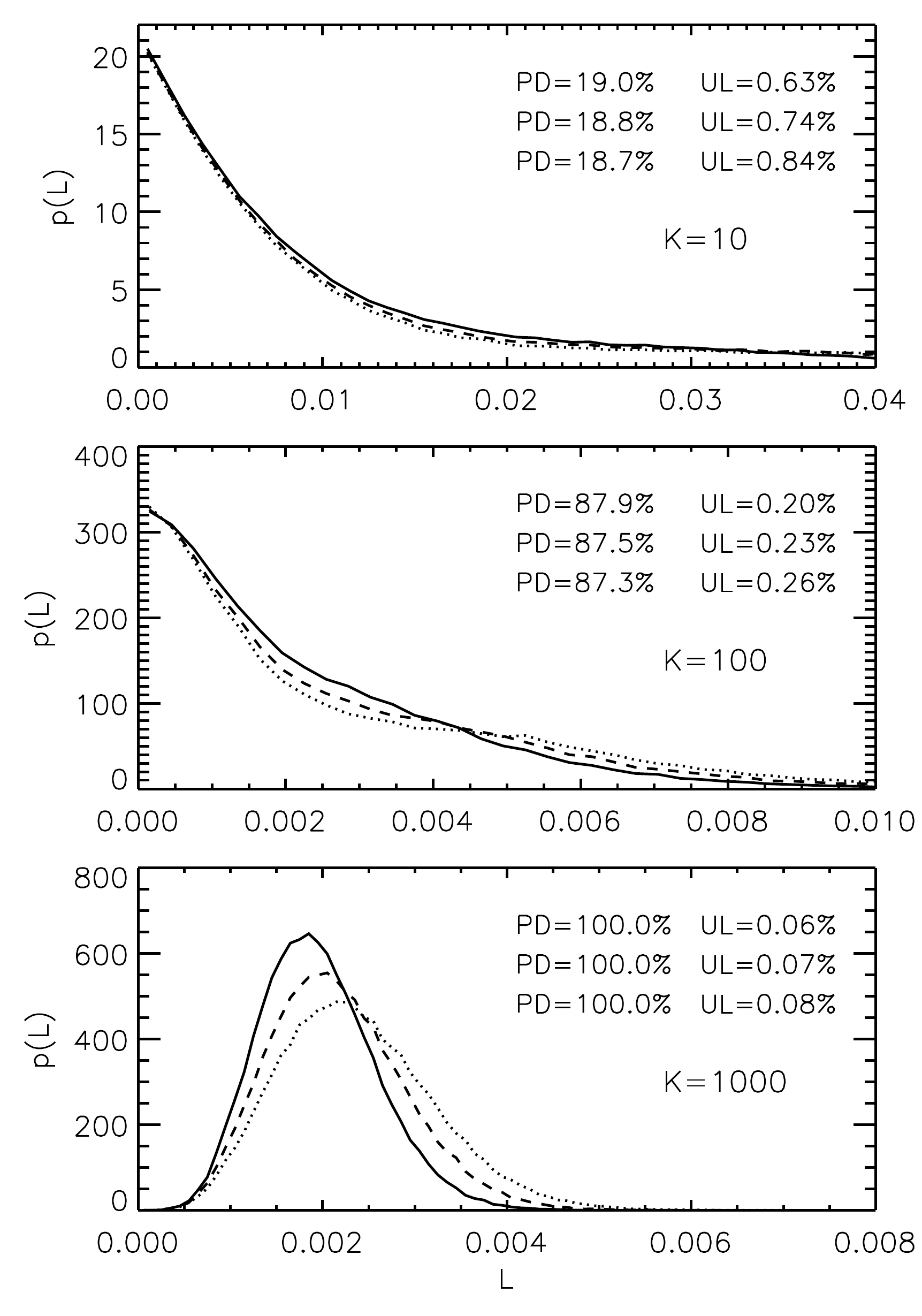}
\caption{\label{fig20f}Loss distributions for portfolios with different
standard deviations $\sigma_J$ of the jump size. $\sigma_J=0.2$ (solid
lines), $\sigma_J=0.3$ (dashed lines) and $\sigma_J=0.4$
(dotted lines) for three different portfolio sizes $K=10,100,1000$,
respectively.  The mean is always $\mu_J=-0.4$. 
The expected loss EL is 0.20\%, 0.22\% and 0.24\%, and the portion of
negative jumps are 96\%, 91\% and 87\% for these $\sigma_J$. 
The insets show, for every portfolio
size, the portfolio default probability PD and the unexpected loss UL
 from top to bottom for $\sigma_J=0.2,0.3,0.4$.}
\end{figure}

As discussed in Sec.~\ref{ourmodel1}, we make the specific assumption
that the distribution of the jump sizes is log--normal. Hence, it is
important to see how the loss distributions depend on the parameters
of the jump size distribution.  The losses grow when the absolute
value of the jump mean $\mu_J$ increases.  This is illustrated in
Fig.~\ref{fig20e}.  As for the jump intensity, the sensitivity of the
loss distribution to the jump mean does not significantly depend on
the portfolio size. Furthermore, as displayed in Fig.~\ref{fig20f},
the credit losses slightly increase, for all portfolio sizes, with the
jump size standard deviation $\sigma_J$.

\subsection{Correlations}
\label{num7}

\begin{figure}
\includegraphics[width=\linewidth]{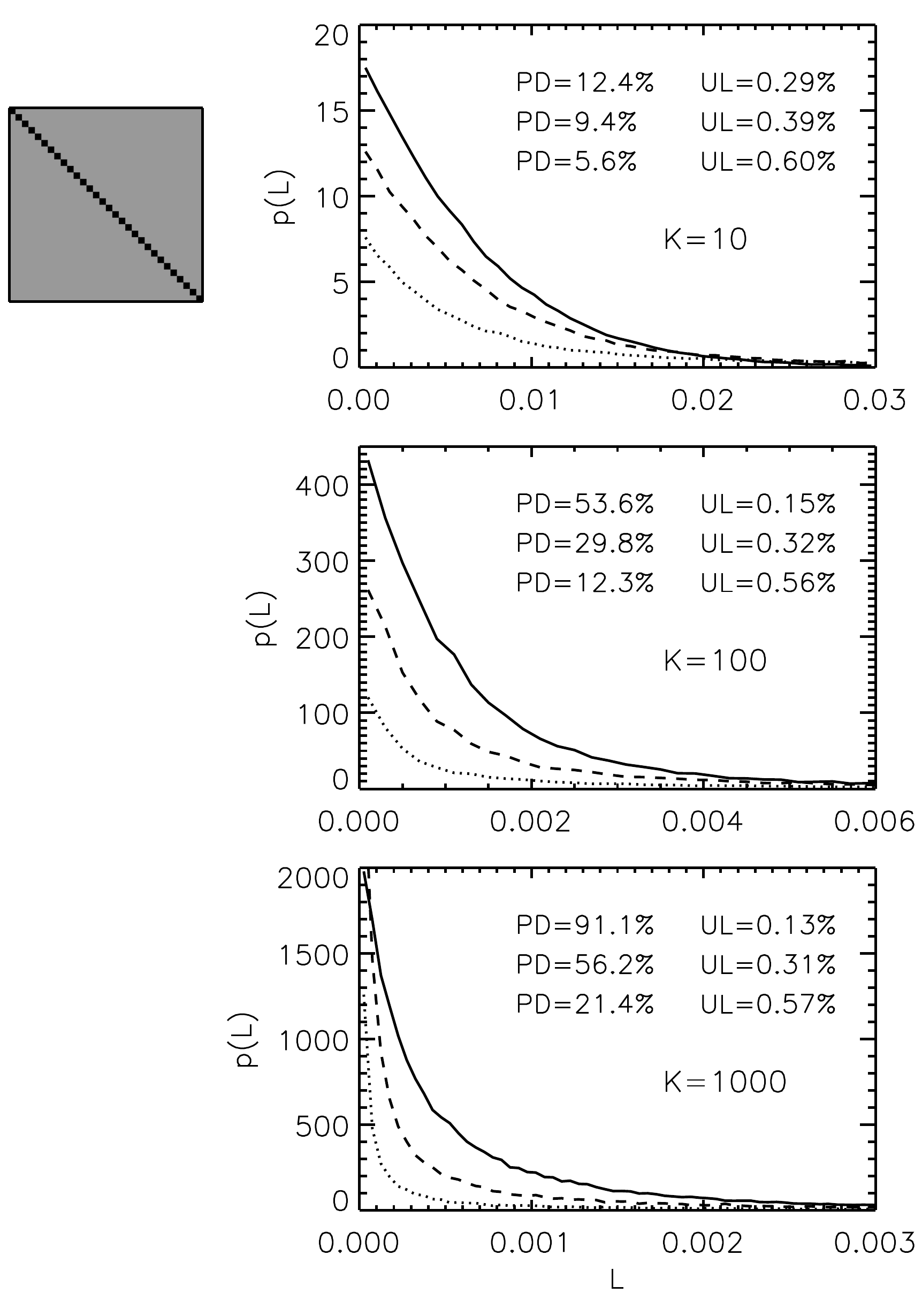}
\caption{\label{fig21} Loss distributions for portfolios with different
branch correlations. The top left plot shows the structure of the
correlation matrix, all companies are in one branch, indicated
in grey. The other three plots show the loss distributions for three
different portfolio sizes $K=10,100,1000$, respectively.  The
correlations are measured by the branch correlation parameter $c=0.2$
(solid lines), $c=0.5$ (dashed lines) and $c=0.8$ (dotted
lines). The expected loss is 0.076\%
for all three branch correlations, i.e.\ it is independent of the
branch correlation. The insets show, for every portfolio size, the
portfolio default probability PD and the unexpected loss UL from top to bottom for
$c=0.2,0.5,0.8$.}
\end{figure}

To begin with, we choose the simplest possible correlation matrix
$C=C(T)$, i.e.~a portfolio where all companies are in the same
branch. Figure~\ref{fig21} shows the structure of the correlation
matrix and the loss distributions for different portfolio sizes and
branch correlations $c=C_b$. In the sequel, the structure of the
correlation matrix is always presented as a gray scale image where the
intensity corresponds to the correlation. Black areas stand for
unit correlation, white areas for zero correlation.
The expected loss EL is 0.076\% for all three branch correlations, 
i.e.\ it is independent of the branch correlation and, as in the case of 
uncorrelated portfolios, it is also independent of the portfolio size.
The unexpected loss UL becomes larger
as the branch correlation grows, while the portfolio default probability PD decreases.
The latter can be understood, because the portfolio increasingly acts as a single company with growing correlation strength. The portfolio default probability makes a transition from $1-(1-P_D)^K$, corresponding to $K$ uncorrelated companies ($c=0$), to $P_D$ for a single company ($c=1$).

\begin{figure}
\includegraphics[width=0.85\linewidth]{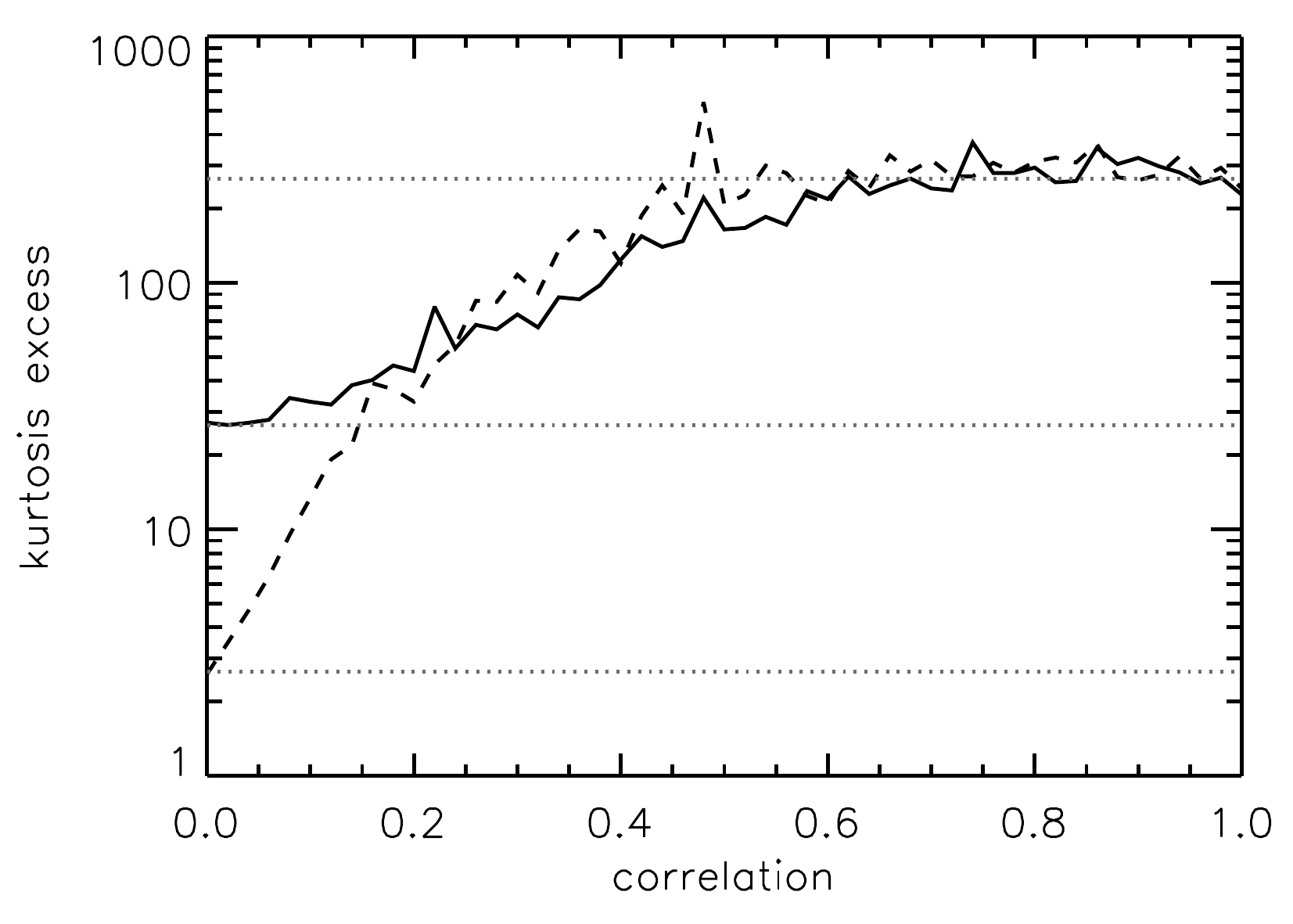}
\caption{\label{fig21a}The kurtosis excess as a function of the
correlation parameter $c$ for portfolio sizes
$K=10$ (solid line) and $K=100$ (dashed line).
The horizontal dotted lines indicate the kurtosis excess of 
the corresponding uncorrelated portfolios, $264.6/K$, 
for $K=1$, 10 and 100, respectively.
}
\end{figure}

\begin{figure}
\includegraphics[width=\linewidth]{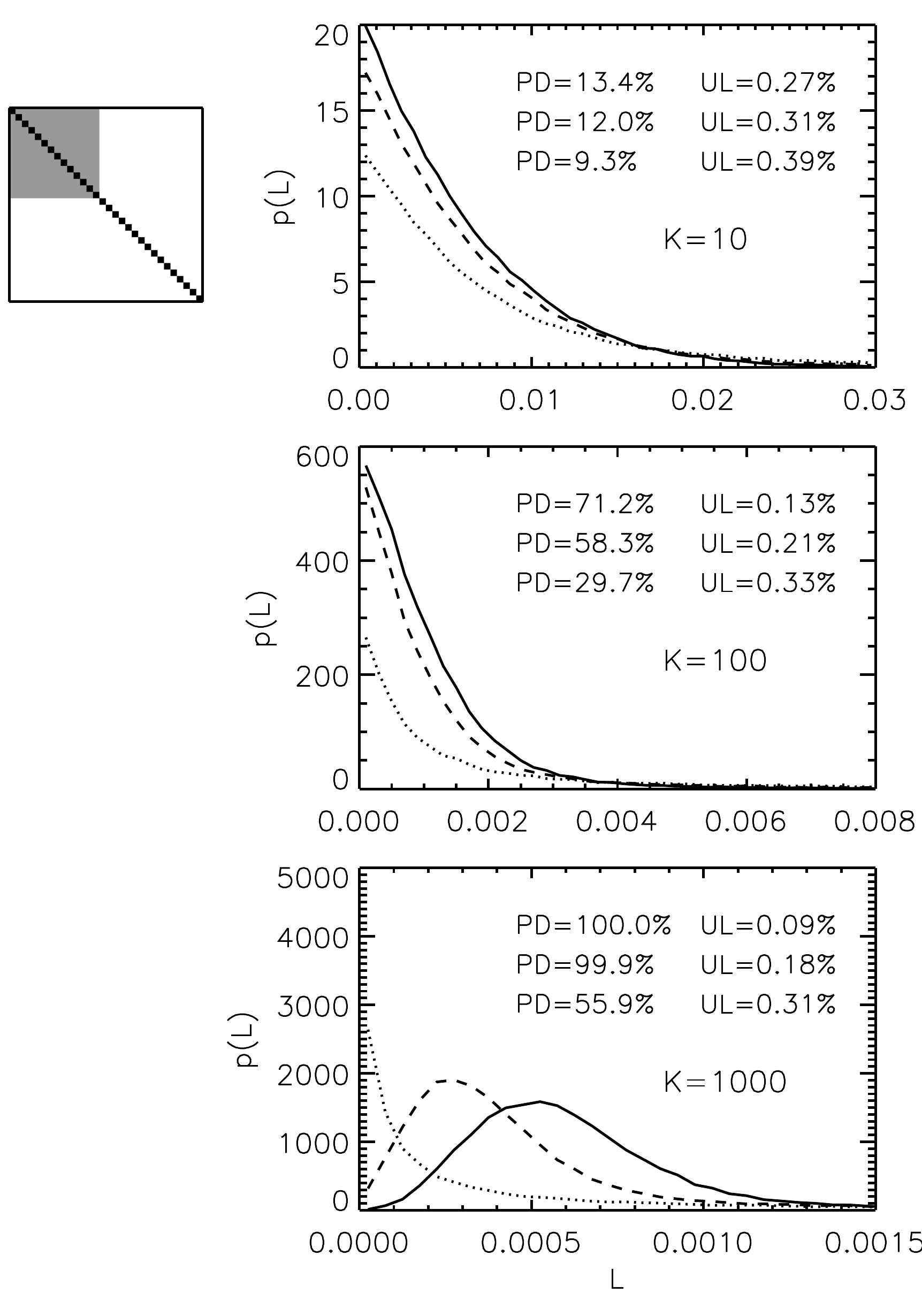}
\caption{\label{fig22} Loss distributions for portfolios with different
branch size. The top left plot shows the structure of the correlation
matrix, only the companies within the grey block are in the
branch. The other three plots show the loss distributions for three
different portfolio sizes $K=10,100,1000$, respectively.  The branch
sizes are (in percent of $K$) $\kappa_1=30\%$ (solid lines),
$\kappa_1=60\%$ (dashed lines) and $\kappa_1=100\%$ (dotted
lines). The expected loss is 0.076\%
for all three branch sizes. The insets show, for every portfolio size, 
the portfolio default probability PD
and the unexpected loss UL from top to bottom for
$\kappa_1=30\%,60\%,100\%$.}
\end{figure}

Figure~\ref{fig21} shows that, as for a majority of the model
parameters, the impact of the correlation on the loss distribution
increases with the portfolio size. Remarkably, there is a sizable
deviation from a Gaussian--type--of shape. 
This is seen in Fig.~\ref{fig21a} which shows the kurtosis
excess of the loss distribution as a function of the branch
correlation $c=C_b$ for this one--branch scenario. 
The kurtosis excess for a portfolio of size $K$ shows a transition 
from the value $264.6/K$ for an uncorrelated portfolio ($c=0$) to 
$264.6$, which corresponds to the value for an individual obligor.
Between $c=0.6$ and $c=1$ the kurtosis excess even exceeds 
this value. 
To gain further insight into the influence of correlations, we now assume
that the size $\kappa_1$ of the branch is smaller than the total size
$K$ of the correlation matrix. Thus, $\kappa_1$ companies are
correlated, $K-\kappa_1$ are not.  In Fig.~\ref{fig22} we plot loss
distributions for different branch sizes $\kappa_1$, the branch
correlation is $c=C_b=0.5$. 
We observe that bigger branch size yields
a higher unexpected loss UL and a lower portfolio default probability PD. 
Importantly,
the loss distributions for $K=1000$ show a transition from a strongly
leptokurtic to a more Gaussian--type--of, but still asymmetric,
distribution as $\kappa_1$ decreases. This clearly visualizes the
effect already mentioned in Sec.~\ref{ana1}.  A symmetric
Gaussian--type--of shape can only be reached without correlations
between the asset processes. 

\begin{figure}
\includegraphics[width=\linewidth]{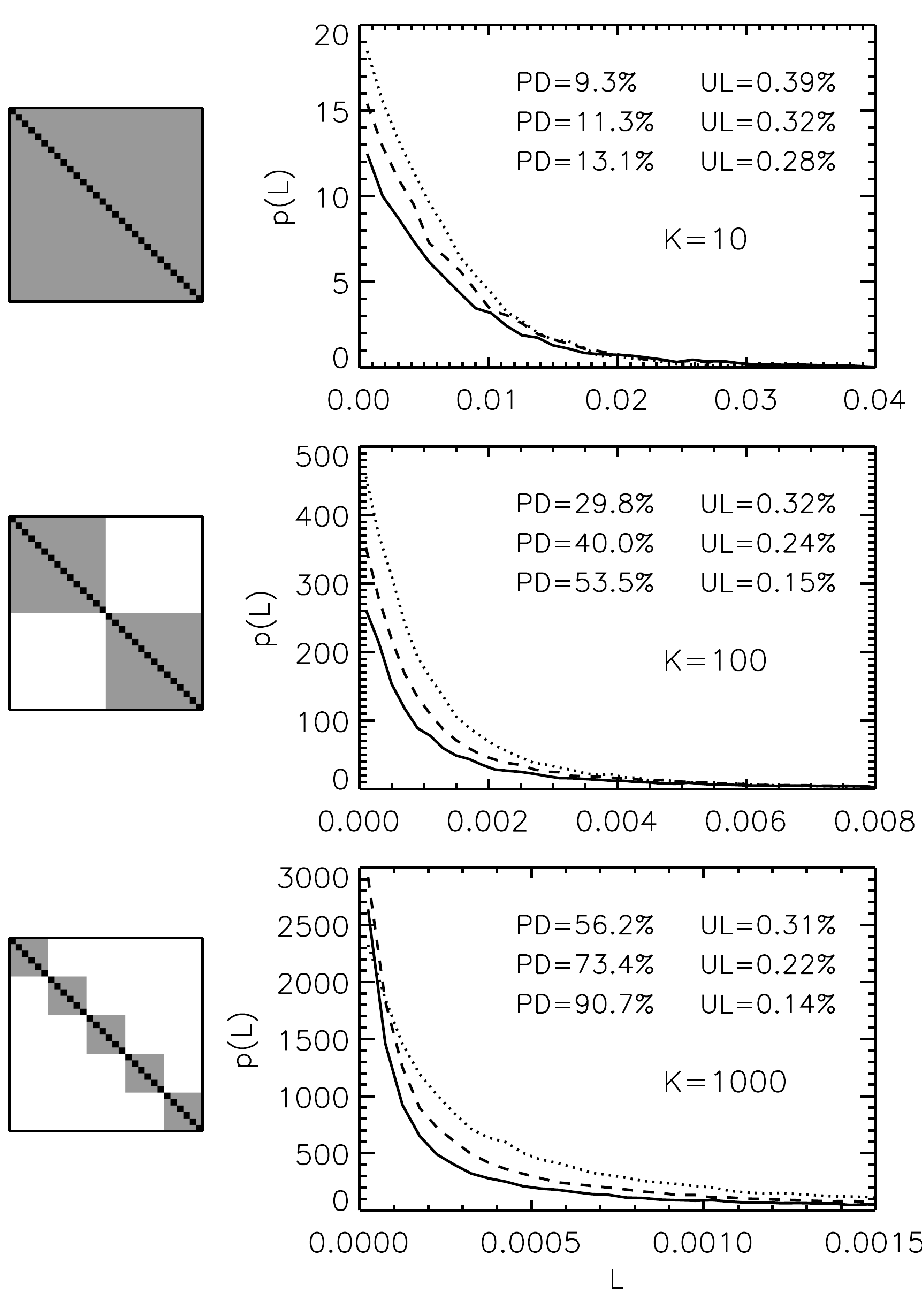}
\caption{\label{fig23} Loss distributions for portfolios with different
number $B$ of branches. The left column shows the structure of the
correlation matrices, labeled C1, C2, C3, respectively.  The right
column shows the loss distributions for three different portfolio
sizes $K=10,100,1000$ for C1 (solid lines), C2 (dashed lines)
and C3 (dotted lines). The branch correlation is $c=C_b = 0.5$. The
default probability is 1.49\%. The insets show, for every portfolio
size, the expected loss EL and the unexpected loss UL from top to
bottom for C1, C2, C3.}
\end{figure}

In a realistic economical scenario, companies will belong to $B$
different branches. In Fig.~\ref{fig23}, results are depicted for
portfolios consisting of companies in $B=1,2,5$ branches, all with the
same branch correlation $c=C_b = 0.5$. The structure of the
correlation matrices is also shown in the figure.  The number of
branches does not significantly affect the loss
distribution. Moreover, the loss distribution does not approach the
normal distribution as the size of the portfolio increases.
This is so, because the high branch correlation $c=C_b = 0.5$ makes
the branches in the portfolio behave like individual companies. Hence,
the $K\times K$ correlation matrix is effectively only $B\times B$.
The curves in the subfigures of Fig.~\ref{fig23} are almost identical,
since for high $c=C_b$ there is no significant difference between
portfolios containing one, two or five companies.

\begin{figure}
\includegraphics[width=\linewidth]{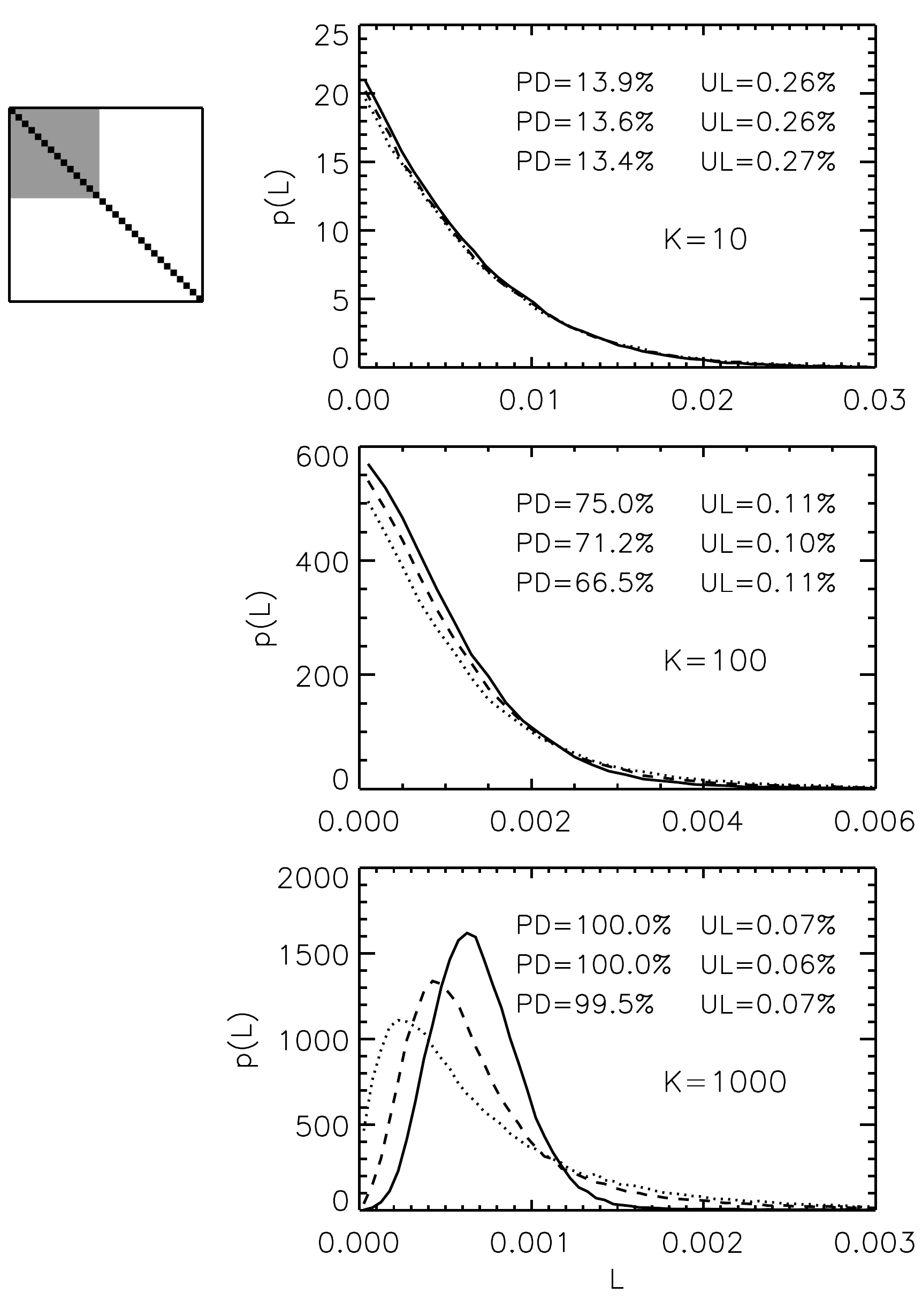}
\caption{\label{fig24} Loss distributions for portfolios where the product
of the branch size and the branch correlation is constant. The top
left plot shows the structure of the correlation matrix.  The other
three plots show the loss distributions for three different portfolio sizes
$K=10,100,1000$, respectively.  The branch sizes are (in percent)
$\kappa_1=10\%$ (solid lines), $\kappa_1=50\%$ (dashed lines)
and $\kappa_1=90\%$ (dotted lines). The expected loss is 0.076\%
for all three branch sizes.  The insets show, for every portfolio size,
the portfolio default probability PD and the unexpected loss UL from top to bottom for
$\kappa_1=10\%,50\%,90\%$.}
\end{figure}

An interesting question is whether a small, highly correlated branch
has more or less effect than a large branch with weakly correlated
companies. To look into this, we investigate how the loss distribution
depends on the product $\kappa_1 c$ of the branch size $\kappa_1$ and
the branch correlation $c=C_b$.  Figure~\ref{fig24} shows loss
distributions for portfolios where 90\%, 50\% and 10\% of the obligors
in the portfolio are in a branch. The correlations for these branches
are $0.1$, $0.18$ and $0.9$ respectively. Thus, the product is
$\kappa_1 c=9$ in all three cases.  We notice that $\kappa_1$ denotes
the branch size in percent of the total number $K$ of obligors.  We
conclude that a small, highly correlated branch and a larger branch
with less correlated companies are equivalent from a creditor's point
of view. The unexpected loss UL is very similar, although not identical, 
for the three different parameter settings, while the portfolio default
probability PD is decreasing with increasing $\kappa_1$. 
The effect on PD is observed to be strongest for intermediate 
size portfolios ($K=100$ in our case).

\subsection{Jumps versus correlations}
\label{num8}

\begin{figure}
\includegraphics[width=0.8\linewidth]{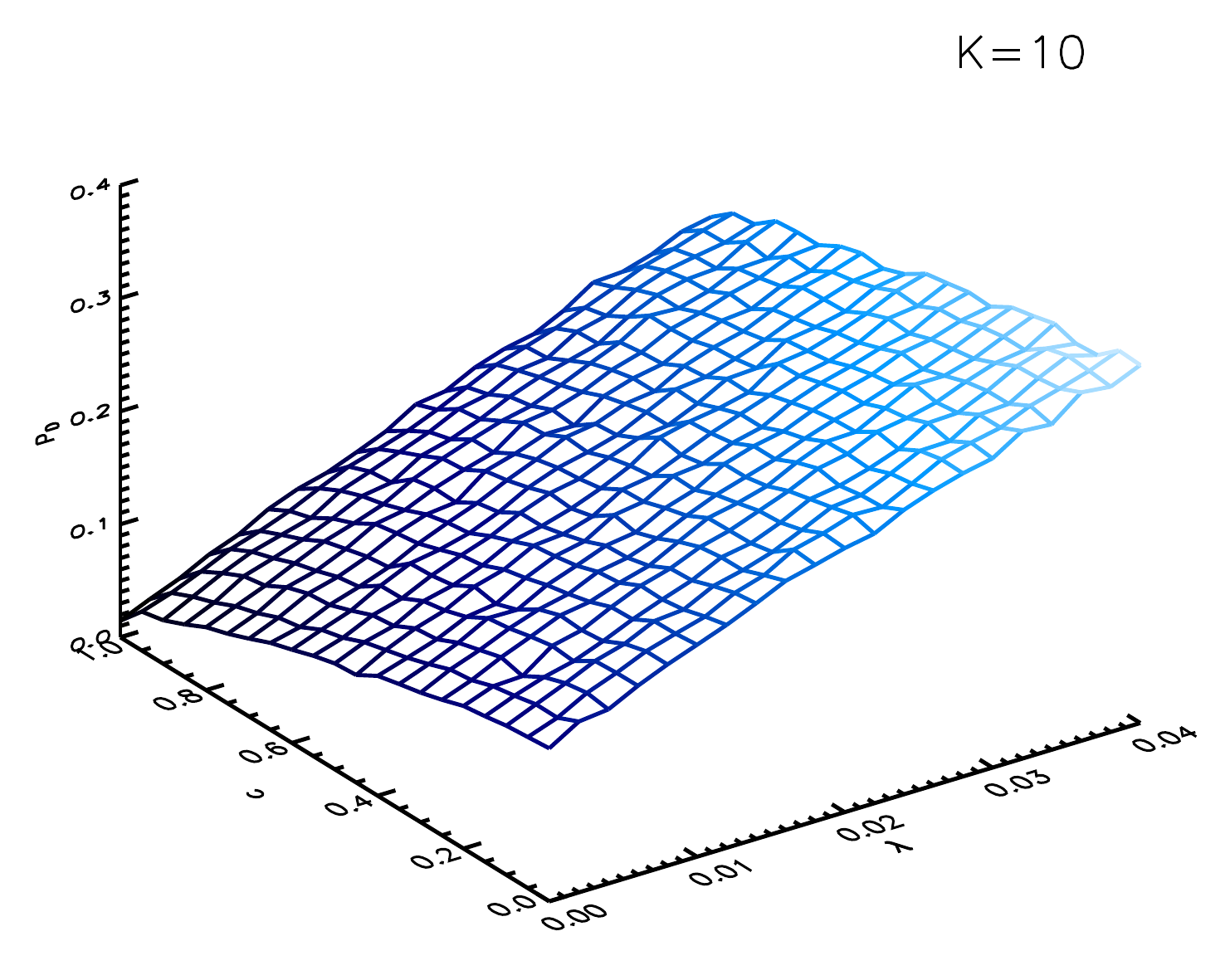}
\includegraphics[width=0.8\linewidth]{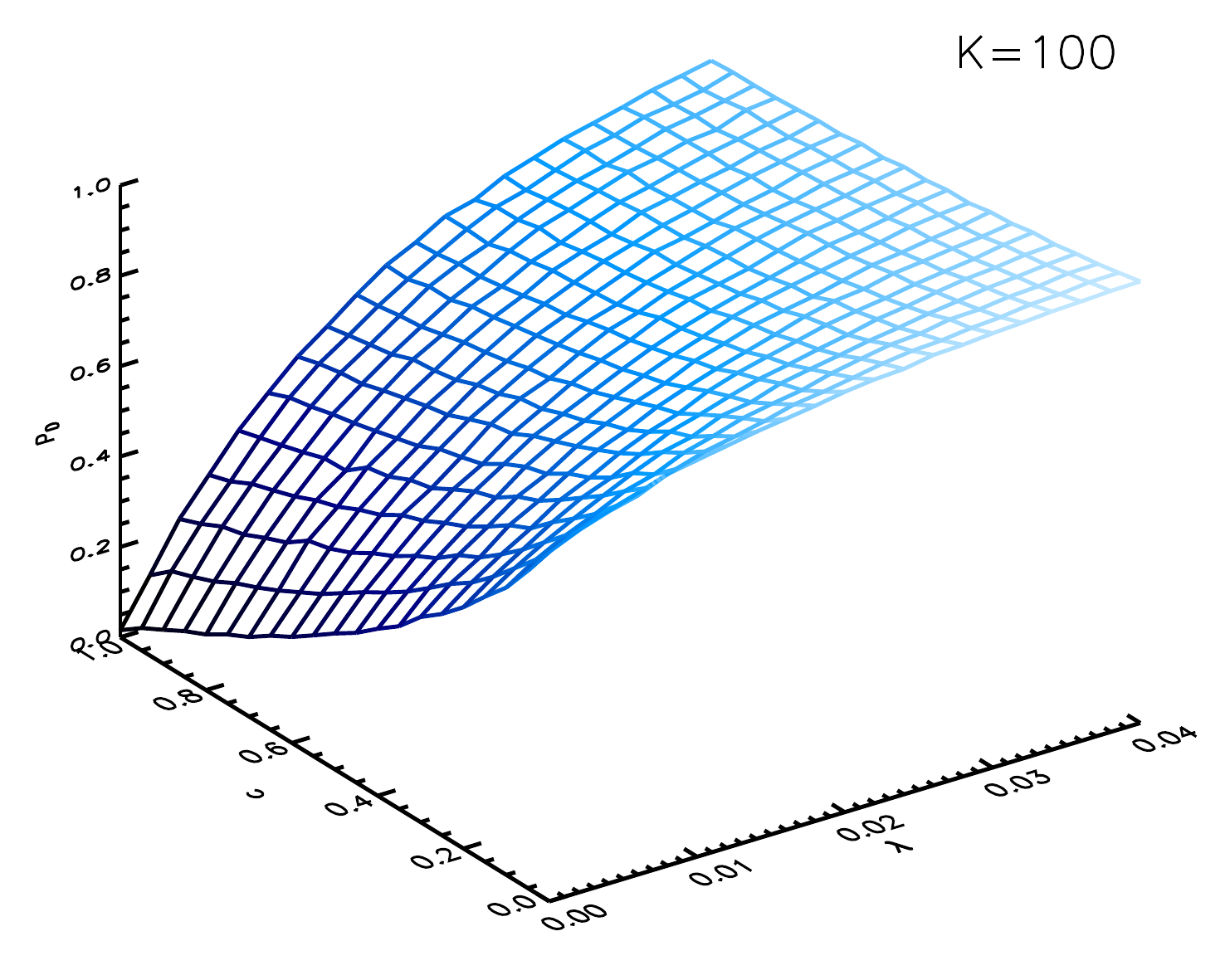}
\includegraphics[width=0.8\linewidth]{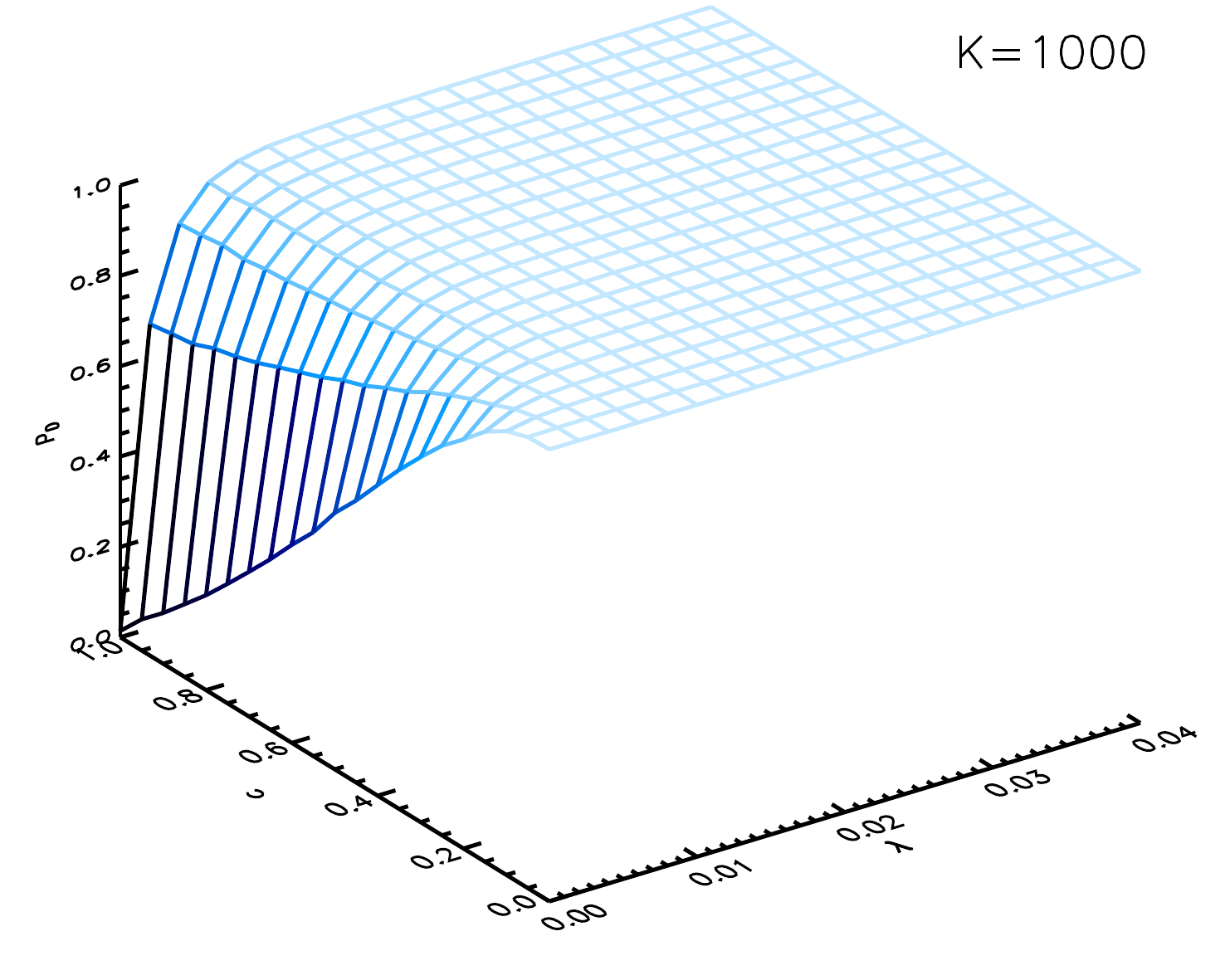}
\caption{\label{fig26} Portfolio default probability PD as a function of the 
jump intensity $\lambda$ and the correlation parameter $c$ for
different portfolio sizes $K=10,100,1000$, respectively. 
}
\end{figure}

\begin{figure}
\includegraphics[width=0.8\linewidth]{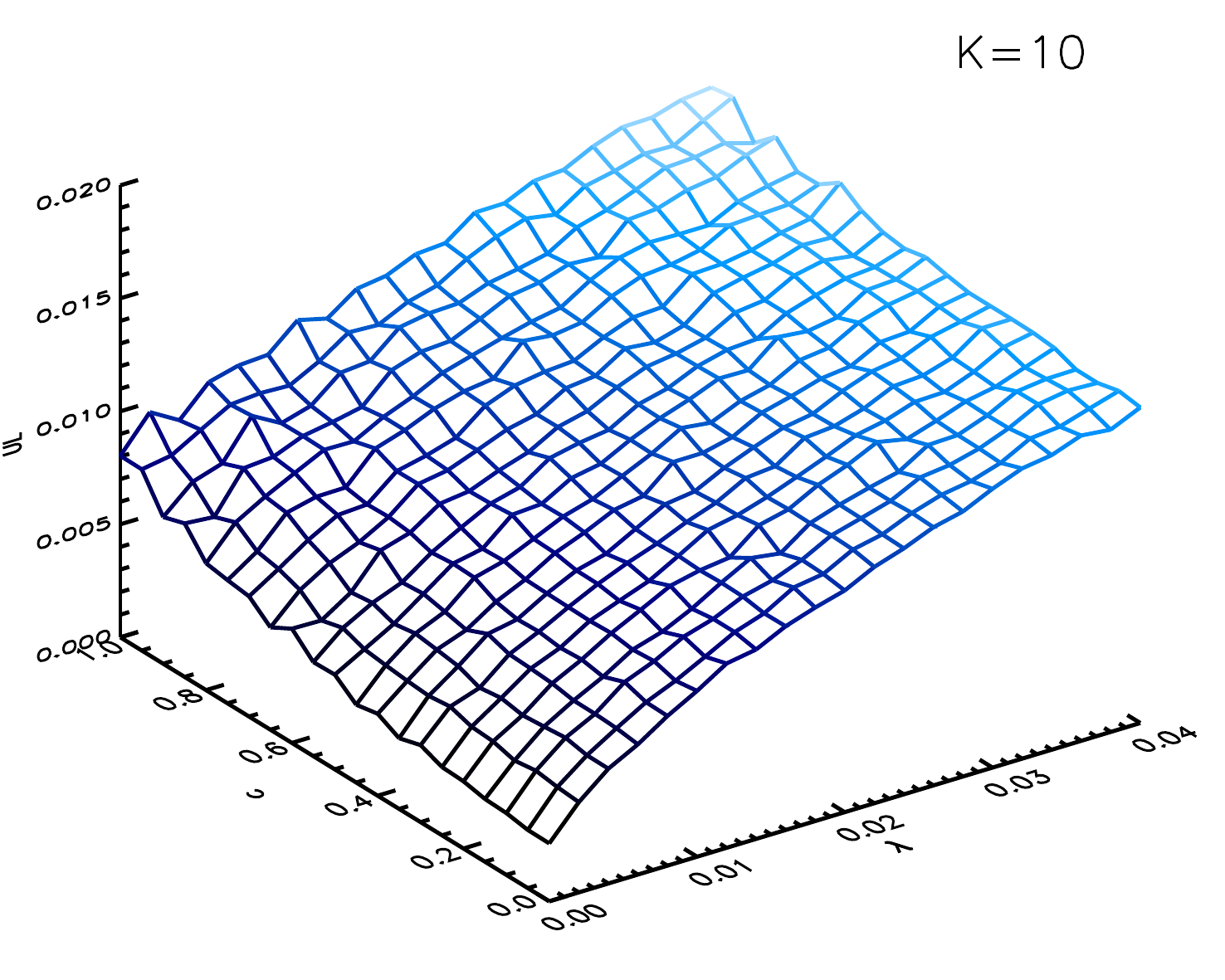}
\includegraphics[width=0.8\linewidth]{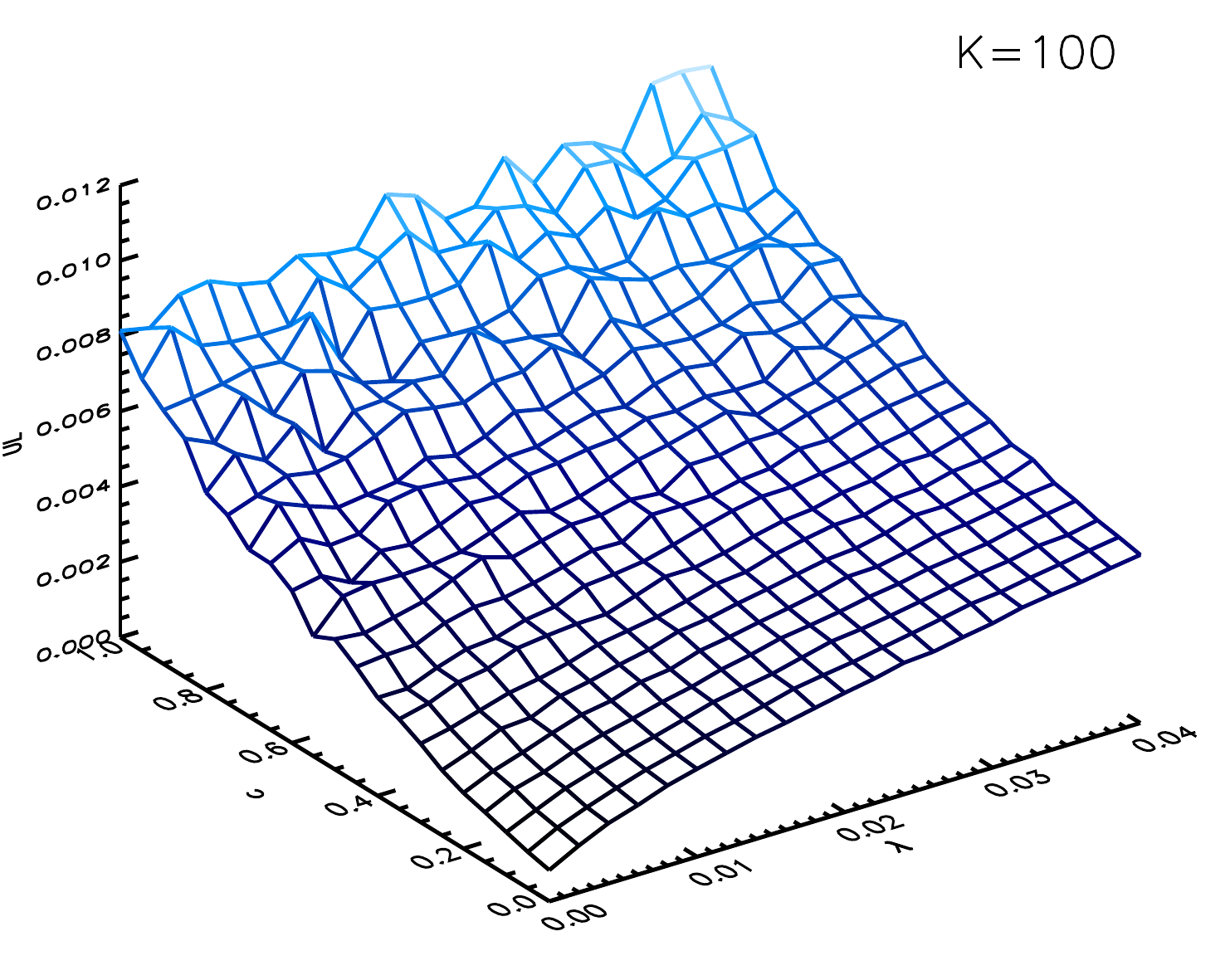}
\includegraphics[width=0.8\linewidth]{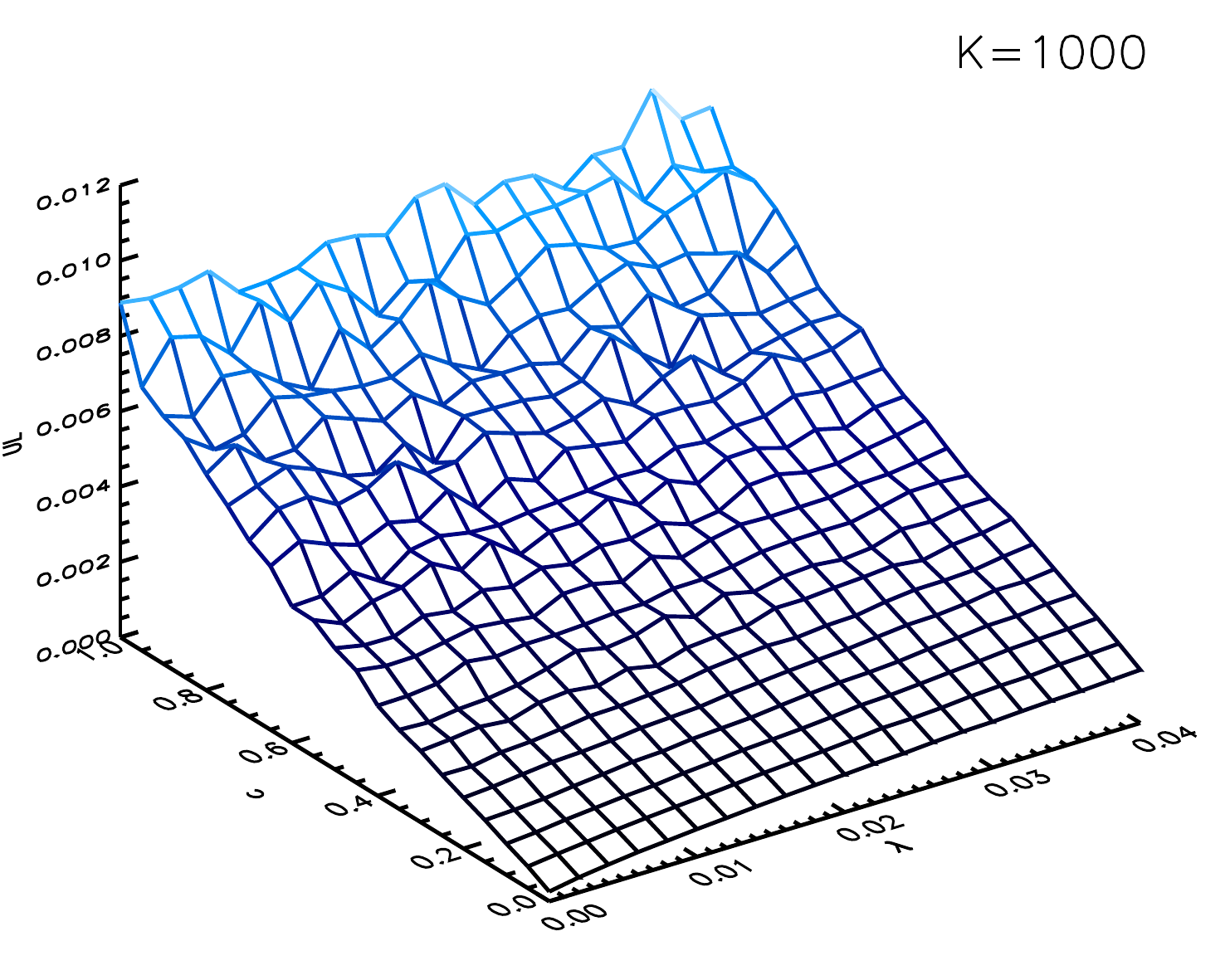}
\caption{\label{fig26a} Unexpected loss UL as a function of the 
jump intensity $\lambda$ and the correlation parameter $c$ for
different portfolio sizes $K=1,10,100,1000$, respectively. 
}
\end{figure}

Here, we study the interplay between jumps and correlations.  In
Figs.~\ref{fig26} and~\ref{fig26a} the portfolio default probability PD and the
unexpected loss UL are displayed as functions of the jump intensity
$\lambda$ and the branch correlation $c$. To keep the discussion
transparent, the correlation structure is a single branch containing
all companies. 
The portfolio default probability PD increases with the jump intensity and decreases with growing correlation. For fixed small jump intensity $\lambda$ both PD and UL show a transition from $K$ uncorrelated obligors to an individual one. However, for larger values of $\lambda$ the influence of the uncorrelated jumps becomes more dominant leading to a less pronounced dependence on the correlation strength $c$. The unexpected loss UL increases both with the jump intensity $\lambda$ and with the correlation $c$. As the correlation changes from 0 to 1, the dependence of UL on the jump intensity changes from a sqare-root-type-of behavior to a linear one.
The expected loss EL (not shown) grows linearly with the
jump intensity and is
independent of the correlation strength and the portfolio size.

\subsection{Correlated jumps}
\label{num9}

\begin{figure}
\includegraphics[width=0.8\linewidth]{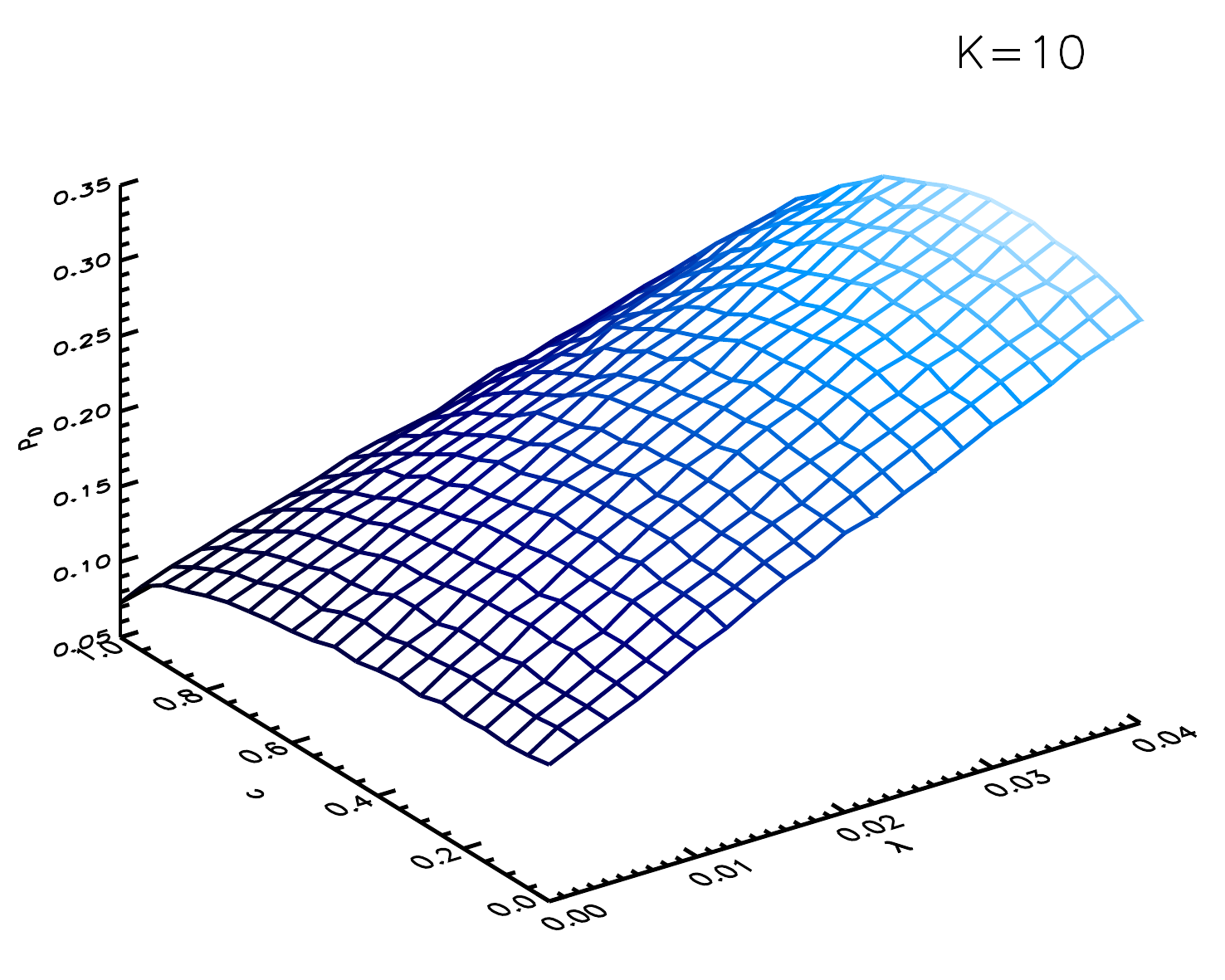}
\includegraphics[width=0.8\linewidth]{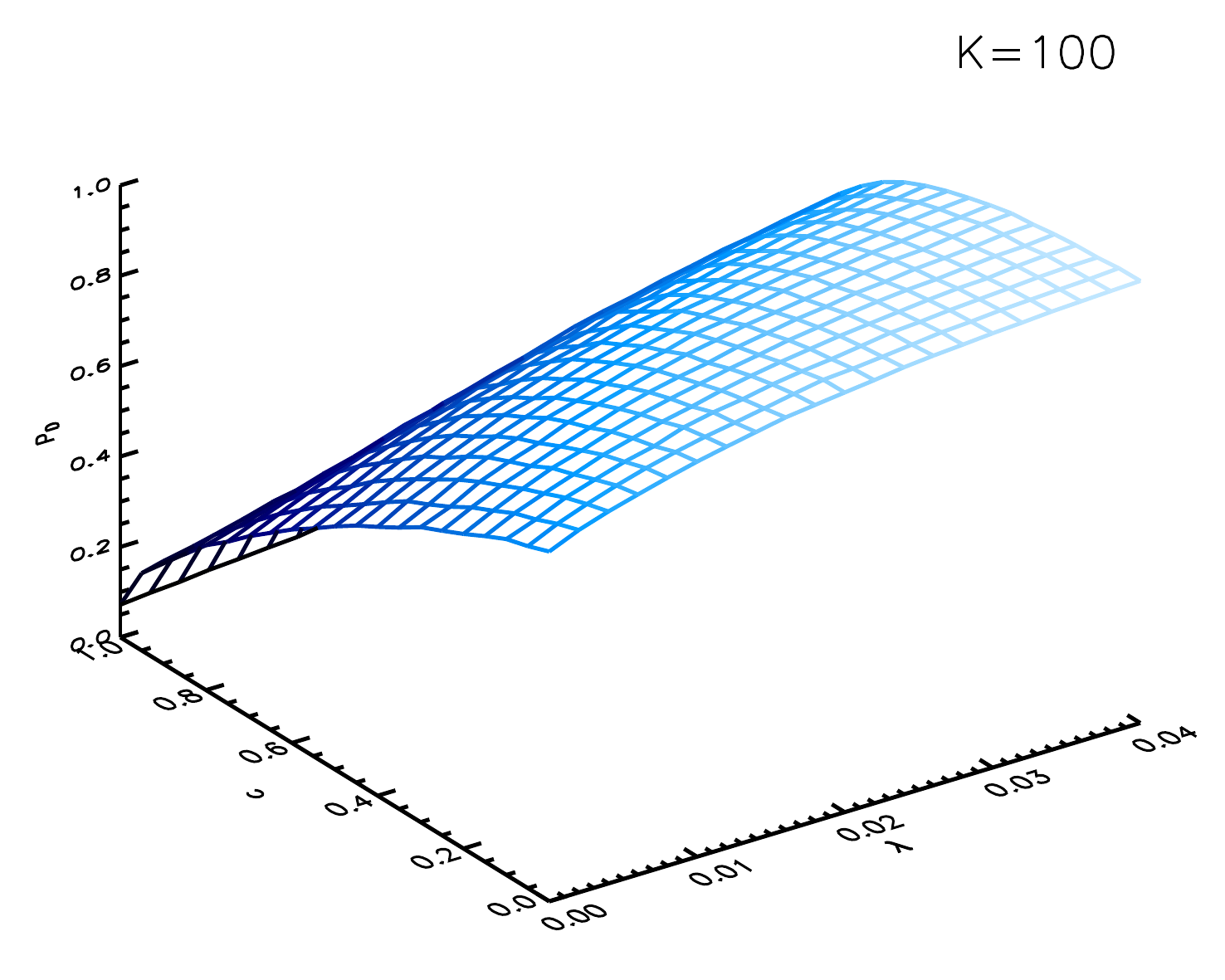}
\includegraphics[width=0.8\linewidth]{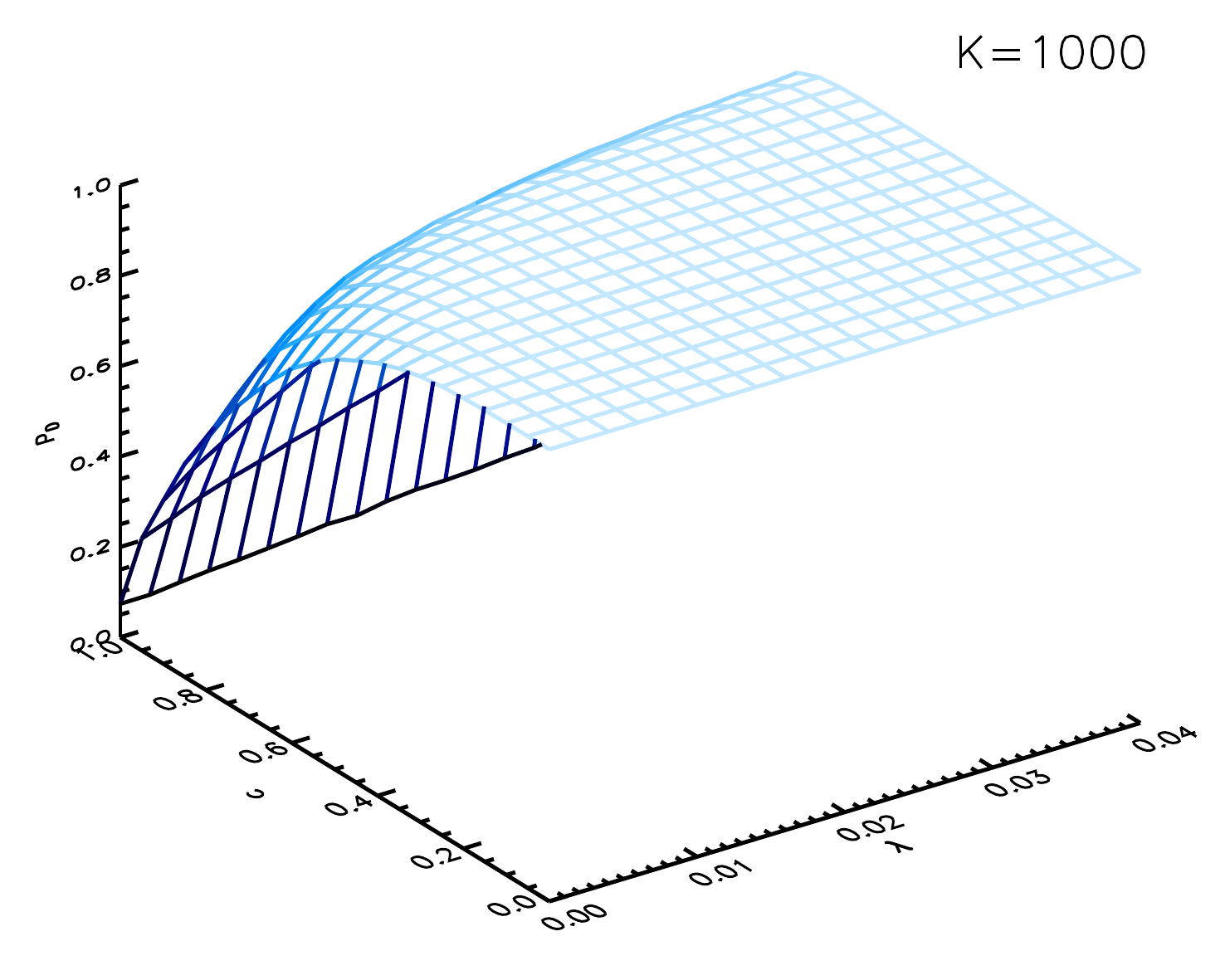}
\caption{\label{fig27} Portfolio default probability PD as a function of the 
jump intensity $\lambda$ and the correlation parameter $c$ for
different portfolio sizes $K=10,100,1000$, respectively. 
}
\end{figure}

\begin{figure}
\includegraphics[width=0.8\linewidth]{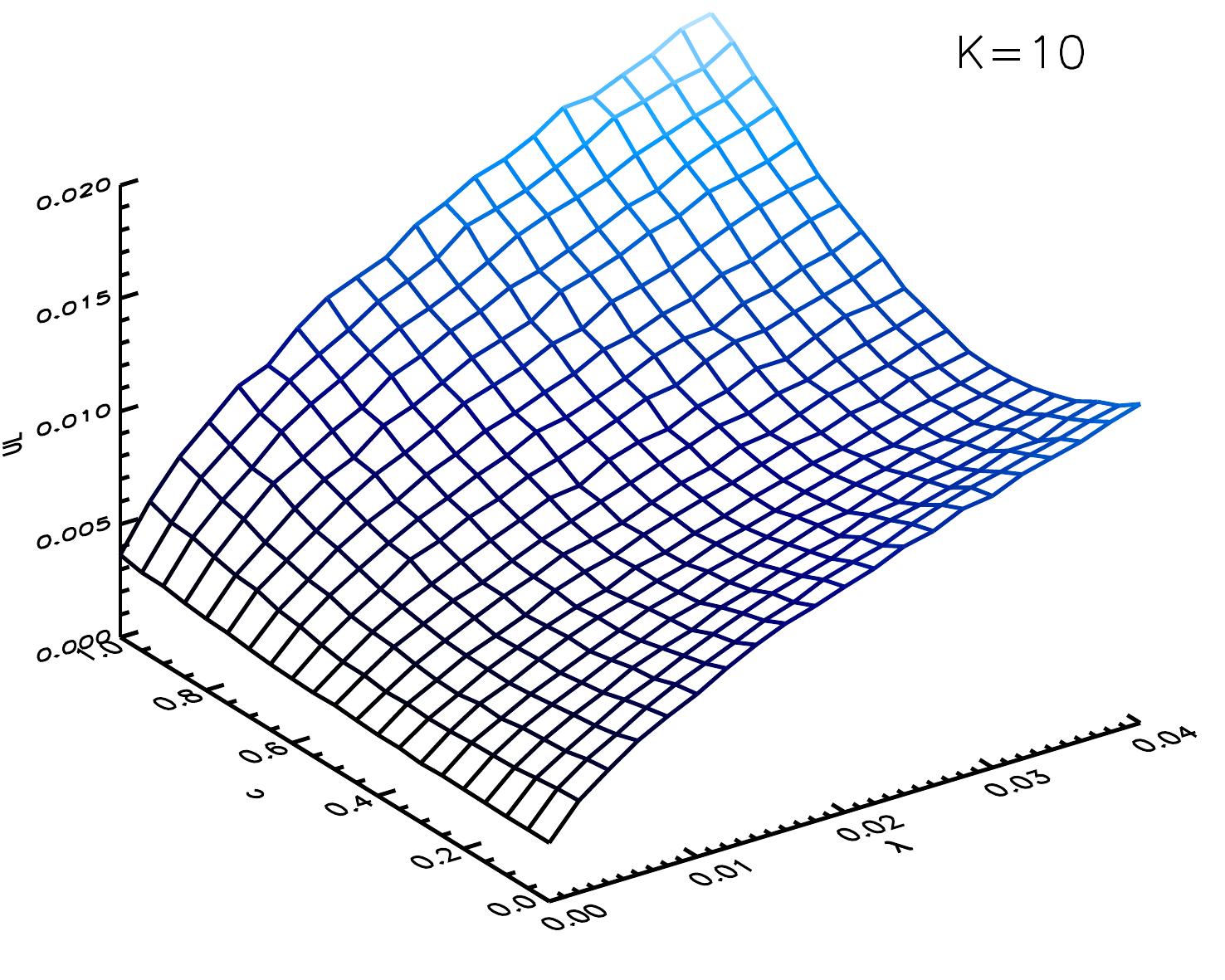}
\includegraphics[width=0.8\linewidth]{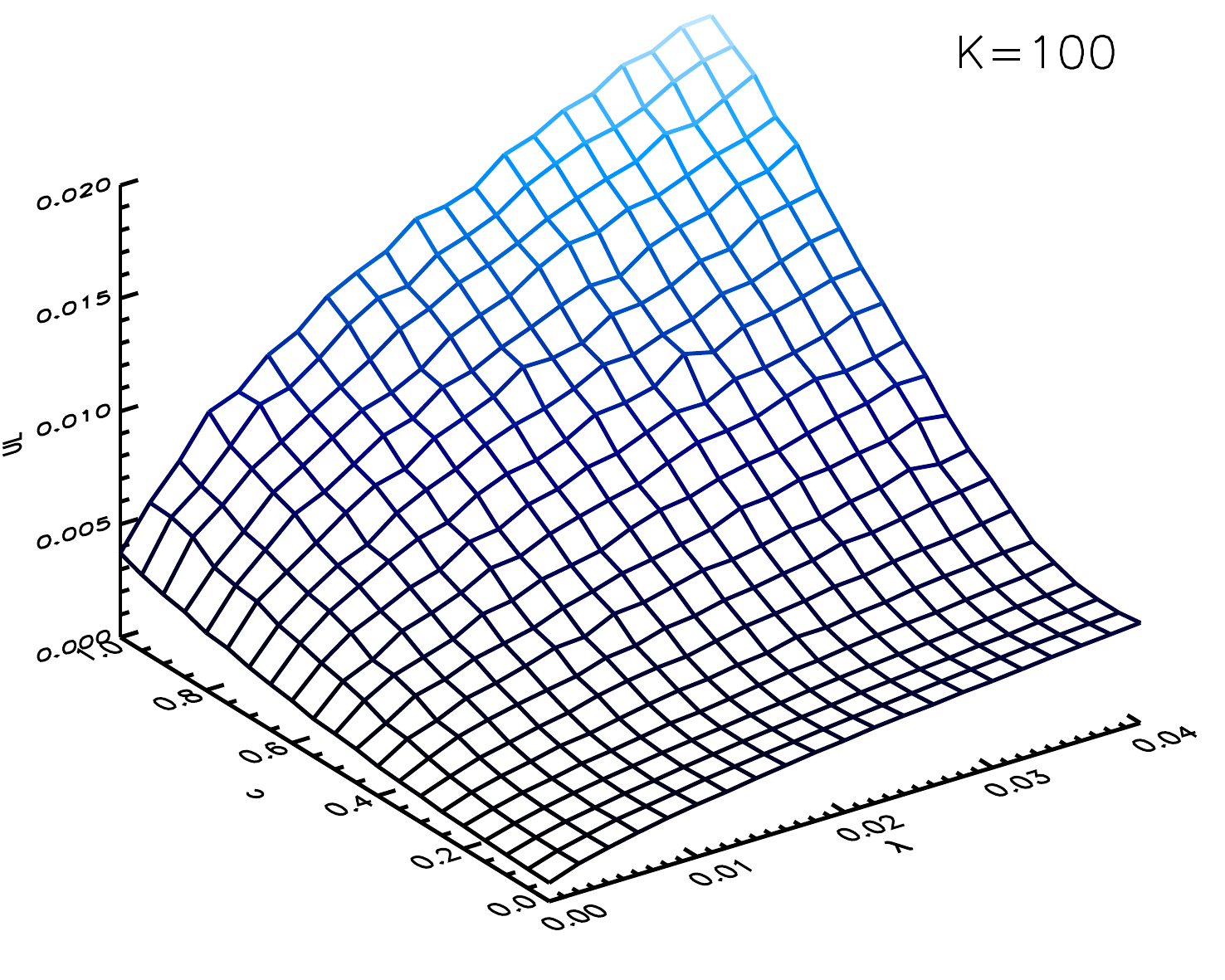}
\includegraphics[width=0.8\linewidth]{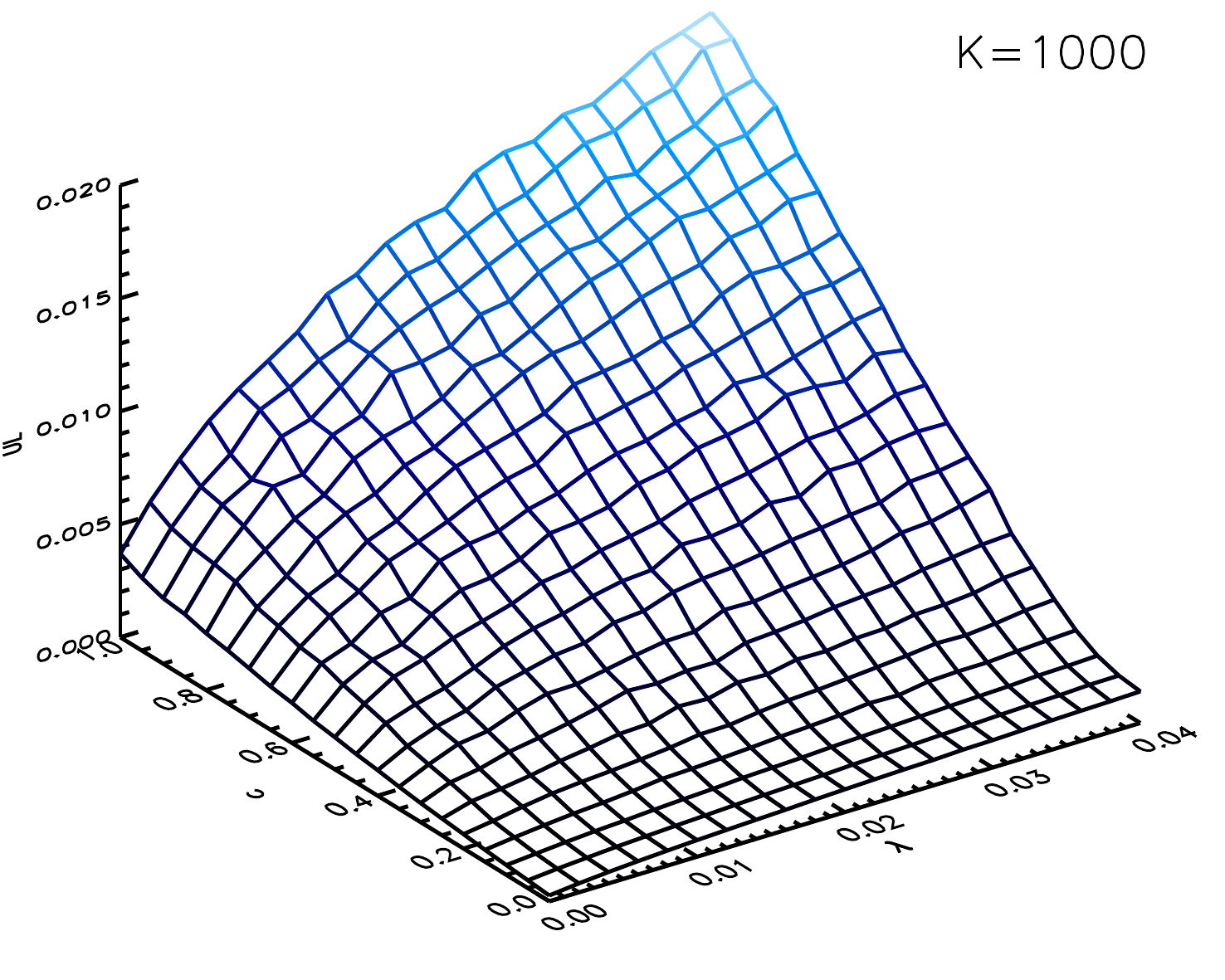}
\caption{\label{fig27a} Unexpected loss UL as a function of the 
jump intensity $\lambda$ and the correlation parameter $c$ for
different portfolio sizes $K=1,10,100,1000$, respectively. 
}
\end{figure}

In the previous section, we considered a correlated diffusion process with uncorrelated jumps. Now we study the case where also the jumps are correlated, i.e.~a jump can occur both in the branch specific time series $\eta_{b(k)}(t)$ and in the branch independent time series $\varepsilon_k(t)$. The jump sizes are scaled with the correlation coefficient according to Eq.~(\ref{serlin}). We consider a correlation structure with five branches of size $K/5$.
In Figs.~\ref{fig27} and~\ref{fig27a} the portfolio default probability PD and the
unexpected loss UL are displayed as functions of the jump intensity
$\lambda$ and the branch correlation $c$. 
As the correlation changes from 0 to 1, both PD and UL show a transition from $K$ uncorrelated obligors to the behavior of the five branches. For higher jump intensities, this transition can be non-monotonous --- for $K=10$ we observe a maximum in PD at $c\approx0.3$ and a minimum in UL at $c\approx0.35$. It is important to note that the unexpected loss UL increases tremendously with the correlation coefficient.
The expected loss EL (not shown) grows linearly with the jump intensity and is independent of the portfolio size. For fixed jump intensity it shows a minimum at $c=0.5$ due to the scaling of jump sizes according to Eq.~(\ref{serlin}).

\begin{figure}
\includegraphics[width=\linewidth]{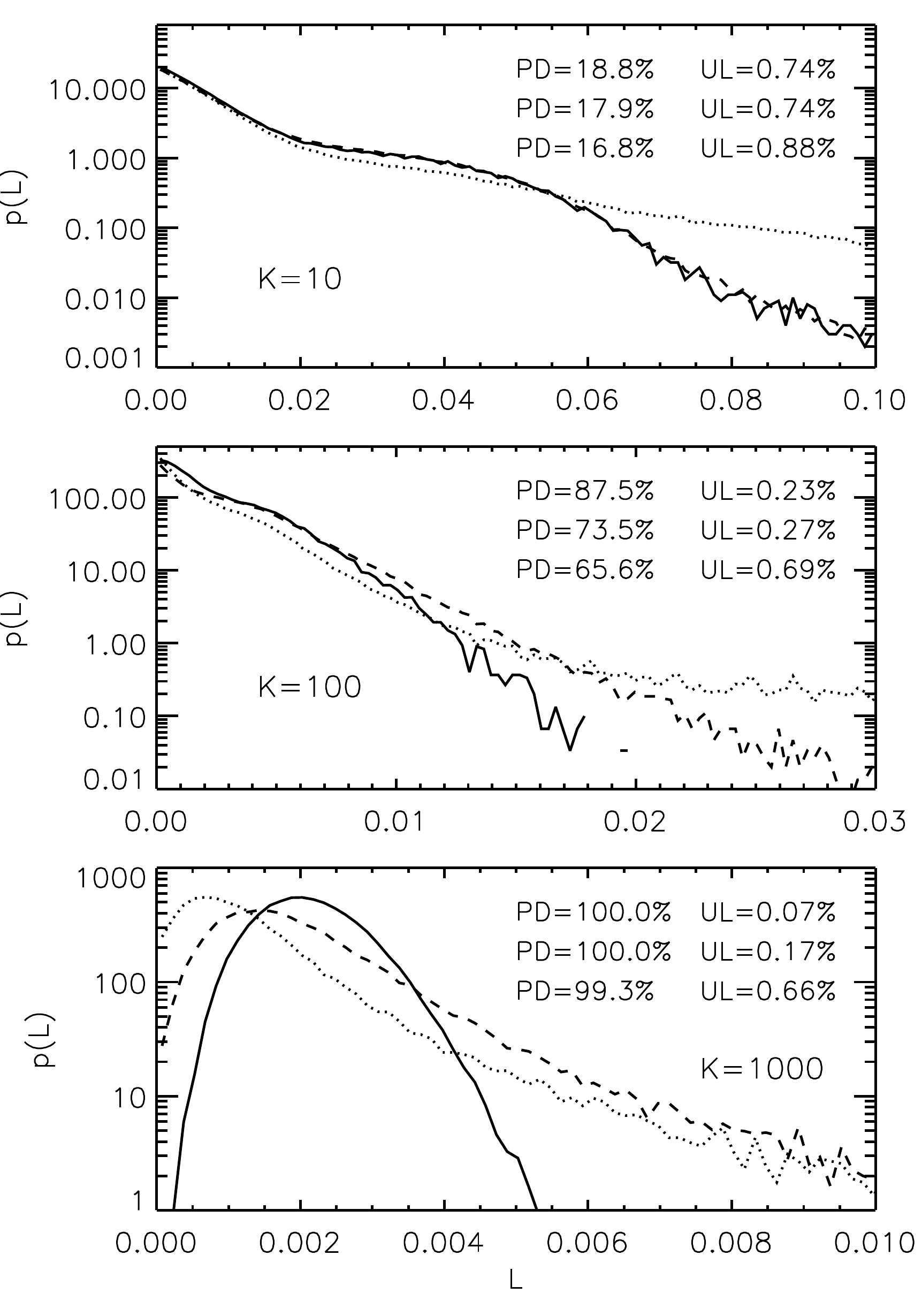}
\caption{\label{fig28}Loss distributions for uncorrelated jump diffusion (solid lines), correlated diffusion with uncorrelated jumps
(dashed lines) and correlated jump diffusion (dotted lines) for three
different portfolio sizes $K=10,100,1000$, respectively.  
The insets show, for every portfolio
size, the portfolio default probability PD and the unexpected loss UL 
from top to bottom for the uncorrelated case, correlated diffusion with jumps and correlated jump diffusion.}
\end{figure}

\begin{figure}
\includegraphics[width=\linewidth]{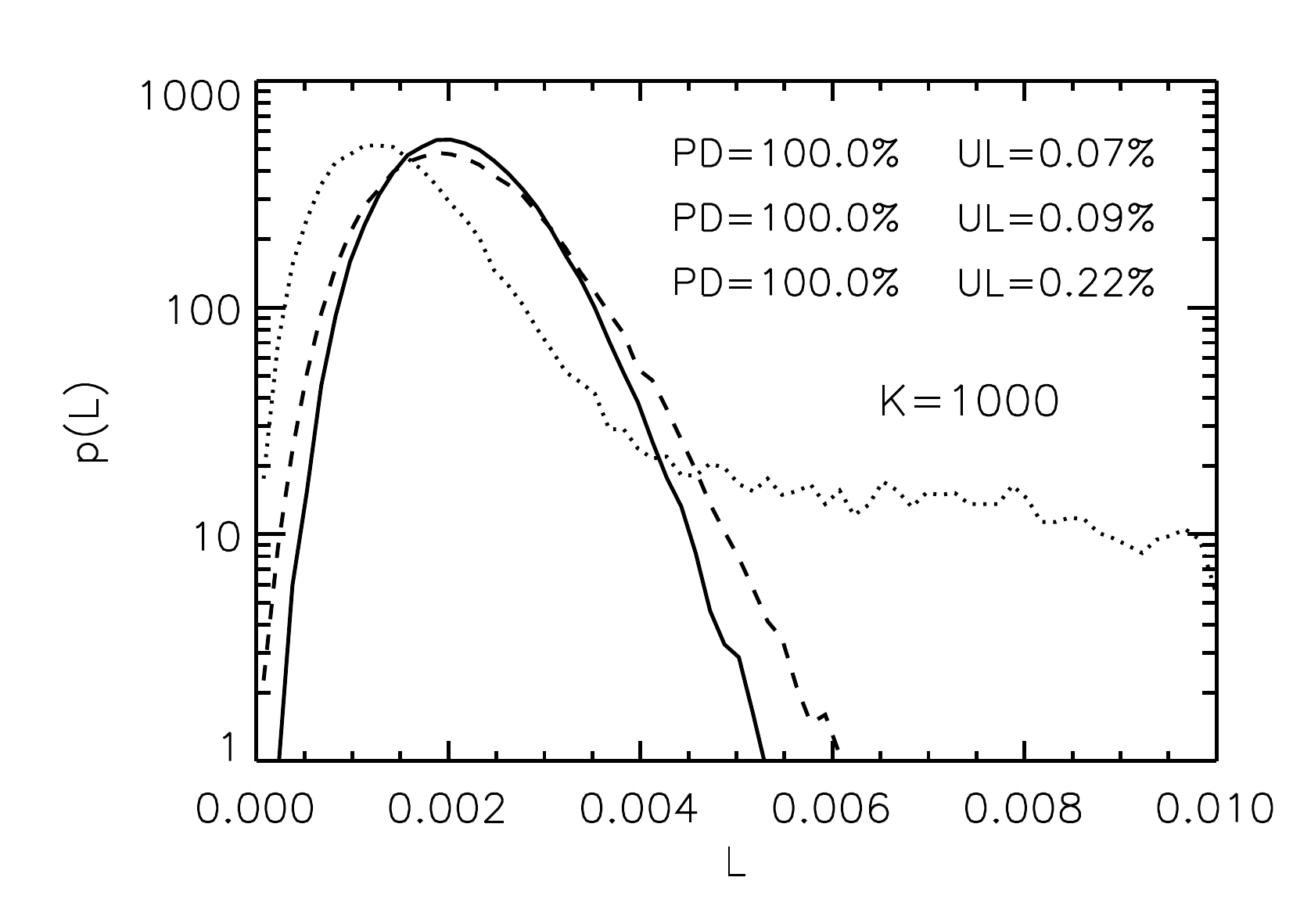}
\caption{\label{fig29}Loss distributions for uncorrelated jump diffusion (solid lines), correlated diffusion with uncorrelated jumps
(dashed lines) and correlated jump diffusion (dotted lines) for portfolio size $K=1000$ and 50 branches.  
The insets show the portfolio default probability PD and the unexpected loss UL 
from top to bottom for the uncorrelated case, correlated diffusion with jumps and correlated jump diffusion.}
\end{figure}

In Figure~\ref{fig28} we show how the loss distributions are affected by introducing correlations. We compare three cases: uncorrelated jump diffusion, correlated diffusion with uncorrelated jumps and fully correlated jump diffusion, as discussed above.
As correlation structure we choose again five branches of size $K/5$, and the correlation coefficient for each branch is set to $c=0.5$.
In the case of correlated jumps, we rescaled the jump intensity and jump sizes, so that they match the uncorrelated case.
The correlations in the diffusion already lead to fatter tails; introducing correlations between the jumps enhances this effect. For the case of only five branches, it is most visible in smaller and medium sized portfolios. 
Figure~\ref{fig29} shows the loss distribution for $K=1000$ and a correlation structure with 50 equally sized branches. The correlation of the diffusion terms leads to an only slightly fatter tail of the distribution, while the correlation of the jumps has a much more pronounced effect on the tail behavior.

\subsection{Drill down risk}
Finally, we consider how different indicators behave when removing one
company from a portfolio of initial size $K$. 
We start with a portfolio which is composed in a rather realistic way. 
It consists of a large portion of low risk bonds, but also contains a
portion of more risky ones. The portfolio was divided into five
different categories $\zeta = 1,2,...,5$. Each category has its own
setup of the initial values $V_0$ and face values $F$. All companies
within a single branch are identical. The portfolio composition is summarized in Tab.~(\ref{table_comp3}). The fraction of the $K$ companies that
belongs to a certain catagory is $\alpha$ and the fraction of
the total amount of money invested in that category is denoted $\gamma$.

\begin{table}[htbp]
\begin{center}\begin{tabular}{|c|c|c|c||c|c|}
\hline 
$\zeta$&
$V_{0}$&
$F$&
$\alpha$&
$F/V_{0}$&
$\gamma$\tabularnewline
\hline
\hline 
1&
75&
50&
0.5&
0.67&
0.355\tabularnewline
\hline 
2&
100&
75&
0.3&
0.75&
0.319\tabularnewline
\hline 
3&
125&
100&
0.1&
0.80&
0.142\tabularnewline
\hline 
4&
150&
125&
0.08&
0.83&
0.142\tabularnewline
\hline 
5&
175&
150&
0.02&
0.86&
0.042\tabularnewline
\hline
\end{tabular}\end{center}
\caption{\label{table_comp3}Portfolio distribution, where $F$ is the face
  value and $V_0$ is the initial value of the company. The fraction of the $K$ companies that
belongs to a certain catagory is $\alpha$ and the fraction of
the total amount of money invested in that category is denoted $\gamma$.}
\end{table}

Next it has to be decided which company should be removed from the
portfolio. A ranking system is established with respect to the product
of the default probability and the mean loss 
\begin{equation}
R_{k}=P_{Dk}\mu_{k}\label{ranking_companies}
\end{equation} 
for company $k$. The company with the highest value of $R_{k}$
is considered to be the worst, and is the one which is removed. By
using this ranking system one of the companies in category five will
be removed, since they have the highest default probability and
expected loss in the entire portfolio.

\begin{figure}
\begin{center}\resizebox{85mm}{!}{\includegraphics{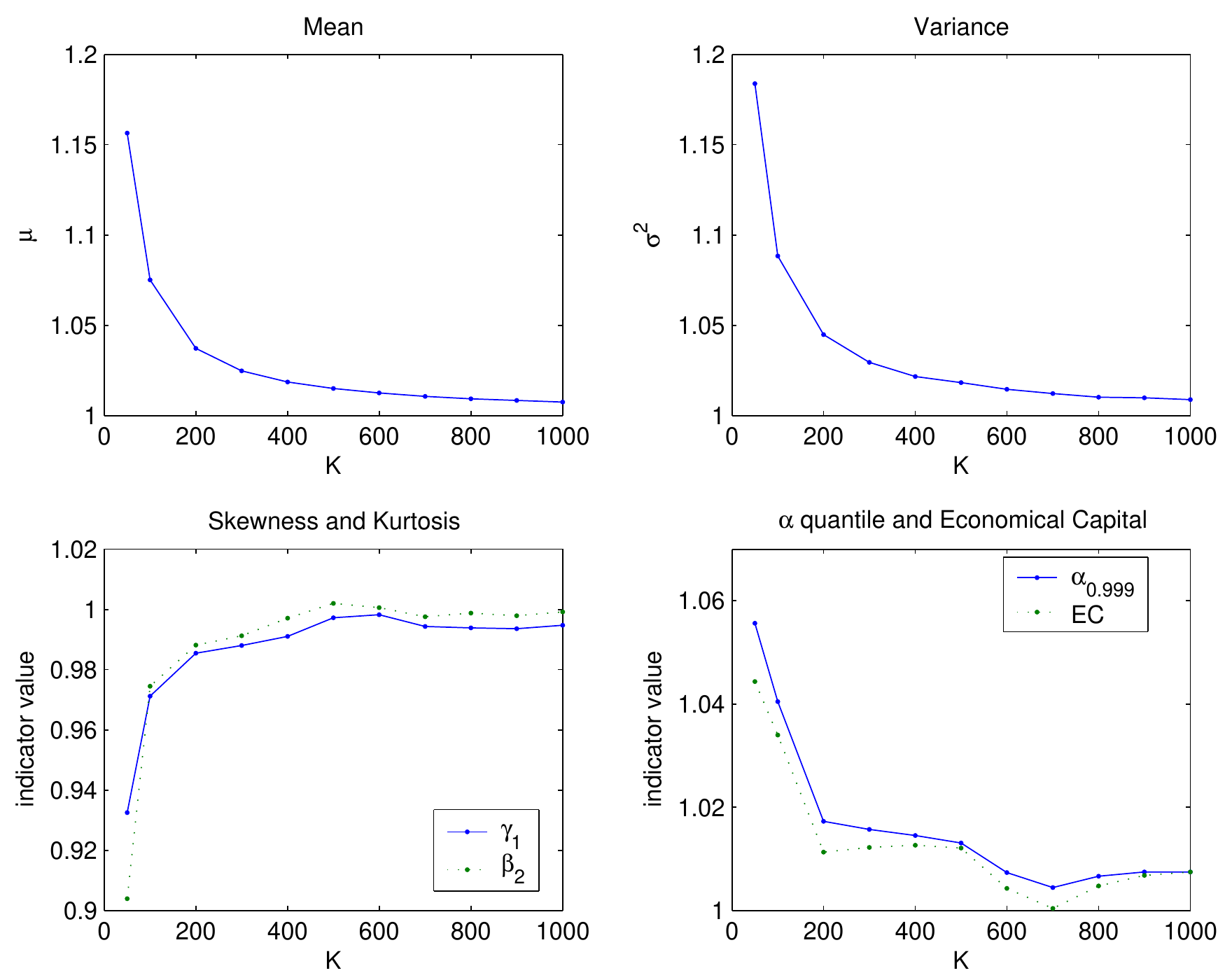}}\end{center}
\caption{\label{cap:indicators_drilldown2}The ratio between the different
indicators as a function of the portfolio size $K$. The
indicators are calculated from the loss distribution using a portfolio
generated by the rule set in Tab.~(\ref{table_comp3}). The ratios
are calculated as $x^{(K)}/x^{(K-1)}$, where $x$ is the current
indicator in focus.}
\end{figure}

The different indicators can be seen in Fig.~(\ref{cap:indicators_drilldown2}),
as a function of $K$. 
Naturally, smaller portfolios are more sensitive to the pruning
procedure. For example, when going from
$K=50$ to $K=49$, the $EL$ is reduced by about 16\% and the $UL$
is reduced by an enormous 18\%! The same comparison when going from
$K=1000$ to $K=999$ is less dramatic but none the less quite significant.
In this case the $EL$ is reduced by 0.75\% and the $UL$ by 0.88\%.

The shape of the loss distribution is also affected slightly when
removing only one single company. Both the skewness and kurtosis are
\emph{increased}, when going from $K$ to $K-1$ companies. On the
other hand, both the 99.9:th percentile $\alpha_{0.999}$ and the economical
capital $EC$ are \emph{decreased}.

\subsubsection{Drill down risk in the presence of correlations}

How are the moments, and other indicators changed when there are correlations
involved? In this section the portfolio, composed by the rule set in Tab.~\ref{table_comp3}, is divided into ten different branches. Each branch is correlated
with correlation strength one half ($C_{b}=0.5$). Note that a branch is not equivalent to a
category. The concept of categories is developed as a tool to generate
portfolios. However, companies which belong to the same branch are generally from different categories.

\begin{figure}
\resizebox{85mm}{!}{\includegraphics{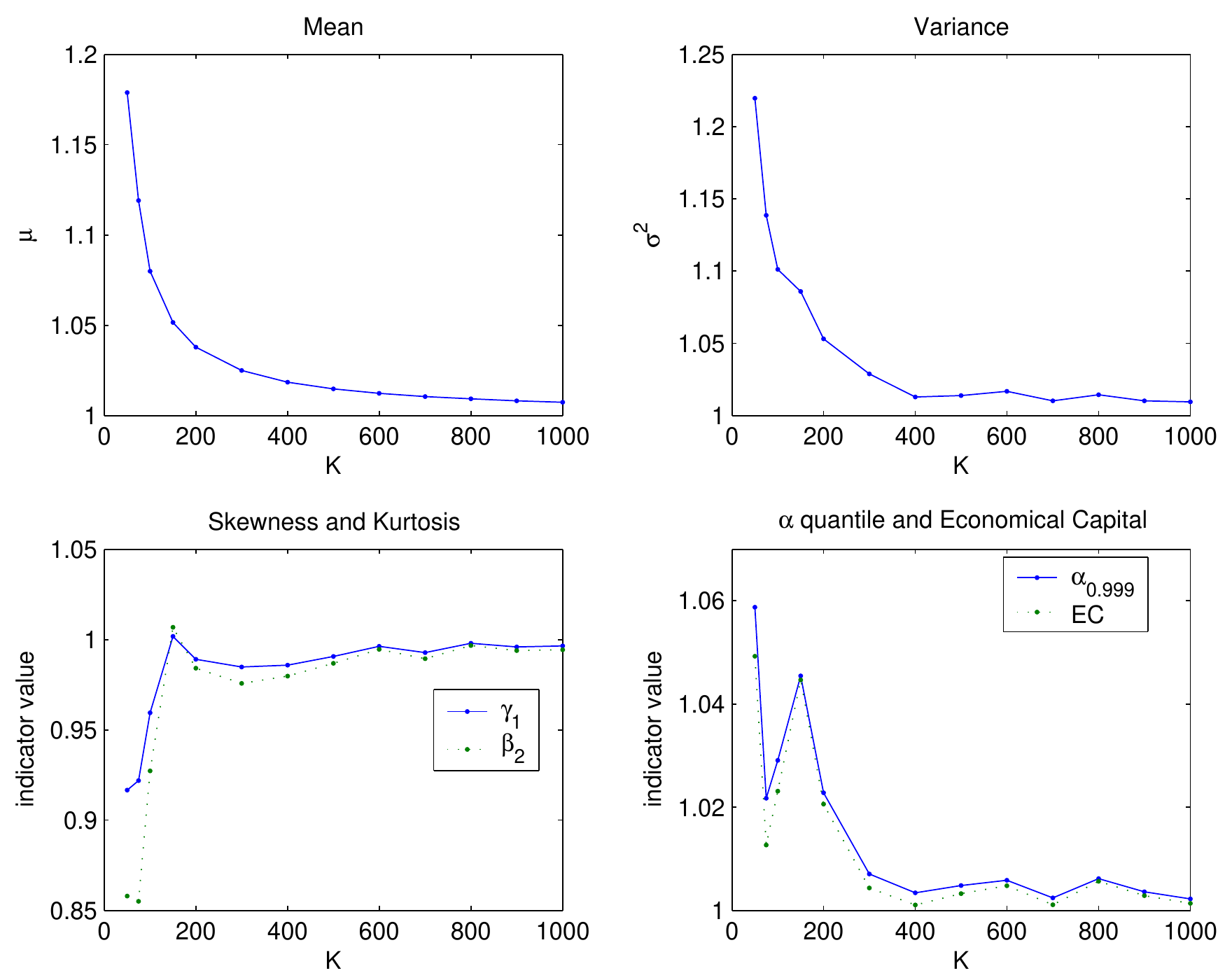}}

\caption{\label{cap:indicators_drilldownCorrPort3}
The ratio between the different
indicators as a function of the portfolio size $K$. The
companies were divided into $B=10$ branches and with correlation strength $C=0.5$ The indicators are calculated from the loss distribution using a portfolio
generated by the rule set in Tab.~\ref{table_comp3}. The ratios are
calculated as $x^{(K)}/x^{(K-1)}$, where $x$ is the current indicator in focus.}
\end{figure}

The companies are assigned to the branches randomly for each iteration. 
This assures that, for example, two of the high
risk companies in the portfolio will not be correlated with each other every time the simulation is run. 
In Fig.~\ref{cap:indicators_drilldownCorrPort3} 
the indicators are plotted as a function of the portfolio size $K$.
For a portfolio of size $K=50$ the $EL$ is improved by about
18\%, and the $UL$ by 22\%. For a large portfolio with $K=1000$
companies one finds the improvement in $EL$ to be 0.74\% and about 0.72\% for the $UL$.
The skewness and kurtosis decrease slightly, and the $\alpha_{0.999}$-quantile and the economical
capital $EC$ are both improved. 

\subsubsection{Drill down risk in the presence of jumps}

Finally, we examine how the jump term $J(t)$ alters the pruning
of a portfolio. To make this effect as clear as possible the correlations
between the bonds are turned off. The jump probability is set to
$\lambda=0.01$, along with the parameters $\mu_{J}=-0.4$ and $\sigma_{J}=0.3$.

\begin{figure}
\begin{center}\resizebox{85mm}{!}{\includegraphics{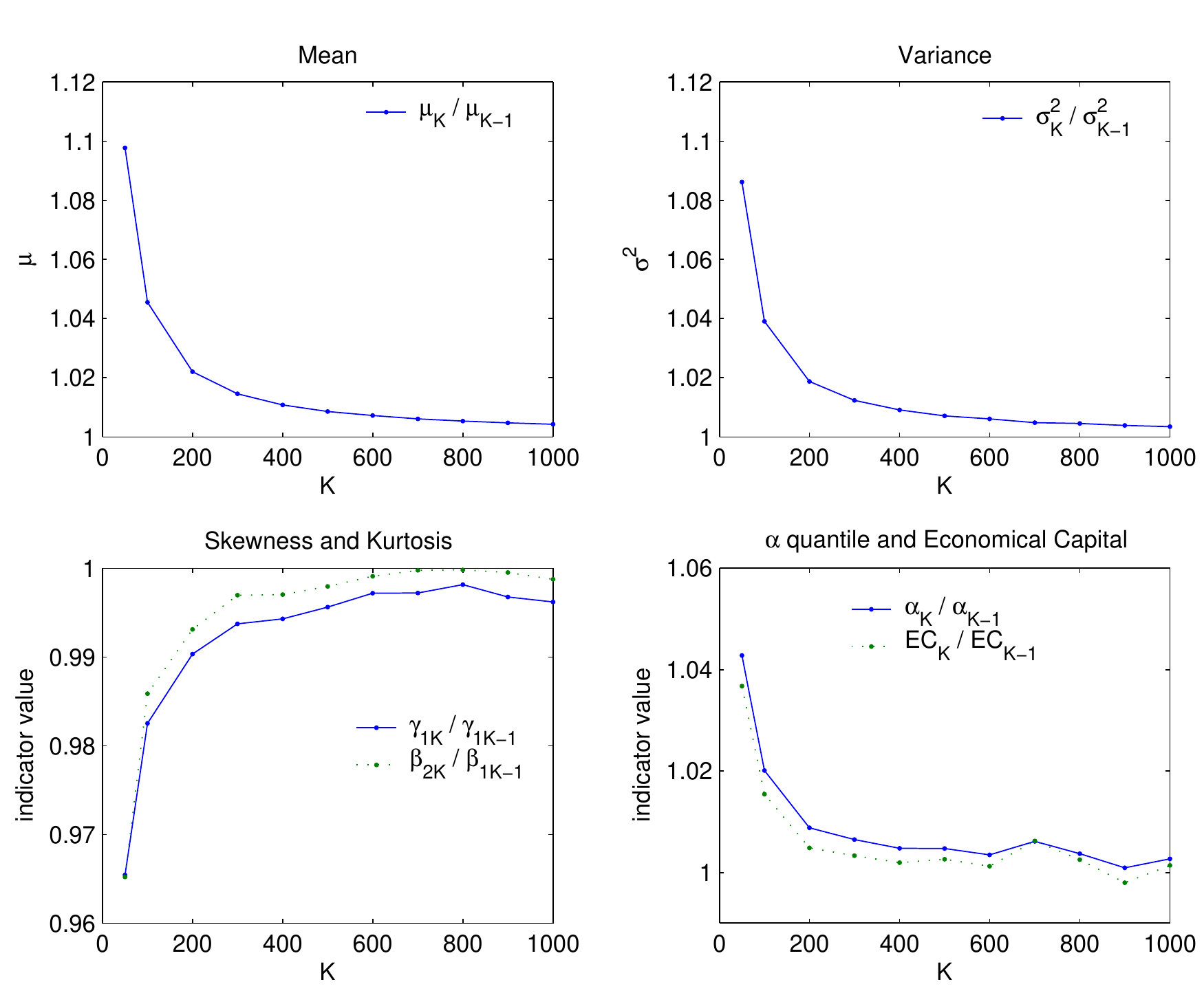}}\end{center}
\caption{\label{cap:indicators_drilldownJumpPort3}The ratio between
 the different indicators as a function of the portfolio
 size $K$, when the jump term $J(t)$ is included. The parameters for
 the jump term are $\sigma_{J}=0.3$, $\mu_{J}=-0.4$ and
 $\lambda=0.01$. The portfolio was generated by the rule set
 in Tab.~\ref{table_comp3}. The ratios are calculated as
 $x^{(K)}/x^{(K-1)}$, where $x$ is the current indicator in focus.} 
\end{figure}

In this simulation, all bonds have the same parameters for the jump
term. The portfolio is again generated by the rule set in Tab.~\ref{table_comp3}. 
In Figure~\ref{cap:indicators_drilldownJumpPort3} the various indicators are shown as a function of the portfolio size $K$.

The benefit of removing the worst company from the portfolio is quite impressive. 
In these settings the $EL$ is reduced by
10\%, when having a portfolio with original size $K=50$. The $UL$ is improved by 9\% for the same portfolio size. Even though these improvements are large,
they are still somewhat lower than in the case without jumps. 
The jump term tends to smear out the differences between the
companies in the portfolio, making them all behave worse. This means
that the company which is removed doesn't stand out as much as it
otherwise would. This fact is reflected in the ratio between the indicators seen in Fig.~\ref{cap:indicators_drilldownJumpPort3}.

As in the previous studies the skewness and kurtosis will increase
when removing one company, whereas the $\alpha_{0.999}$-quantile and
the $EC$ will improve by as much as four percent for portfolios of
initial size $K=50$.

\section{Summary and conclusions}
\label{sec:sumcon}

The microscopic and dynamical character make structural credit risk
models well suited for a variety of different applications in physics
and complex systems.
We formulated a structural model for credit risk which includes jumps
in the stochastic processes for the asset prices and correlations
between them. As a first step, we solved a simplified version of the
model, i.e.~without jumps and correlations, analytically 
and obtained an asymptotic approximation.  Thereafter,
we numerically evaluated loss distributions and the corresponding
expected loss EL and the unexpected loss UL. To this end, we performed
detailed Monte Carlo simulations for the full model.  For theoretical
and practical purposes, the asymmetric and leptokurtic character of
the loss distribution is highly important.  Hence, we carefully
investigated how the shape of the loss distribution depends on the
various model parameters.  First, we illustrated that a large
portfolio is less risky than a small one, in accordance to a basic
concept of investment theory known as diversification. Second, we
demonstrated the influence of growing maturity and the interplay between drift and volatility. 

Having presented the basic features of our model, we turned to more
advanced issues which are not only of theoretical interest, but might
also be of direct relevance for practitioners.  The portion of the
company's assets which is financed by loans, the leverage, affects the
default probability and the losses rather dramatically. A high
leverage induces a high default probability and large losses. Since
the default probability strongly depends on the leverage, the
sensitivity of the loss distribution to the leverage considerably grows with the portfolio size.

Taking the jumps into account, we demonstrated, first, that an
increase of the jump intensity, the jump size mean as well as the jump
size standard deviation induced larger losses and higher default
probabilities. Second, we showed that the impact of the jump on the
loss distribution does not depend on the portfolio size in the same
way as, for example, on the drift and the volatility.

Correlations heavily influence the loss distribution and the default
probability. Although the expected loss is
independent of the branch correlation, the correlations do affect the
features of the loss distribution.  We carefully illustrated how the
correlations hamper a convergence to a symmetric Gaussian--type--of
shape and showed that the kurtosis excess develops a
maximum. Moreover, the impact of the correlation structure depends
significantly on the size of the correlations. Strong correlations
make a branch act like a single company.  We also showed that a small,
highly correlated branch gives essentially the same model outcome as a
large branch with a modest correlation. 

There is a subtle interplay between jumps and correlations. The
uncorrelated jumps can partly neutralize the
correlations. Economically, it is also meaningful to include correlations between the jumps themselves. We have demonstrated that correlated jumps have a pronounced effect on the loss distribution, leading to extremly fat tails.

We investigated how the moments of the loss distribution will
be influenced when adding a new company to a given portfolio. It was
shown that it is possible to express the moments of the loss
distribution for a portfolio that consists of $K$ companies in terms
of the moments of the $K-1$ sized portfolio.  
In addition, we studied numerically how different indicators, such as the
expected loss, unexpected loss, skewness, kurtosis, economical capital
and the $\alpha_{0.999}$-quantile, behave when removing one company
from a given portfolio. 
We examined three portfolio setups, one with and one without correlations, 
and the third one including jumps.
For all three setups it was found that both the EL and UL was
decreased whereas the skewness and kurtosis was increased. Furthermore, 
the $\alpha_{0.999}$-quantile and the economical capital
decreased. These effects were more pronounced for smaller portfolios.

When comparing the ratios of the different indicators used, it was
found that, in general, correlations~\emph{ increase} the indicator
ratios, whereas jumps tend to~\emph{decrease} the indicator ratios.

\begin{acknowledgments}
We thank P.~Jochumzen and A.~M\"uller--Groeling for fruitful 
discussions and helpful comments. We acknowledge financial support from Det Svenska Vetenskapsr\aa det. RS is grateful for support from Deutsche Forschungsgemeinschaft under grant no.~SCHA 1462/1-1.
\end{acknowledgments}

%\appendix*
\appendix
\section{Loss distribution for a homogenous portfolio}
\label{appendix}

We express the distribution~(\ref{pfdis}) in terms of its Fourier
transform,
\begin{equation} \label{pfchr}
p(L) = \frac{1}{2\pi}\int\limits_{-\infty}^{+\infty} 
           r(\omega) \exp(- iL \omega) d\omega \, ,
\end{equation}
where
\begin{eqnarray} \label{eq:pport}
r(\omega) &=& \prod_{k=1}^{K}
\int\limits_{0}^{1} dL_k p_k(L_k) \nonumber \\
  & & \int\limits_{-\infty}^{+\infty} dI_k \tilde{p}_k(I_k) 
      \exp\left(i\omega L_k I_k/K\right) \ .
\end{eqnarray}
is referred to as characteristic function.  If we assume that the face
values and the parameters of the geometric Brownian motion are the
same for all companies $k$, the distributions $p_k(L_k)$ for the individual
losses become the same for all companies. This also holds then for the
distributions $\tilde{p}_k(I_k)$ of the default indicators. Hence, we find
\begin{eqnarray}
r(\omega) &=& \prod_{k=1}^{K}
                         \int\limits_{0}^{1} dL_k p_k(L_k) \nonumber \\ 
                     & & \left((1-P_D)
                + P_D\exp\left(i\omega L_k/K\right)\right) \nonumber \\
  &=& \Biggl(\int\limits_{0}^{1} dL_k p_k(L_k) \bigl((1-P_D)\bigr. + \Biggr. \nonumber \\
  & & \qquad\qquad 
            \Biggl. \bigl. P_D \exp\left(i\omega L_k/K\right) \bigr) \Biggr)^K 
                                          \nonumber \\
  &=& \exp\bigl(K\ln\big((1-P_D)\bigr.\bigr. \nonumber \\
  & & \qquad\qquad  \bigl.\bigl. +P_DQ(\omega/K)\bigr)\bigr)
\end{eqnarray}
with
\begin{equation}\label{qchr}
Q(\omega/K) = \int\limits_{0}^{1} dL_k p_k(L_k) \exp\left(i\omega L_k/K\right) \ .
\end{equation}
As we aim at an approximation for large $K$, we expand the exponential
to obtain an asymptotic series in $1/K$.  Up to third order we have
\begin{eqnarray}
Q(\omega/K) &=& 1+\frac{i\omega}{K}\langle L_k \rangle
 - \frac{\omega^2}{2K^2}\langle L_k^2 \rangle \nonumber \\
  & & \qquad - \frac{i\omega^3}{6K^3}\langle L_k^3 \rangle
                             + \mathcal{O}(1/K^4) \ ,
\end{eqnarray}
where $\langle L_k^n \rangle$ is the $n$--th moment~(\ref{momres}) of
the distribution $p_k(L_k)$. This yields
\begin{eqnarray}\label{apprend}
&& K\ln\left((1-P_D)+P_DQ(\omega/K)\right) \nonumber \\
&& \quad =  i \omega P_D\langle L_k \rangle - \frac{\omega^2}{2K} \left(
P_D \langle L_k^2 \rangle - P_D^2 \langle L_k \rangle^2  \right)
\nonumber \\
& & \qquad \quad - \frac{i\omega^3}{6K^2} \bigl( P_D \langle L_k^3 \rangle +
3P_D^2 \langle L_k \rangle \langle L_k^2\rangle \bigr. \nonumber \\
& & \qquad \quad \bigl. +2P_D^3 \langle L_k \rangle^3 \bigl) + \mathcal{O}(1/K^3) \ .
\end{eqnarray}
Collecting everything, we arrive at the
expression~(\ref{eq:analyticportapprox}).

\section{Loss distribution for an inhomogenous portfolio}
\label{appendix2}

We generalize approximation~(\ref{eq:analyticportapprox}) to the case of an inhomogenous portfolio with varying face values $F_k$ and individual loss distributions $p_k(L_k)$. To this end we write the portfolio loss as
\begin{equation}
L=\frac{\sum_{k=1}^{K} F_k L_k I_k}{\sum_{k=1}^{K}F_{k}}
=\sum_{k=1}^{K} \gamma_k L_k I_k
 \, ,\label{eq:LossDef}
\end{equation}
where $\gamma_k=F_k / \sum_{j=1}^{K}F_{j}$ is the fraction of money invested in obligor $k$. 
Following App.~\ref{appendix}, the characteristic function is then
\begin{equation} \label{eq:charf}
r(\omega) =  \prod_{k=1}^{K}
                         \Biggl( (1-P_{D,k}) + P_{D,k} Q_k(\omega) \Biggr) \; ,
\end{equation}
with
\begin{equation}
Q_k(\omega)=\int\limits_{0}^{1} dL_k p_k(L_k) \exp\left(i\omega \gamma_k L_k \right) \; .
\end{equation}
Expanding the exponential in $Q_k(\omega)$ yields
\begin{eqnarray}
Q_k(\omega) &=& 1+i \omega \gamma_k \langle L_k \rangle
 - \frac{1}{2} \omega^2 \gamma_k^2 \langle L_k^2 \rangle \nonumber \\
  & & \qquad - \frac{i}{6} \omega^3 \gamma_k^3 \langle L_k^3 \rangle
                             + \mathcal{O}(\gamma_k^4) \ ,
\end{eqnarray}
which finally leads to
\begin{eqnarray} \label{eq:analyticportapprox2} 
p(L) & \approx &
\frac{1}{2\pi} \int\limits_{-\infty}^{+\infty}d\omega 
\exp\left(-i\omega\left(L- \sum\limits_{k=1}^K \gamma_k P_{D,k} \langle L_k \rangle\right)\right) \nonumber \\
& & \quad \exp\left(- \frac{\omega^2}{2}  \sum\limits_{k=1}^K \gamma_k^2  \left( P_{D,k} \langle L_k^2 \rangle
             - P_{D,k}^2 \langle L_k \rangle^2 \right)\right) \nonumber \\
& & \quad \exp\biggl(-\frac{i\omega^3}{6}  \sum\limits_{k=1}^K 
              \gamma_k^3 \bigl(P_{D,k} \langle L_k^3\rangle \bigr. \biggr. \nonumber \\
& & \qquad \biggl. \bigl. + 3P_{D,k}^2 \langle L_k \rangle \langle L_k^2\rangle 
                  + 2P_{D,k}^3 \langle L_k \rangle^3\bigr)\biggr)
\end{eqnarray}
as a generalized approximation for the loss distribution.

\section{Drill down risk}
\label{drillDownRiskAppendix}
We want to express the moments of the loss distribution for a
portfolio which consists of $K$ companies in terms of the moments of a
portfolio which consists of $K-1$ companies. The loss of an individual company is
given by Eq.~(\ref{lossDef2}) and the total loss for a portfolio containing $K$ companies by Eq.~(\ref{totalLossDef}).
To obtain the distribution of the
entire portfolio loss, one needs to average over all
distributions
of the individual losses $p_{k}(L_{k})$, and over the indicator
distribution
$\widetilde{p}_{k}(I_{k})$. This is done in the following way, c.f.
Eq.~(\ref{pfdis}):

\begin{eqnarray}
p^{(K)}(L^{(K)}) & = & \int_{-\infty}^{+\infty}dI_{1}\widetilde{p}_{1}(I_{1})\cdots\int_{-\infty}^{+\infty}dI_{K}\widetilde{p}_{K}(I_{K})\nonumber \\
 & \times & \int_{0}^{F_{1}}d\Gamma_{1}p_{1}(\Gamma_{1})\cdots\int_{0}^{F_{K}}d\Gamma_{K}p_{K}(\Gamma_{K})\nonumber\\
 & \times &
 \delta\left(L^{(K)}-\frac{\sum_{j=1}^{K}\Gamma_{j}I_{j}}{\sum_{j=1}^{K}F_{j}}\right),\label{lossGivenDistrDef}
 \end{eqnarray}
where the upper index $(K)$ indicates that the total number of companies
that are averaged over is $K$. 

The moments of the loss distribution are defined as
\begin{equation}
\left\langle \left(L^{(K)}\right)^{n}\right\rangle
_{K}=\int_{0}^{{1}}\left(L^{(K)}\right)^{n}p^{(K)}(L^{(K)})dL^{(K)},\label{momentDef2}
\end{equation}
where $L^{(K)}$ is the loss for a portfolio containing $K$ companies,
$p^{(K)}(L^{(K)})$ is the corresponding probability density function
and $n$ is the order of the moment. The sub index $K$ in the bracket
indicates that the distribution for $K$ companies, $p^{(K)}$, is
used. Plugging equation~(\ref{lossGivenDistrDef})
into~(\ref{momentDef2}) we find
\begin{eqnarray}
\left\langle \left(L^{(K)}\right)^{n}\right\rangle _{K} & = &
 \int_{0}^{{1}} dL^{(K)}   \left(L^{(K)}\right)^{n}   \label{LK_moment0}
 \\
 & \times & \int_{-\infty}^{+\infty}dI_{1}\widetilde{p}_{1}(I_{1})\cdots\int_{-\infty}^{+\infty}dI_{K}\widetilde{p}_{K}(I_{K})\nonumber \\
 & \times & \int_{0}^{F_{1}}d\Gamma_{1}p_{1}(\Gamma_{1})\cdots\int_{0}^{F_{K}}d\Gamma_{K}p_{K}(\Gamma_{K})\nonumber \\
 & \times &
 \delta\left(L^{(K)}-\frac{\sum_{j=1}^{K}\Gamma_{j}I_{j}}{\sum_{j=1}^{K}F_{j}}\right).
 \nonumber
\end{eqnarray}

Integration over $dL^{(K)}$ yields
\begin{eqnarray}
\left\langle \left(L^{(K)}\right)^{n}\right\rangle _{K} & = &
 \left(\frac{\sum_{j=1}^{K}\Gamma_{j}I_{j}}{\sum_{j=1}^{K}F_{j}}\right)^{n}\label{LK_moment1} \\
 & \times & \int_{-\infty}^{+\infty}dI_{1}\widetilde{p}_{1}(I_{1})\cdots\int_{-\infty}^{+\infty}dI_{K}\widetilde{p}_{K}(I_{K})\nonumber \\
 & \times &
 \int_{0}^{F_{1}}d\Gamma_{1}p_{1}(\Gamma_{1})\cdots\int_{0}^{F_{K}}d\Gamma_{K}p_{K}(\Gamma_{K}).
 \nonumber
\end{eqnarray}
We introduce the sum over all the face values
\begin{equation}
F^{(K)}=\sum_{j=1}^{K}F_{j}  \label{F_KDefAppendix}
\end{equation}
and rewrite the sum in Eq.~(\ref{LK_moment1}) as
\begin{equation}
\frac{\sum_{j=1}^{K}\Gamma_{j}I_{j}}{F^{(K)}}=\frac{\Gamma_{K}I_{K}}{F^{(K)}}+\frac{F^{(K-1)}}{F^{(K)}}\underbrace{\frac{1}{F^{(K-1)}}\sum_{j=1}^{K-1}\Gamma_{j}I_{j}}_{\equiv
  L^{(K-1)}} \, .   \label{sum_rewrite}
\end{equation}
Using the binomial theorem 
to rewrite the $n$:th power sum, we find
\begin{eqnarray}
\left\langle \left(L^{(K)}\right)^{n}\right\rangle _{K} & = & \left(\frac{F^{(K-1)}}{F^{(K)}}\right)^{n}\nonumber \\
 & \times & \int_{-\infty}^{+\infty}dI_{1}\widetilde{p}_{1}(I_{1})\cdots\int_{-\infty}^{+\infty}dI_{K}\widetilde{p}_{K}(I_{K})\nonumber \\
 & \times & \int_{0}^{F_{1}}d\Gamma_{1}p_{1}(\Gamma_{1})\cdots\int_{0}^{F_{2}}d\Gamma_{K}p_{K}(\Gamma_{K})\nonumber \\
 & \times & \sum_{\nu=0}^{n}\left(
\begin{array}{c}
n\\
\nu
\end{array}
\right)\left(\frac{1}{F^{(K-1)}}\sum_{j=1}^{K-1}\Gamma_{j}I_{j}\right)^{n-\nu}\nonumber \\
& \times
&\left(\frac{I_{K}\Gamma_{K}}{F^{(K-1)}}\right)^{\nu}.\label{LK_moment2}
\end{eqnarray}
This can now be expressed in terms of the moments of the corresponding
$K-1$~distribution, $\left\langle \left(L^{(K-1)}\right)^{n}\right\rangle _{K-1}$, and the moments of the new distributions, $\widetilde{p}_{K}(I_{K})$ and $p_{K}(\Gamma_{K})$, in the following way
\begin{eqnarray}
\left\langle \left(L^{(K)}\right)^{n}\right\rangle _{K} & = & \left(\frac{F^{(K-1)}}{F^{(K)}}\right)^{n}\nonumber \\
 & \times & \sum_{\nu=0}^{n}\left(\begin{array}{c}
n\\
\nu\end{array}\right)\left\langle
 \left(L^{(K-1)}\right)^{n-\nu}\right\rangle _{K-1}\nonumber\\
& \times & \frac{\left\langle I_{K}^{\nu}\right\rangle \left\langle \Gamma_{K}^{\nu}\right\rangle
 }{(F^{(K-1)})^{\nu}} \; ,    \label{LK_moment3}
\end{eqnarray}
where $\nu$ is an integer. It is important to remember which distribution
the bracket notation $\left\langle \cdots\right\rangle $ refers to.
The moments of $I_{K}$ and $\Gamma_{K}$ are taken over the distributions
$\widetilde{p}_{K}(I_{K})$ and $p_{K}(\Gamma_{K})$ respectively,
and the moments of $L^{(K-1)}$ are taken over the distribution $p^{(K-1)}(L^{(K-1)})$.


%\bibliography{credit_paper_refs}

\end{document}